%% file: ssrp.tex
\documentclass[10pt]{article}
\usepackage[utf8]{inputenc}
\usepackage[a4paper, total={6.7in, 9.2in}]{geometry}
\usepackage[english]{babel}
\usepackage{csquotes}
\usepackage[numbers]{natbib}
\usepackage{subfiles}
\usepackage{lmodern}
\usepackage{amsmath}
\usepackage{amsfonts}
\usepackage{amssymb}
\usepackage{caption}
\usepackage{hyperref}
\usepackage{float}
\usepackage{amsthm}
\usepackage{mathtools}
\usepackage{tikz}
\usepackage{titlesec}
\usepackage{thmtools, thm-restate}
\usepackage{comment}
\usepackage{algorithm}
\usepackage[noend]{algpseudocode}
\usepackage{algorithmicx}
\usepackage{float}
\newcommand{\kindaSqrtV}{\sqrt{n_H}}

\newtheorem{theorem}{Theorem}[section]
\newtheorem{claim}{Claim}[section]
\newtheorem{lemma}{Lemma}[section]
\theoremstyle{definition}
\newtheorem{definition}{Definition}[section]
\newtheorem{observation}{Observation}[]

\titlespacing\subsubsection{0pt}{0pt}{1pt}

\usetikzlibrary{arrows.meta}
\usetikzlibrary{backgrounds}
\usetikzlibrary{decorations.pathreplacing}
\tikzstyle{marked edge} = [draw,line width=7.5pt,-,red!40]

\newcounter{step}
\newenvironment{step}[1][]{
    \refstepcounter{step}\par\medskip\noindent
    \textbf{\normalsize Step~\thestep: #1 \newline}
    }{\medskip}

\DeclareCaptionFormat{algor}{
  \hrulefill\par\offinterlineskip\vskip1pt
    \textbf{#1#2}#3\offinterlineskip\hrulefill}
\DeclareCaptionStyle{algori}{singlelinecheck=off,format=algor,labelsep=space}
\newcommand{\partTitle}[1]{\State{\hrulefill}\State{\textbf{#1}}\\}

\newcommand{\snl}{\vskip 4pt}
\newcommand{\NL}{\vskip 10pt}
\newcommand{\mynewpage}{}

\newcommand{\mathtitlewrapper}[1]{\texorpdfstring{#1}{Lg}}
\newcommand{\depart}[3][{{s}}]{\textup{Depart}({#1}, {#3} , {#2})}
\newcommand{\pivot}[2]{\textup{Pivot}({{s}}, {#2},  {#1})}
\newcommand{\etal}{\textit{et al.}}

\newcommand{\N}{\mathbb{N}}
\newcommand{\simple}{$K$-simple}
\newcommand{\lgV}{\lfloor\log(n_H)\rfloor}
\newcommand{\sqrtV}{\lfloor\sqrt{n_H}\rfloor}
\newcommand{\Nfty}{\N\cup\{\infty\}}
\newcommand{\whp}{\text{w.h.p}}
\newcommand{\WLOG}{\text{WLOG}}

\newcommand{\RP}[4]{path from #1 to #2 in the graph #4 $-$ #3}

\newcommand{\R}[5][H]{R(#3,#4,#1_{#2} - #5)}
\newcommand{\Rnm}[3][H]{R(#2,#3,#1)}

\newcommand{\Dhat}[5][H]{\widehat{d}(#3,#4,#1_{#2} - #5)}
\newcommand{\DRZ}[4][H]{\widehat{d}_{RZ}(#2,#3,#1 - #4)}
\newcommand{\D}[5][H]{d(#3,#4,#1_{#2} - #5)}
\newcommand{\Dnm}[3][H]{d(#2,#3,#1)}
\newcommand\eqdef{\mathrel{\stackrel{\makebox[0pt]{\mbox{\normalfont\tiny def}}}{=}}}
\newcommand{\myendalg}{\vskip0pt \noindent\rule{\textwidth}{0.4pt} \vskip 10pt}

\def\dnsparagraph{\vspace{-2pt}\paragraph}

\def\dnsparagraph{\vspace{-2pt}\paragraph}

\def\shiri#1{\textcolor{magenta}{\textcolor{magenta}{{[Shiri: #1]}}}}

\title{Near Optimal Algorithm for the Directed Single Source Replacement Paths Problem}
\author{
    Shiri Chechik
    \footnote{Tel Aviv University, Israel. E-mail:
    \href{mailto:shiri.chechik@gmail.com}{shiri.chechik@gmail.com}}
    \and Ofer Magen
    \footnote{Tel Aviv University, Israel. E-mail:
    \href{mailto:ofermagen98@gmail.com}{ofermagen98@gmail.com}}
}

\begin{document}

\maketitle

\begin{abstract}
In the Single Source Replacement Paths (SSRP) problem we are
given a graph $G = (V, E)$, and a shortest paths tree $\widehat{K}$ rooted at a node $s$, and the goal is to
output for every node $t \in V$
and for every edge $e$ in $\widehat{K}$
the length of the shortest path from $s$ to $t$ avoiding $e$.
\snl
We present an $\tilde{O}(m\sqrt{n} + n^2)$
time randomized combinatorial algorithm for unweighted directed graphs \footnote{
    As usual, $n$ is the number of vertices, $m$ is the number of edges and the $\tilde{O}$ notation
    suppresses poly-logarithmic factors.
}.
Previously such a bound was known in the directed case only for the seemingly easier problem of replacement path where both the source and the target nodes are fixed.

Our new upper bound for this problem matches the existing conditional combinatorial lower bounds.
Hence, (assuming these conditional lower bounds) our result is essentially optimal and completes the picture of the SSRP problem
in the combinatorial setting.
\snl
Our algorithm extends to the case of small, rational edge weights.
We strengthen the existing conditional lower bounds in this case by
showing that any $O(mn^{1/2-\epsilon})$ time (combinatorial or algebraic) algorithm for some fixed $\epsilon >0$
yields a truly subcubic algorithm for the weighted All Pairs Shortest Paths problem (previously such a bound was known only for the combinatorial setting).

\end{abstract}

\thispagestyle{empty}
\newpage
\setcounter{page}{1}

\subfile{sections/definitions}

\subfile{sections/overview}

\subfile{sections/algorithm}

\mynewpage
\subfile{sections/pre_analysis}
\mynewpage
\subfile{sections/DEPART}
\mynewpage
\subfile{sections/DSTE}
\mynewpage

\subfile{sections/PS}

\mynewpage
\subfile{sections/PT}

\mynewpage

\subfile{sections/TT}

\mynewpage

\subfile{sections/SS}

\mynewpage
\subfile{sections/post_analysis}

\mynewpage
\subfile{sections/running_time}

\mynewpage
\subfile{sections/reductions}

\mynewpage
\section*{Acknowledgements}
This publication is part of a project that has received funding from the
European Research Council (ERC) under the European Union’s Horizon 2020 research and 
innovation programme (grant agreement No 803118 UncertainENV)
\bibliographystyle{plain}
\bibliography{ssrp}
\newpage
\section{Appendix}
\subsection*{Figures}

\subfile{sections/appendix_figures}

\subfile{sections/DEPART_proofs}

\mynewpage
\subfile{sections/DSTE_proofs}
\end{document}

%% file: sections/definitions.tex
\section{ Introduction}
In the replacement paths (RP) problem, we are given a graph $G$ and a shortest path $P$ between two vertices $s$ and
$t$ and the goal is to return for every edge $e$ in $P$
the length $\D[G] {} s t e$,
where $G-e$ is the graph obtained by removing the edge
$e$ from $G$, and $\D[G] {} s t e$ is the distance between $s$ and $t$ in the resulted
graph.
In some cases the goal is to provide the shortest path itself
and not only its length.
The interest in replacement path problems stems from the fact that
failures and changes in real world networks are inevitable, and in many cases
 we would like to have a solution or a data structure that can adapt to these failures.
The replacement paths problem is a notable example where we would like to have backup paths
between two distinguished vertices in the event of edge failures.
The replacement paths problem is also very well motivated as it is used as an important
ingredient in other applications such as
the Vickrey pricing of edges owned by selfish
agents from auction theory  \cite{NiRo99,HeSu01}.
Another application of the replacement path problem is finding the $k$ shortest simple paths between a pair of vertices.
The $k$ shortest simple paths problem can be solved by invoking the replacement paths algorithm $k$ times
and adding a very small weight to the path found in each invocation.
The $k$ shortest simple paths problem has many applications by itself \cite{Eppstein98}.
The replacement paths problem has been extensively studied and the literature covers many aspects of this problem with many near optimal solutions in many of the cases
(see e.g. \cite{MMG89,Nardelli2003,NaPrWi01,Roditty2005,Emek08,Klein10,WuNi10,WilliamsRP11,CC19}).

In this paper we consider a natural and important generalization of the replacement paths problem,
referred to as the single source replacement paths (SSRP) problem, which is defined as
follows.
Given a graph $G$
and a shortest paths tree $\widehat{K}$ rooted at
a node $s$,
the SSRP problem is to compute the values of $\D[G] {} s t e$ for every vertex $t \in V(G)$ and for every edge $e \in E(\widehat{K})$.
Note that as the number of edges in $\widehat{K}$
is $n-1$, there are $O(n^2)$ different distances we need to evaluate.
It follows that the size of the SSRP output is $O(n^2)$.

Despite of its natural flavor, the picture of the SSRP problem is not yet complete in many of the cases.
To the best of our knowledge the first paper that considered the SSRP problem is by Hershberger \etal\ \cite{HSB07}
who referred to the problem as edge-replacement shortest paths trees
and showed that in the path-comparison
model of computation of Karger \etal\ \cite{KaKoPh93}, there is a lower bound of $\Omega(mn)$
comparisons in order to solve the SSRP problem for arbitrarily weighted directed graphs.

For the directed weighted case it was shown by Vassilevska Williams and Williams \cite{WW10}
that any truly sub-cubic algorithm for the simpler problem
of RP in \textbf{directed}, arbitrarily weighted graph
admits a truly sub-cubic
algorithm for the arbitrarily weighted All Pairs Shortest Paths (APSP)  problem.
The conditional lower bound from \cite{WW10} holds only for the directed case,
and quite interestingly for the undirected arbitrarily weighted case,
the classical RP problem admits a near linear time algorithm
\cite{MMG89,Nardelli2003,NaPrWi01}.
However, the SSRP problem in undirected graphs
appears to be much harder than the RP problem.
In \cite{CC19} it was shown by Chechik and Cohen that any truly sub-cubic solution for the
SSRP problem in \textbf{undirected} arbitrarily weighted graphs,
admits a truly sub-cubic algorithm for the arbitrarily weighted APSP problem.
Therefore, it seems there is no hope to solve the SSRP problem in weighted graphs,
both in the directed and undirected case.
Meaning that if we seek for truly sub-cubic algorithms for the SSRP problem
we must either consider unweighted graphs or restrict the edge weights in some other way.

One way to restrict the weights is to consider only bounded \textbf{integer} edge weights.
This restriction was considered by  Grandoni and Vassilevska Williams \cite{GW12},
who were also the ones to name this problem the single source replacement paths problem.
Grandoni and Vassilevska Williams \cite{GW12} gave the first non trivial upper bound for the SSRP problem.
They showed that one can bypass the cubic lower bounds
by using fast matrix multiplications and by restricting the weights to be integers in a bounded range.
More precisely, they showed that for graphs with positive integer edge weights in the range $[1,M]$, SSRP can be computed in $\tilde{O}(Mn^\omega)$ time
(here $\omega<2.373$ is the matrix multiplication exponent \cite{williams2012multiplying,LeGall14}).
This matches the current best known bound for the simpler problem of RP for
directed graph with weights $[-M,M]$, by Vassilevska Williams \cite{WilliamsRP11}.
Quite interestingly, for integer edge weights in the range $[-M,M]$,
the authors of \cite{GW12} gave a higher upper bound of
$\tilde{O}(M^{\frac{1}{4-\omega}}n^{2+\frac{1}{4-\omega}})$ time,
which creates an interesting gap between the SSRP problem and the RP problem
for negative integer weights.
Grandoni and Vassilevska Williams \cite{GW12} conjectured that the gap between these two problems
is essential and in fact they conjectured that the SSRP problem with negative weights is as hard
as the directed APSP problem.

The algorithm described in \cite{GW12} uses fast matrix multiplication tricks in order
to break the trivial cubic upper bound,
such algorithms are known as "algebraic algorithms".
Algorithms that do not use any matrix multiplication tricks
are known as "combinatorial algorithms".
The interest in combinatorial algorithms mainly stems from the assumption that in practice
combinatorial algorithms are much more efficient since
the constants and sub-polynomial factors hidden in the matrix multiplication bounds
are considered to be very high.

The SSRP problem was also recently considered in the combinatorial setting by Chechik and Cohen in \cite{CC19} for undirected unweighted graphs.
Specifically, Chechik and Cohen in \cite{CC19} gave an $\tilde{O}(m\sqrt{n} + n^2)$ time randomized algorithm for SSRP in undirected unweighted graphs.
Moreover, using conditional lower bounds Chechik and Cohen also showed that under some reasonable assumptions any combinatorial
algorithm for the SSRP problem in unweighted undirected graphs requires  $\tilde{\Omega}(m\sqrt{n})$ time.

Since there is little hope to solve the weighted case, the only missing piece in the picture of combinatorial SSRP
is the case of directed unweighted graphs.

For the directed unweighted case it was shown earlier by Vassilevska Williams and Williams \cite{WW10},
using a conditional combinatorial lower bound that under some reasonable assumptions any combinatorial algorithm
for the directed unweighted RP (and hence SSRP) problem
requires  $\tilde{\Omega}(m \sqrt{n})$ time.
For the seemingly easier problem of replacement paths Roditty and Zwick \cite{Roditty2005} showed a near optimal solution of
$\tilde{O}(m\sqrt{n})$ time for directed unweighted graphs.

Note that in the undirected unweighted case there is an essential gap between the RP and the SSRP problems.
A natural question is whether such a gap also exists in the directed unweighted case.
In this paper we show that this is not the case by providing a combinatorial near optimal $\tilde{O}(m\sqrt{n}+n^2)$ time algorithm
for the case of directed unweighted graphs, which up to the $n^2$ factor (that is unavoidable as 
the output itself is of size $O(n^2)$) matches the running time of the algorithm in \cite{Roditty2005} (and also matches the running time of the undirected case in   \cite{CC19}).
We therefore (up to poly-logarithmic factors) complete the picture of combinatorial SSRP.
\NL
Our main result is as follows.

\begin{theorem}
\label{thm:main}
There exists an $\tilde{O}(m\sqrt{n} +n^2)$ time combinatorial algorithm for the SSRP problem on unweighted directed graphs.
Our randomized algorithm is Monte Carlo with a one-sided error, as we always output distances which are at least the exact distances,
and with high probability (of at least $1 -n^{-C}$ for any constant $C>0$) we output the exact distance.
\end{theorem}
Note that for unweighted directed graphs where $m = \tilde{O}(n^{1.5})$
our algorithm runs in $\tilde{O}(n^2)$ time, which is
the time it takes just to output the result.
Namely, in this range of density our algorithm
surpasses the current best algebraic SSRP algorithm \cite{GW12}
(which has a running time complexity of $\tilde{O}(n^\omega)$)
as long as  $\omega > 2$.
\snl
We will note that while we focus on the case of edge failures,
in the directed case
there is a well known reduction showing that
edge failures can be used to simulate vertex failures.
The reduction is as follows,
replace every vertex $v$ with two vertices $v_{\text{in}}$ and  $v_{\text{out}}$,
and connect them by a direct edge $(v_{\text{in}},v_{\text{out}})$.
Then, for every incoming edge $(u,v)$ add the edge $(u,v_{\text{in}})$,
and for every outgoing edge $(v,u)$ add the edge $(v_{\text{out}},u)$.
The failure of the vertex $v$ is now simulated by the failure
of the edge $(v_{\text{in}}, v_{\text{out}})$.
\NL
Our main novelty is in the introduction of a tool which we refer to as \textit{weight functions}.
This tool proved to be very useful in order to apply a divide and conquer approach
and could perhaps be utilized in other related problems.
\subsection{ Rational Weights}
While we describe an algorithm for the
problem of SSRP in unweighted graphs,
our algorithm (much like the directed RP algorithm \cite{Roditty2005})
can be easily generalized to solve the case of
weighted graphs for \textbf{rational} edge weights
in the range $[1,C]$,
for every constant $C \ge 1$,
in the same time complexity.
This is because the only place
our algorithm (and the algorithm from \cite{Roditty2005})
uses the fact that
the graph is unweighted is in the claim that a path of length
$l$ contains $\Theta(l)$ vertices, which is used in order to utilize sampling techniques.
As this is also true for rational weights in the range
$[1,C]$, our algorithm generalizes for this case trivially.

Algebraic algorithms inherently can not perform on graphs with
rational weights.
This is since algebraic algorithms use a reduction from
a problem known as min-plus product \footnote{
    Also known as funny matrix multiplication or distance product,
    see \cite{AGM97,G76}
}
to the problem of matrix product,
and this reduction works only for \textbf{integer} weights.
Since in some use-cases (like the $k$-simple paths problem)
it is very useful to have rational weights,
this shows another potential interest in combinatorial algorithms.
\snl
We note that in order to store \textbf{rational} numbers, we must make some common assumptions
regarding the model of computation.
More specifically, we assume that computing the summation of
$n$ edge weights can be performed
in $\tilde{O}(n)$ time and that all numbers we are dealing with can be stored in one (or $O(1)$) space unit.
A realistic option
is working in a word-RAM model,
and considering only rational edge weights which are of the form
$\frac{m}{2^k}$, where the two integers $m$ and $2^k$ fit in the size of $O(1)$ computer words.
This way, the summation of $O(n)$ numbers also fits in $O(1)$
computer words.
This way of representing rational numbers is
reminiscent of the floating-point representation, that is commonly used in practical
applications.
\snl
In Section \ref{sec:reductions} we show that any algorithm (combinatorial or not)
for the SSRP problem for graphs
with rational edge weights from the range $[1,2)$,
that runs in $O(mn^{1/2-\epsilon})$ time for any fixed $\epsilon> 0$ implies a truly
sub-cubic algorithm for
APSP over graphs with arbitrary integer weights.
The claim is formally stated in Theorem \ref{theorem:lowerbound}.
Previously such a conditional lower bound was only known
for combinatorial algorithms using a reduction from Boolean Matrix Multiplication (see \cite{CC19}).

\section{ Preliminaries}
We will use the following notation:
$\N = \{0,1,2,...\},\N^+ = \{1,2,3,...\}, \forall a\in \N^+: [a] = \{1,2,...,a\}$.
Let $G$ be a weighted directed graph then $E(G)$ denotes the set of edges in $G$ and $V(G)$ the set of nodes.
For a vertex $v$ we say that $v\in G$ if $v\in V(G)$ and for an edge $e$ we say that $e\in G$ if $e\in E(G)$.
Let $u,v \in V(G)$ be two vertices,
we denote by $\Dnm[G] u v$ the distance from $u$ to $v$ in the graph $G$,
and denote by $\Rnm[G] u v$ \textbf{some} shortest path from $u$ to $v$ in $G$.
Let $P \subseteq G$ be a path from $u$ to $v$, we define $|P| = |E(P)| = |V(P)| - 1$.
We also denote the length of $P$ by $d(P)$.
Note that $d(\Rnm[G] u v) = \Dnm[G] u v$.
For a set of edges $A \subseteq E(G)$ we denote the graph $(V(G),E(G)\setminus A)$ by $G-A$.
For an edge $e$ we shortly denote $G-\{e\}$ by $G-e$,
and for a path $P$ we shortly denote $G-E(P)$ by $G-P$.
\snl
We denote by $G^R$ the graph obtained
by reversing the directions of all edges -
that is the graph obtained by replacing each edge $(v,u) \in E(G)$
with the edge $(u,v)$ with the same weight.
Given a sub-graph $H \subseteq G$ we denote by $G[H]$ the sub-graph of $G$ induced by the nodes in $V(H)$.
\snl
Let $P \subseteq G$ be a shortest path from a node $s$ to a node $t$.
Let $u,v \in V(P)$ be two nodes in $P$,
we say that $u$ is \textbf{before} $v$ in $P$ if $\Dnm[G] u t \ge \Dnm[G] v t$
and that $u$ is \textbf{after} $v$ in $P$ if $\Dnm[G] u t \le \Dnm[G] v t$,
For an edge $e=(u,v)\in E(P)$ and a node $a\in V(P)$
we say that $a$ is \textbf{before} $e$ in the path $P$ if $a$ is \textbf{before} $u$ in $P$
and say that $a$ is \textbf{after} $e$ in $P$ if $a$ is \textbf{after} $v$ in the path $P$.
\NL
The following sampling Lemma is a folklore.
\begin{lemma}[Sampling Lemma]
    \label{lemma:sampling}
    Consider $n$ balls of which $R$ are red and $n-R$ are blue.
    Let $C > 0, N > 0$ be two numbers such that $R > C \cdot \ln(N)$.
    Let $B$ be a random set of balls such that each ball is chosen to
    be in $B$ independently at random with probability $\frac{C\cdot \ln(N)}{R}$.
    Then \whp\ (with probability at least $1-\frac{2}{N^c}$)
    there is a red ball in $B$ and the size of $B$ is $\tilde{O}(n/R)$.
\end{lemma}
The following separation Lemma was used extensively
in many divide and conquer algorithms on graphs including
the algebraic SSRP algorithm from \cite{GW12}.
\begin{lemma}[Separator Lemma]
    \label{lemma:seperator}
    Given a tree $K$ with $n$ nodes rooted at a node $s$,
    one can find in $O(n)$ time a node $t$
    that separates the tree $K$ into 2 edge disjoint sub-trees $S,T$
    such that $E(S) \cup E(T) = E(K)$, $V(T) \cap V(S) = \{t\}$
    and $\frac{n}{3} \le |V(T)|,|V(S)| \le \frac{2n}{3}$
\end{lemma}
\WLOG\ we always assume that ${{s}}\in V(S)$,
which implies that $t$ must be the root of $T$.
Note that it might be the case that $t={{s}}$.

\subsection{ The Generalized SSRP Problem}
\label{subsec:problem_def}
We next describe a generalization of the directed-SSRP problem that our algorithm works with.
We start by describing the notation of weight functions, a new concept we developed that 
allows us to compress a lot of information into one recursive call of the algorithm.
In the next section we will give more intuition about the weight functions and this specific generalization.

\begin{definition}[Weight Function]
    Let $G$ be an \textbf{unweighted} directed graph.
    Let ${s} \in V(G)$ be some special source node.
    A function $w : V(G) \rightarrow \Nfty$ is a weight function
    (with respect to the source node ${s}$)
    if $w(v) \ge \Dnm[G] {{s}} v$ for every vertex $v \in V(G)$.
    We refer to this requirement as the weight requirement.
\end{definition}

For a source node $s \in V(G)$ and a weight function $w$
(with respect to the source node ${s}$) 
we define the \textbf{weighted} directed graph $G_w$
by taking the unweighted graph $G$,
and assigning each edge the weight 1.
We then add for every node $v \in V(G)$ the edge $({s},v)$
and assign to it the weight $w(v)$.
Note that $G$ is a sub-graph of $G_w$.
Also, note that by the weight requirement, for every two
nodes $u,v \in V(G) : \Dnm[G] u v = \Dnm [{G_w}] u v$.

The generalized SSRP problem is now defined as follows.
The input consists of the following:

\begin{itemize}
    \item An unweighted directed graph $H$ and a source vertex ${{s}}\in V(H)$
    \item A BFS tree $K$ in $H$ rooted at the source ${{s}}$ ($E(K)\subseteq E(H)$)
    \item A set of weight functions $W$ (with respect to source node ${s}$)
    \item A set of queries $Q\subseteq E(K) \times V(H) \times W$
\end{itemize}

The goal is to output for every $(e,x,w)\in Q$ the distance $\D[H] w {{s}} x e$.
Note that this problem is indeed a generalization of the classic SSRP problem.
In order to solve the SSRP problem on the initial graph $G$ and the BFS tree $\widehat K$,
we simply define a single weight function $w : V(G) \rightarrow \Nfty$
that is defined to be $w\equiv\infty$.
We then invoke our algorithm with the graph $G$,
the BFS tree $\widehat{K}$, the  set of weight functions $W = \{w\}$,
and the query set $Q = E(\widehat{K}) \times V(G) \times W$.
Note that
$G$ and $G_w$ are the same graph in the sense
that for every edge $e \in E(G)$ and destination
$x \in V(G)$ we have that $\D[G] {} s x e = \D[G] w s x e$.
Hence, invoking our algorithm for the generalized SSRP would suffice.
As we will only work with the generalized SSRP problem,
we here after refer to it as the SSRP problem for simplicity.

%% file: sections/overview.tex
\section{Overview}
Our algorithm uses a divide and conquer approach.
Each recursive call works on a different
sub-tree ($K$) of the original BFS tree ($\widehat{K}$),
where both the destination and the edge failure are within this sub-tree
(for the case when the edge failure and the destination are not in the same sub-tree
our algorithm solves this in a non recursive manner to be described later in case 1 of the algorithm overview).
The vertices of the sub-tree $K$ induces a sub-graph $H$ of the original graph $G$.
Denote by $n = |V(G)|, m = |E(G)|, n_H = |V(H)|, m_H = |E(H)|$.
\NL
The first step of our algorithm
is to separate the input BFS tree $K$
into two edge disjoint sub-trees $S$ and $T$ using a balanced tree separator
(see Lemma \ref{lemma:seperator}).
We denote the root of the BFS tree $K$ by $s$.
We assume \WLOG\ that the root of $S$ is $s$ and the root of $T$ is some
node $t$. It might be the case that $s=t$.
We define $P$ as the path from $s$ to $t$ in the BFS tree $K$.
Note that $P \subseteq S$. An illustration of this separation
can be found in Figure \ref{fig:separation} in the appendix.
\snl
Let $K'$ be one of the two sub-trees of $K$ (that is $S$ or $T$).
If a replacement path is fully contained in the graph induced by $K'$ then
simply using the recursive call is enough in order to compute its length.
The more challenging case is when the replacement path contains vertices that are not in $K'$.
\NL
In \cite{GW12} the authors used a somewhat similar divide and conquer approach.
Consider a recursive call on a sub-tree $K'$ and consider the case when the edge failure
and destination node are both in $K'$.
In their algorithm, the authors of \cite{GW12} used sampling techniques
and a truncated version of the algebraic APSP algorithm (as presented in \cite{Zwick2002})
in order to create a compressed version of the subgraph induced over $K'$ (by adding shortcuts between vertices in $K'$),
which (\whp) preserves all information needed in order to compute the
true distance.
\snl
However, in the combinatorial setting,
one cannot use this sort of compression process for several reasons.
Firstly after the first call to the compression step (as described in the algorithm in \cite{GW12}),
the resulted graph could be very dense, maybe even complete.
Since the conditional lower bound of $\tilde{\Omega}({m\sqrt{n} + n^2})$
for a combinatorial SSRP (\cite{CC19,WW10})
depends on the number of edges, we do not want to receive such
dense graphs.
Secondly, as we are in the combinatorial setting, we cannot use fast
matrix multiplication in the compression step, which is a critical part of the algorithm
described in \cite{GW12}.
Lastly, after the compression step the resulted graph is weighted which leaves us with a substantially
more difficult problem. In fact in the combinatorial setting there is no sub-cubic time algorithm
that solves the even seemingly easier problem of
weighted RP
(see \cite{WW10} for conditional lower bounds).
\NL
So in the combinatorial setting we must devise a new,
more restricted, compression technique.
We will essentially show that if we add weighted edges only from the source $s$ to all other vertices,
and restrict the weights to be such that the weight of the edge $(s,v)$ is at least
$\Dnm s v$,
then solving replacement path on such a graph still requires only $\tilde{O}(m_H\sqrt{n_H} + n_H^2)$ time.
We therefore would like to add only edges between $s$ and all other nodes.
However, this quickly proves to be difficult,
and it seems that if we add only weighted edges from $s$
to the compressed graph we either ''under-shoot"
and do not represent all replacement paths,
or we ''over-shoot" and represent replacement paths that does not really exist in
the graph $H-e$ (for some edge failure $e$) - such paths will be called
untruthful paths.
\NL
We have devised a technique to fix the over-shooting.
That is, we give the recursive call weights that may represent untruthful replacement
paths in $H-e$, but we force the recursive call to restrict
the replacement paths it searches for, so we will be able to fix them before the algorithm outputs them,
while maintaining optimality.
The way we do so is by a novel concept we call weight functions.
The idea is that the unweighted graph $H$ will come equipped with a set of functions $W$,
such that every $w \in W$ is a function from $V(H)$ to $\Nfty$ and
for every vertex $v$ it holds that $w(v) \ge \Dnm s v$.
For every weight function $w \in W$ the weighted graph $H_w$ is defined by adding for every
node $v \in V$ the edge $(s,v)$ with weight $w(v)$.
The goal of the algorithm is then outputting $\D w s v e$ for every triplet
$x \in V,w \in W, e \in E(K)$.
By restricting the algorithm to only use a single, specific weight function
we achieve enough "control" to fix the untruthful paths.
In order
to maintain the desired running time it will be critical to keep
the number of weight functions ($|W|$) at most $\tilde{O}(\sqrt{n})$
(where $n$ is the number of nodes of the original graph $G$).

\subsection{Algorithm Overview}
In the remaining of this section we sketch the ideas of our algorithm in high level.
For the sake of simplicity, the algorithm in this section runs in $\tilde{O}(n^{2.5})$
time rather than $\tilde{O}(m\sqrt{n} + n^2)$ time. At the end of this section
we will briefly describe how one can use some simple techniques
to reduce the running time to the near optimal of $\tilde{O}(m\sqrt{n} + n^2)$.
While sketching the algorithm, we also ignore the query set $Q$, as
it is only necessary when reducing the running time of the algorithm to $\tilde{O}(m\sqrt{n} + n^2)$.
So the goal of the algorithm in this section is to estimate $\D w s x e$ \textbf{for every} $e \in E(K) , x \in V(H) \text{ and } w \in W$.
The complete algorithm and proof of correctness can be found in Sections \ref{sec:algorithm} and \ref{sec:correctness} correspondingly.
\NL
In our algorithm we distinguish between a few cases according to where the edge failure and the destination are with respect to $S$ and $T$.
Note that for each edge failure $e$ and destination node $x$ we clearly know
in which case we are.
In each such case we distinguish between different sub-cases according
to different properties of the replacement path.
Clearly we do not know the replacement path a-priori,
meaning that we do not know in which sub-case we are.
So when proving the correctness of our algorithm in Section \ref{sec:correctness},
we show that the estimation created for every sub-case
is always at least the real value of $\D w s x e$, that is, we do not underestimate.
Then we show that for the true sub-case (the sub-case describing the true replacement path)
our estimation matches the true value of $\D w s x e$ \whp.
By returning the minimum estimation from all of the sub-cases
we are guaranteed to return the true distance \whp.
To distinguish between the different sub-cases
we first define two useful characterization of replacement paths in $H_w$.
\begin{restatable}[Weighted paths]{definition}{DefWeightedPaths}
  Let $e \in E(K), x \in V(H), w \in W$, and let
  $R$ be a path from ${s}$ to $x$ in the graph $H_w-e$.
  The path $R$ will be called weighted if it uses some edge
  from $E(H_w) - E(H)$.
  $R$ will be called unweighted if it is fully contained in $H-e$.
\end{restatable}

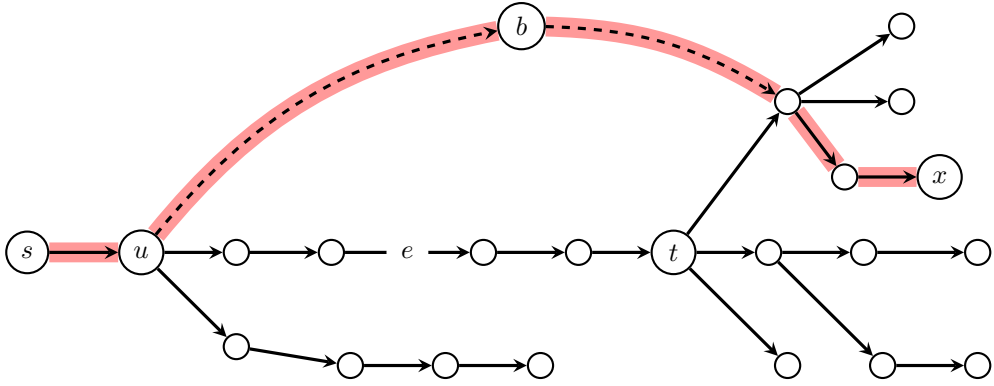
\begin{figure}[H]
  \centering
  \captionsetup{justification=centering}
  \begin{tikzpicture}[scale=1]
  \begin{scope}[every node/.style={circle,thick,draw}]
      \node (V41) at (-1.5,0) {${{{s}}}$} ;
      \node (V42) at (0,0) {$u$} ;
      \node (V43) at (1.25,0) {};
      \node (V44) at (2.5,0) {};
      \node (V445) at (4.5,0) {};
      \node (V45) at (5.75,0) {};
      \node (V46) at (7,0) {$t$};
      \node (V47) at (8.25,0) {};
      \node (V48) at (9.5,0) {};
      \node (V49) at (11,0) {} ;

      \node (V21) at (9.25,1) {} ;
      \node (V22) at (10.5,1) {$x$} ;

      \node (V11) at (8.5,2) {} ;
      \node (V12) at (10,2) {} ;

      \node (V01) at (10,3) {} ;

      \node (V51) at (1.25,-1.25) {};
      \node (V52) at (2.75,-1.5) {};
      \node (V53) at (4,-1.5) {};
      \node (V54) at (5.25,-1.5) {};

      \node (V55) at (8.5,-1.5) {} ;
      \node (V56) at (9.75,-1.5) {} ;
      \node (V57) at (11,-1.5) {} ;

      \node (B) at (5,3) {$b$};

  \end{scope}

  \begin{scope}[>={stealth},
                every edge/.style={draw=black,very thick}]

      \path [->] (V41) edge (V42);
      \path [->] (V42) edge  (V43);
      \path [->] (V43) edge (V44);
      \path [->] (V44) edge  node[style={fill=white,circle}] {$e$} (V445);
      \path [->] (V445) edge  (V45);

      \path [->] (V45) edge (V46);
      \path [->] (V46) edge  (V47);
      \path [->] (V47) edge (V48);
      \path [->] (V48) edge (V49);

      \path [->] (V46) edge (V11);
      \path [->] (V11) edge (V21);
      \path [->] (V21) edge (V22);

      \path [->] (V11) edge (V12);
      \path [->] (V11) edge (V01);
      \path [->] (V46) edge (V55);

      \path [->] (V47) edge (V56);
      \path [->] (V56) edge (V57);

       \path [->] (V42) edge (V51);
       \path [->] (V51) edge (V52);
       \path [->] (V52) edge (V53);
       \path [->] (V53) edge (V54);

      \path [->] (V42) edge[bend left=20,dashed] (B);
      \path [->] (B) edge[bend left=15,dashed] (V11);

  \begin{scope}[on background layer,every edge/.style=marked edge]

      \path (V41) edge (V42);
      \path (V42) edge[bend left=20] (B);
      \path (B) edge[bend left=15] (V11);
      \path (V11) edge (V21);
      \path (V21) edge (V22);

  \end{scope}

  \end{scope}
  \end{tikzpicture}
  \caption{Departing replacement path in $H$}
  \label{fig:departing}
\end{figure}

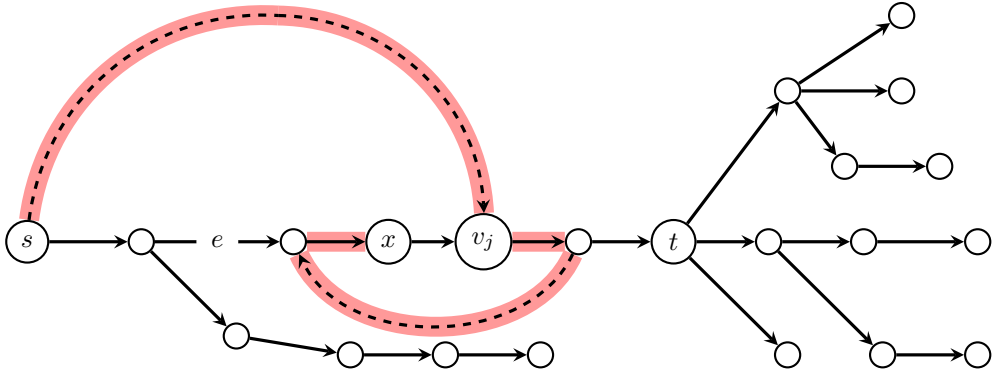
\begin{figure}[H]
  \centering
  \captionsetup{justification=centering}
  \begin{tikzpicture}
  \begin{scope}[every node/.style={circle,thick,draw}]
  \node (V41) at (-1.5,0) {${{s}}$} ;
  \node (V42) at (0,0) {} ;
  \node (V43) at (2,0) {};
  \node (V44) at (3.25,0) {$x$};
  \node (V445) at (4.5,0) {$v_j$};
  \node (V45) at (5.75,0) {};
  \node (V46) at (7,0) {$t$};
  \node (V47) at (8.25,0) {};
  \node (V48) at (9.5,0) {};
  \node (V49) at (11,0) {} ;

  \node (V21) at (9.25,1) {} ;
  \node (V22) at (10.5,1) {} ;

  \node (V11) at (8.5,2) {} ;
  \node (V12) at (10,2) {} ;

  \node (V01) at (10,3) {} ;

  \node (V51) at (1.25,-1.25) {};
  \node (V52) at (2.75,-1.5) {};
  \node (V53) at (4,-1.5) {};
  \node (V54) at (5.25,-1.5) {};

  \node (V55) at (8.5,-1.5) {} ;
  \node (V56) at (9.75,-1.5) {} ;
  \node (V57) at (11,-1.5) {} ;

  \end{scope}

  \begin{scope}[>={stealth},
              every edge/.style={draw=black,very thick}]

  \path [->] (V41) edge (V42);
  \path [->] (V42) edge node[style={fill=white,circle}] {$e$} (V43);
  \path [->] (V43) edge (V44);
  \path [->] (V44) edge  (V445);
  \path [->] (V445) edge  (V45);

  \path [->] (V45) edge (V46);
  \path [->] (V46) edge  (V47);
  \path [->] (V47) edge (V48);
  \path [->] (V48) edge (V49);

  \path [->] (V46) edge (V11);
  \path [->] (V11) edge (V21);
  \path [->] (V21) edge (V22);

  \path [->] (V11) edge (V12);
  \path [->] (V11) edge (V01);
  \path [->] (V46) edge (V55);

  \path [->] (V47) edge (V56);
  \path [->] (V56) edge (V57);

  \path [->] (V42) edge (V51);
  \path [->] (V51) edge (V52);
  \path [->] (V52) edge (V53);
  \path [->] (V53) edge (V54);

  \path (V41) edge[bend left=43,dashed] (1.8,3);
  \path [->] (1.8,3) edge[bend left=43,dashed] (V445) ;
  \path [->](V45) edge [bend left=65,dashed]  (V43);

  \begin{scope}[on background layer,every edge/.style=marked edge]

  \path (V41) edge[bend left=43] (1.8,3);
  \path [->] (1.8,3) edge[bend left=43] (V445) ;

  \path (V445) edge (V45);
  \path (V45) edge [bend left=66]  (V43);
  \path (V43) edge (V44);

  \end{scope}

  \end{scope}
  \end{tikzpicture}
  \caption{Case 4.2: $R(s,x,H_w - e)$ is weighted,
  it uses no nodes from $T$ after $v_j$}
  \label{fig:jumpingC}
\end{figure}

The following crucial observation allows us to handle many cases involving weighted replacement paths
\begin{observation}
  \label{observation:weighted_simple}
  Let $e \in P$ be an edge failure, $x \in H$ be a destination node and $w \in W$ be a weight function.
  If the replacement path $R(s,x,H_w - e)$ is weighted
  then it leaves $P$ at $s$ and does not intersect with $P$ until \textbf{after} the edge failure.
\end{observation}
To see why this observation is true,
first note that all the edges in $E(H_w) - E(H)$ begin at $s$ by definition,
so $R(s,x,H_w - e)$ indeed leaves $P$ at $s$.
Also, $R(s,x,H_w - e)$ does not intersect with $P$ until after the edge failure as
otherwise $R(s,x,H_w - e)$ could have used the path $P$
to get from $s$ to the intersection point,
which is a shortest path by the weight requirements. In other words, the use of the weighted edge is unnecessary.
An illustration of such path can be seen in Figure \ref{fig:jumpingC}.
\NL
For edge failures from $P$ we also define the following useful characterization
\begin{restatable}[Jumping and Departing Paths]{definition}{DefJumpPaths}
  Let $e \in E(P), x \in V(H), w \in W$ ,
  and let $R$ be a path from $s$ to $x$ in the graph $H_w-e$.
  The path $R$
  will be called \textit{jumping}
  if it uses some node $u$
  such that $u \in P$ and $u$ is \textbf{after} the edge
  failure $e$ in the path $P$.
  A path that is not jumping will be called \textit{departing}.
\end{restatable}
\mynewpage
\dnsparagraph{First case - the failure is in $P$ and the destination is in $T$:\;}
This case can be solved in a non-recursive manner,
using observation \ref{observation:weighted_simple}
and somewhat similar observations to those that were used in \cite{GW12}.
We distinguish between 3 different forms
the path $R(s,x,H_w - e)$ can take:
\NL
\textbf{Case 1.1:}
$R(s,x,H_w - e)$ is departing and weighted.
Using observation \ref{observation:weighted_simple},
we can conclude that $R(s,x,H_w - e)$ is edge-disjoint from $P$
as it does not intersect with $P$ before the edge failure nor after
(since it is departing).
This implies that the length of $R(s,x,H_w - e)$ is $d(s,x,H_w - P)$.
This value can easily be computed by running Dijkstra's
algorithm from $s$ in the graph $H_w - P$ for every weight function $w$.
\NL
\textbf{Case 1.2:} $R(s,x,H_w - e)$ is departing and unweighted.
An illustration of this case can be seen in Figure \ref{fig:departing}.
In this case one can use a technique similar to the one used in \cite{GW12} in order to compute
length of the replacement path \whp.
That is, if $e$ is among the last $\kindaSqrtV$ edges of $P$, then we can use a brute force solution to compute
$\D {} s x e$. If $e$ is of distance at least $\kindaSqrtV$ from $t$, then
the length of the detour of $R(s,x,H_w - e)$ is at least $\kindaSqrtV$
as this path departs before $e$ and gets to $T$.
So by sampling a set of nodes $B$ of size $\tilde{O}(\kindaSqrtV)$, we hit every such detour \whp.
Assuming we hit the detour using the pivot node $b \in B$,
we can compute $\D {} s b e$ rather easily, and have that $\D {} s x e = \D {} s b e + \D {} b x P$.
\snl
In the full algorithm we denote the estimation obtained by the pivots sampling by
$\depart e x$.
We show how to compute this estimation in step \ref{step:depart}
of the algorithm
and prove its correctness in Claims \ref{claim:depart_compleness}, \ref{claim:sound_depart_unweighted}.
\NL
\textbf{Case 1.3:} $R(s,x,H_w - e)$ is jumping.
As observed by the authors of \cite{GW12},
taking care of jumping replacement paths in the case when $x\in T$
essentially reduces to solving the RP
problem, where the source node is $s$ and the destination node is $t$.
This is since a jumping replacement path passes \WLOG\
through the separator node $t$.
\snl
So we focus on computing the length of $\R w s t e$ for every $e \in P$ and $w \in W$.
Using observation \ref{observation:weighted_simple}, if $R(s,t,H_w - e)$ is weighted then its length
is  $\min\limits_{u \text{ after } e \text{ in } P}
\{ \D w {{s}} u P + \Dnm u t \}$ as it does not intersect with $P$ until after the edge failure and from the intersection node the replacement path can go to $t$ using the shortest path $P$ (as this subpath does not contain the edge failure $e$).
Computing this value naively for every $w \in W$ and $e \in P$
takes $\tilde{O}(|W|{n_H}^2)$ time.
\snl
If $R(s,t,H_w - e)$ is unweighted, the algorithm of \cite{Roditty2005}
can be used to compute its length.

\begin{figure}[H]
  \centering
  \captionsetup{justification=centering}
  \begin{tikzpicture}
  \begin{scope}[every node/.style={circle,thick,draw}]
      \node (V41) at (-1.5,0) {${{s}}$} ;
      \node (V42) at (0,0) {$v_i$} ;
      \node (V43) at (1.5,0) {};
      \node (V44) at (3,0) {};
      \node (V45) at (4.5,0) {};
      \node (V46) at (6,0) {$t$};
      \node (V47) at (8,0) {};
      \node (V48) at (9.5,0) {};
      \node (V49) at (11,0) {} ;

      \node (V21) at (8.75,1) {} ;
      \node (V22) at (10.25,1) {} ;
      \node (V23) at (11.75,1) {} ;

      \node (V11) at (7.25,2) {} ;
      \node (V12) at (8.75,2) {} ;

      \node (V01) at (8.75,3) {} ;

      \node (V51) at (1.25,-1.25) {};
      \node (V52) at (2.75,-1.5) {};
      \node (V53) at (4,-1.5) {$v$};
      \node (V54) at (5.25,-1.5) {$x$};

      \node (V55) at (6.75,-1.5) {} ;
      \node (V56) at (8.75,-1.5) {$u$} ;
      \node (V57) at (10.25,-1.5) {} ;

  \end{scope}

  \begin{scope}[>={stealth},
                every edge/.style={draw=black,very thick}]

      \path [->] (V41) edge (V42);
      \path [->] (V42) edge (V43);
      \path [->] (V43) edge (V44);
      \path [->] (V44) edge  (V45);

      \path [->] (V45) edge (V46);
      \path [->] (V46) edge  (V47);
      \path [->] (V47) edge (V48);
      \path [->] (V48) edge (V49);

      \path [->] (V46) edge (V11);
      \path [->] (V11) edge (V21);
      \path [->] (V21) edge (V22);
      \path [->] (V22) edge (V23);
      \path [->] (V11) edge (V12);
      \path [->] (V11) edge (V01);
      \path [->] (V46) edge (V55);

      \path [->] (V47) edge (V56);
      \path [->] (V56) edge  (V57);

       \path [->] (V42) edge (V51);
       \path [->] (V51) edge  node[style={fill=white,circle}] {$e$} (V52);
       \path [->] (V52) edge (V53);
       \path [->] (V53) edge (V54);

      \path [->] (V56) edge[bend left=35,dashed] (V53);

  \begin{scope}[on background layer,every edge/.style=marked edge]
      \path (V45) edge (V46);

      \path (V41) edge (V42);
      \path (V42) edge (V43);
      \path (V43) edge (V44);
      \path (V44) edge (V45);
      \path (V45) edge (V46);
      \path (V46) edge (V47);
      \path (V47) edge (V56);
      \path [->] (V56) edge[bend left=35.5] (V53);
      \path (V53) edge (V54);

  \end{scope}

  \end{scope}
  \end{tikzpicture}
  \caption{$R(s,x,H_w - e)$ is a "help from below" path}
  \label{fig:help_below}
\end{figure}
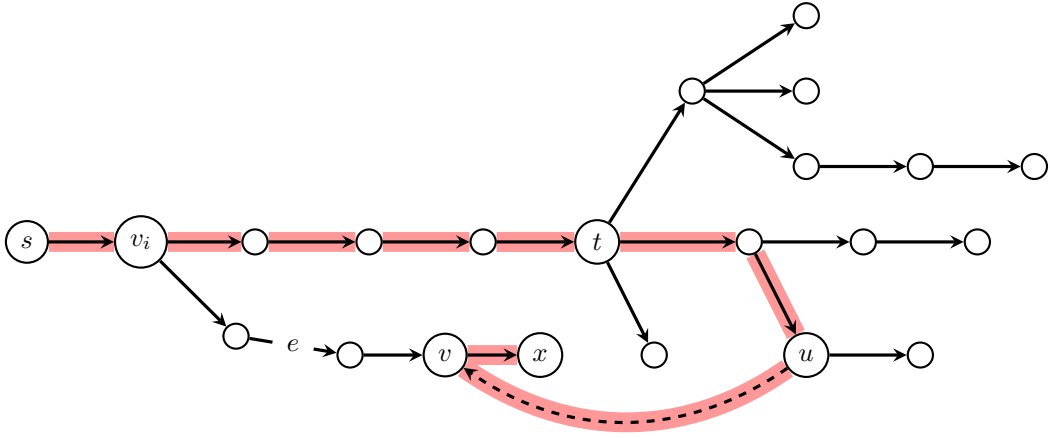

\mynewpage
\dnsparagraph{Second case - the failure is in $T$ and the destination is in $T$:\;}
We solve this case recursively. The recursive call will be invoked over
the subgraph $H[T]$.
Because the root of the tree $T$ is $t$ and not $s$,
we must change the source of our SSRP.
This implies that replacement paths that use the path $P$ to get from
$s$ to $t$ will be $\Dnm s t$ units shorter in the recursive call
than
they truly are.
So when we compress different forms of replacement paths using weight functions,
for normalization reasons we must also subtract $\Dnm s t$ from the weight function.
For simplicity, we ignore this issue in the overview,
but keep in mind that we always need to subtract $\Dnm s t$
from every weight function before the algorithm passes them to the recursion call,
and add this value back when it receives the recursion's estimation.
\snl
We distinguish between two possible forms of the replacement path: weighted and unweighted.
Rather interestingly we will see that this separation provides enough information about the structure
of the replacement path in order to compress it, and find its length recursively.
\NL
\textbf{Case 2.1:} The path $R(s,x,H_w-e)$ is weighted.
We claim that in this case, the only node from $S$ that $R(s,x,H_w-e)$ uses is $s$.
To see this note that for every node $u$ from $S$ that is not $s$,
the path from $s$ to $u$ in the BFS tree $K$ does not contain the edge failure $e$,
since $e \in T$.
By the weight requirement this path is a shortest path in $H_w$.
Hence, if a weighted replacement path uses a node $u$ from $S$,
it can use the path  from $s$ to $u$ in $K$. In other words, the use of a weighted edge was unnecessary.
So in this case the replacement path is almost completely contained within $H[T]$.
Therefore, in order to take care of this case,
we simply need to pass the weight function $w$ to the recursive call.
We formally prove the correctness of this case in Claim \ref{claim:sound_TT_weighted}.
\NL
\textbf{Case 2.2:}  The path $R(s,x,H_w-e)$ is unweighted.
Let $u$ be the last node of $R(s,x,H_w-e)$ that is from $S$.
If $u$ is $t$, then we can separate $R(s,x,H_w-e)$ into two subpaths:
a path from $s$ to $t$ - that is the shortest path $P$ \WLOG,
and the shortest path from $t$ to $x$ in $H[T]-e$.
We can use the recursive call over $H[T]$ to compute the length of
the second sub-path, and when we add $\Dnm s t$ we will get the length of $R(s,x,H_w-e)$.
\snl
The more interesting case is when $u \ne t$.
Let $v$ denote the node right after $u$ on $R(s,x,H_w-e)$.
In this case we say that $v$ gets "helped from above" by $u$,
as illustrated in Figure \ref{fig:help_above} in the appendix.
Since $u$ is the last node in $R(s,x,H_w-e)$ that belongs to $S$
the sub-path of $R(s,x,H_w-e)$ from $v$ to $x$ is fully contained in $H[T]-e$.
So we only need to compress the sub-path of $R(s,x,H_w-e)$ from $s$ to $v$.
In order to do so
we define a new weight function $c_T : V(T) \rightarrow \Nfty$
where for every vertex $v$ , $c_T(v)$ is defined to be $\min\limits_{u \in V(S) - \{t\} : (u,v) \in E(H)} \{\Dnm s u + 1\}$.
The sub-path of $R(s,x,H_w-e)$ from $s$ to $v$ is represented in the graph $H[T]_{c_T}-e$
as the weighted edge $(t,v)$
So by passing $c_T$ to the recursion and computing $\D[{H[T]}] {c_T} t x e$ we will be able
to obtain the length of the replacement path.
We formally prove the correctness of this case in Claim
\ref{claim:sound_TT_unweighted}.
\snl
We note that $c_T$ is truthful,
in the sense that for every edge failure $e \in T$, $c_T(v)$ is the length of some path from $s$ to $v$ in $H-e$.
This is since the path from $s$ to $u$ in $K$ is of length $\Dnm s u$
and does not contain $e$ (as previously claimed), and the edge $(u,v)$
is of length $1$ and
is not in $T$ because $u$ is not in $T$.
We formally prove that $c_T$ is truthful as part of Claim \ref{claim:completeTT}.
\mynewpage
\dnsparagraph{Third case - the failure is in $E(S)-E(P)$ and the destination is in $S$:\;}
We handle this case similarly to the way we handled the second case.
However we still sketch the algorithm for this case as it will introduce
the notation of a "help from bellow" replacement path, which will be useful
in the fourth case.
We solve this case recursively. The recursive call will be invoked over
the subgraph $H[S]$.
In order to take care of this case we distinguish between two forms of the replacement
path $R(s,x,H_w-e)$.
\NL
\textbf{Case 3.1:} The path $R(s,x,H_w-e)$ uses only nodes from $S$. In this case
simply passing the weight function $w$ to the recursive call would suffice in order
to compute the length of $R(s,x,H_w-e)$.
\snl
\textbf{Case 3.2:} $R(s,x,H_w-e)$ uses a node from $V(T) - \{t\}$.
Let $u$ be the last node in $R(s,x,H_w-e)$ that belongs to $V(T) - \{t\}$.
Note that the path from $s$ to $u$ in the BFS tree $K$ uses only edges
from $P$ and $T$, meaning that it does not use the edge failure $e \in E(S) - E(P)$.
Hence, \WLOG\ we may assume that the sub-path from $s$ to $u$ in $R(s,x,H_w-e)$
is the  path from $s$ to $u$ in the BFS tree $K$ as this is a shortest
path by the weight requirements.
Note that this implies in particular that $R(s,x,H_w-e)$ is unweighted.
An illustration of this case can be found in Figure \ref{fig:help_below}.
We name this kind of paths "help from bellow" replacement paths.
Let $v$ be the node right after $u$ in $R(s,x,H_w-e)$.
Since $u$ was the last node in $R(s,x,H_w-e)$ that belongs to $V(T)-\{t\}$
the sub-path of $R(s,x,H_w-e)$ from $v$ to $x$ is fully contained in $H[S]-e$.
\snl
So we only need to compress the sub-path of $R(s,x,H_w-e)$ from $s$ to $v$.
In order to do so
we define a new weight function $c_S : V(S) \rightarrow \Nfty$
where for every vertex $v$ , $c_S(v)$ is defined to be $\min\limits_{u \in V(T) - \{t\} : (u,v) \in E(H)} \{\Dnm s u + 1\}$.
The sub-path of $R(s,x,H_w-e)$ from $s$ to $v$
is represented in the graph $H[S]_{c_S}$ by the weighted edge $(s,v)$.
So by passing the weight function $c_S$ to the recursive call over $H[S]$,
and computing $\D[{H[S]}] {c_S} s x e$, we will be able to compute the length of $R(s,x,H_w-e)$.
We formally prove the correctness of this case in Claim
 \ref{claim:sound_SS_yesT}.
\snl
We also claim that this function is truthful for edge failures from $E(S) - E(P)$
in the sense that for every $e \in E(S)-E(P)$,  $c_S(v)$ is the weight of some path from $s$ to $v$ in $H-e$.
This is since the path from $s$ to $u$ in $K$ is of length $\Dnm s u$
and does not contain $e$ (as previously claimed), and the edge $(u,v)$
is of length $1$ and
is not in $S$ because $u$ is not in $S$.
We formally prove this fact as part of Claim \ref{claim:completeSS}.
\NL
Note that the weight function $c_S$ is untruthful for edge failures from $P$,
as the path from $s$ to $u$ in $K$ contains the entire path $P$.
But if we consider the recursion's estimation for $\D[{H[S]}] {c_S} s x e$ \textbf{only}
for an edge failure $e \in E(S) - E(P)$, we are promised that this estimation
represents the length of a true path in $H-e$.
If we were to add weighted edges instead of weight functions, we would
lose the ability to consider $\D[{H[S]}] {c_S} s x e$ as an estimation
for $\D w s x e$ only for specific edge failures.

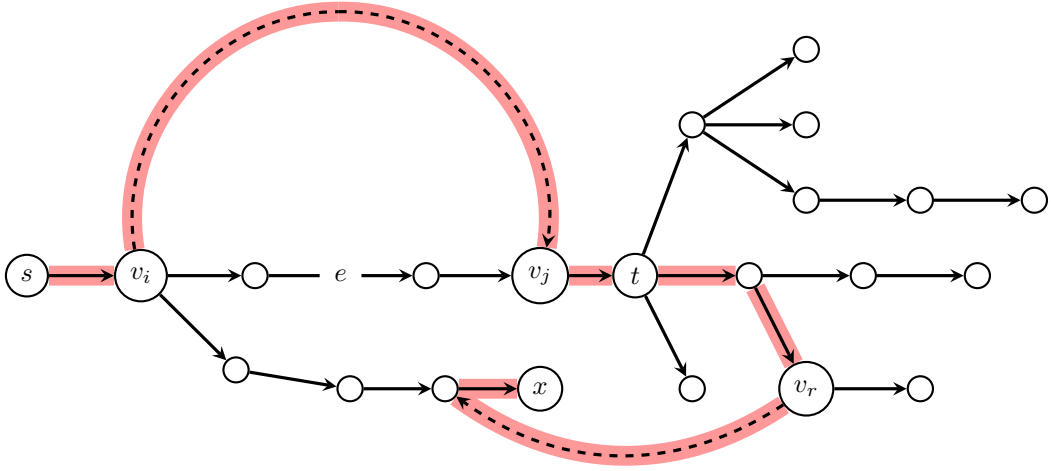
\begin{figure}[H]
  \centering
  \captionsetup{justification=centering}
  \begin{tikzpicture}
  \begin{scope}[every node/.style={circle,thick,draw}]
      \node (V41) at (-1.5,0) {${{s}}$} ;
      \node (V42) at (0,0) {$v_i$} ;
      \node (V43) at (1.5,0) {};
      \node (V44) at (3.75,0) {};
      \node (V45) at (5.25,0) {$v_j$};
      \node (V46) at (6.5,0) {$t$};
      \node (V47) at (8,0) {};
      \node (V48) at (9.5,0) {};
      \node (V49) at (11,0) {} ;

      \node (V21) at (8.75,1) {} ;
      \node (V22) at (10.25,1) {} ;
      \node (V23) at (11.75,1) {} ;

      \node (V11) at (7.25,2) {} ;
      \node (V12) at (8.75,2) {} ;

      \node (V01) at (8.75,3) {} ;

      \node (V51) at (1.25,-1.25) {};
      \node (V52) at (2.75,-1.5) {};
      \node (V53) at (4,-1.5) {};
      \node (V54) at (5.25,-1.5) {$x$};

      \node (V55) at (7.25,-1.5) {} ;
      \node (V56) at (8.75,-1.5) {$v_r$} ;
      \node (V57) at (10.25,-1.5) {} ;

  \end{scope}

  \begin{scope}[>={stealth},
                every edge/.style={draw=black,very thick}]

      \path [->] (V41) edge (V42);
      \path [->] (V42) edge (V43);
      \path [->] (V43) edge node[style={fill=white,circle}] {$e$}(V44);
      \path [->] (V44) edge  (V45);

      \path [->] (V45) edge (V46);
      \path [->] (V46) edge  (V47);
      \path [->] (V47) edge (V48);
      \path [->] (V48) edge (V49);

      \path [->] (V46) edge (V11);
      \path [->] (V11) edge (V21);
      \path [->] (V21) edge (V22);
      \path [->] (V22) edge (V23);
      \path [->] (V11) edge (V12);
      \path [->] (V11) edge (V01);
      \path [->] (V46) edge (V55);

      \path [->] (V47) edge (V56);
      \path [->] (V56) edge (V57);

       \path [->] (V42) edge (V51);
       \path [->] (V51) edge (V52);
       \path [->] (V52) edge (V53);
       \path [->] (V53) edge (V54);

      \path  (V42) edge[bend left=50,dashed] (2.6,3.5);
       \path [->] (2.6,3.5) edge[bend left=50,dashed] (V45) ;

      \path [->] (V56) edge[bend left=35,dashed] (V53);

  \begin{scope}[on background layer,every edge/.style=marked edge]
      \path (V45) edge (V46);
      \path  (V42) edge[bend left=50] (2.6,3.5);
      \path [->] (2.6,3.5) edge[bend left=50] (V45) ;;
      \path (V41) edge (V42);
      \path (V46) edge (V47);
      \path (V47) edge (V56);
      \path [->] (V56) edge[bend left=35.5] (V53);
      \path (V53) edge (V54);

  \end{scope}

  \end{scope}
  \end{tikzpicture}
  \caption{Case 4.1: $R(s,x,H_w - e)$ uses a node from $T$ after it uses $v_j$}
  \label{fig:jumpingA}
\end{figure}

\mynewpage
\dnsparagraph{Fourth case - the failure is in $P$ and the destination is in $S$:\;}
This case is the most complicated case in our algorithm.
Since we cannot allow three recursive calls (in order to obtain the desired running time)
and because we see no efficient way to solve this case in a non-recursive manner,
we will need to use the same recursive call over $H[S]$ as in the previous case (the third case).
We will do so by adding more weight functions.
\NL
We begin by making two simple observations that take
care of some easy cases, so we could focus on the more involved ones.
\begin{itemize}
  \item If the replacement path $R(s,x,H_w - e)$ uses no nodes from $T$ then one can simply use
  a recursive call over the graph $H[S]$ to compute its length.
  \item If $R(s,x,H_w - e)$ is departing, then since we may assume it contains
  nodes from $T$, we can use observations similar to those made in
  cases (1.1) and (1.2)
  in order to compute its length.
\end{itemize}
\NL
So we now focus on the more interesting case when $R(s,x,H_w - e)$ uses nodes from
$T$ and is jumping.
Note that since $R(s,x,H_w - e)$ is jumping it must leave the path $P$
at some node $v_i$ before the edge failure and return to $P$ at some node $v_j$
after the edge failure.
\NL
We will in fact still need to separate this case into 3 further sub-cases, depending on the
order $R(s,x,H_w - e)$ uses nodes from $T$.
These 3 cases present the true power of weight functions,
and their ability to compress graphs in a way that is sometimes untruthful but fixable.
\NL
\textbf{Case 4.1:}
$R(s,x,H_w - e)$ uses a node from $T$ after it uses $v_j$.
An illustration for this case can be found in Figure \ref{fig:jumpingA}.
\snl
We claim that in this case the length of $R(s,x,H_w - e)$ is
$\D[{H[S]}] {c_S}  s x e - \Dnm s t +$ $\D w s t e$.
While formally proving the correctness of this claim is rather technical
we attempt to give some intuition for this claim.
Note that since $R(s,x,H_w - e)$ uses a node from $T$ after it uses $v_j$,
it passes \WLOG\ through $t$
(as $v_j$ is after the edge failure).
So we can split $R(s,x,H_w - e)$ into two sub-paths:
the replacement path from $s$ to $t$ - which is of length $\D w s t e$,
and the path from $t$ to $x$ - which we denote by $R[t,x]$.
\NL
Lets us consider the path $P \circ R[t,x]$.
We claim that even though the path $R[t,x]$ contains nodes from $T$,
the recursive call over $H[S]$ can evaluate the length of the path $P \circ R[t,x]$.
This is because, roughly speaking, the path $P \circ R[t,x]$
is a sort of "help from bellow" replacement path
- as described in the the third case in which $e \in E(S) - E(P)$.
So like in the "help from bellow" case,
the path $P \circ R[t,x]$ would be represented
in $H[S]_{c_S} - e$ as a weighted replacement path.
When we receive the length of this weighted replacement path
we remove $P$
and replace it with the replacement path from $s$ to $t$.
That is, we subtract $\Dnm s t$ and add $\D w s t e$.
We formally prove the correctness of this estimation in Claim \ref{claim:sound_PS_jumpingA}.
As stated in the beginning of the overview, we do not know
a-priori if the replacement path indeed falls in this sub-case,
so we have to make sure that we never underestimate $\D w s x e$.
We formally prove this in Claim \ref{claim:complete_PS_cS}.
In this case we see that
weight functions allow us to assign weights that are untruthful for
some edge failures, but give us enough control in order to fix the
untruthful replacement paths. 
\NL
Note that $\D [{H[S]}] {c_S} s x e$ is used regardless 
of which weight function $w$ the true
replacement path uses. 
The fact that we use one recursive call over
all weight functions, 
allows us to compute this term only once, 
which we could not do if the algorithm would have used a
different recursive call for each weight function.

\NL \textbf{Case 4.2:} $R(s,x,H_w - e)$
is weighted and it uses no nodes from $T$ after $v_j$. An
illustration for this case can be found in Figure
\ref{fig:jumpingC}.

Let $(s,v)$ be the weighted edge in the replacement path $R(s,x,H_w - e)$. Note that $(s,v)$ is not in $P$ (as $P$ contains only unweighted edges from $H$).
Hence, by definition the replacement path leaves the path $P$ at $s$, that is, $v_i =s$.

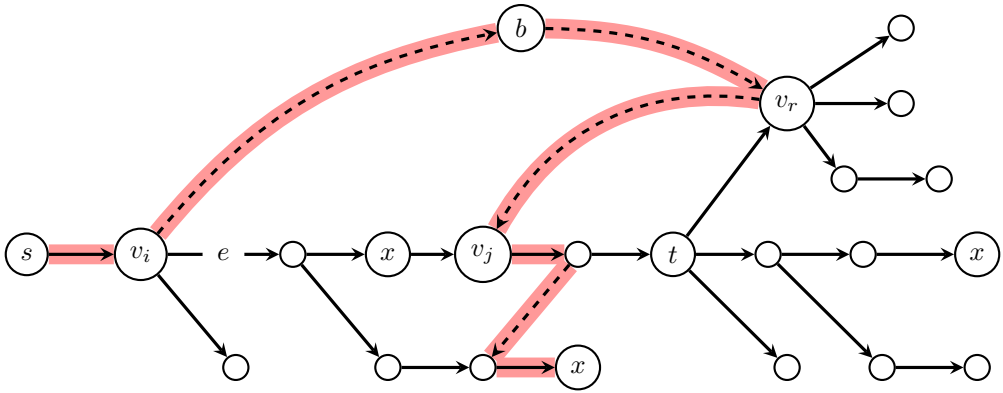
\begin{figure}[H]
  \centering
  \captionsetup{justification=centering}
  \begin{tikzpicture}
  \begin{scope}[every node/.style={circle,thick,draw}]
      \node (V41) at (-1.5,0) {${{s}}$} ;
      \node (V42) at (0,0) {$v_i$} ;
      \node (V43) at (2,0) {};
      \node (V44) at (3.25,0) {$x$};
      \node (V445) at (4.5,0) {$v_j$};
      \node (V45) at (5.75,0) {};
      \node (V46) at (7,0) {$t$};
      \node (V47) at (8.25,0) {};
      \node (V48) at (9.5,0) {};
      \node (V49) at (11,0) {$x$} ;

      \node (V21) at (9.25,1) {} ;
      \node (V22) at (10.5,1) {} ;

      \node (V11) at (8.5,2) {$v_r$} ;
      \node (V12) at (10,2) {} ;

      \node (V01) at (10,3) {} ;

      \node (V51) at (1.25,-1.5) {};
      \node (V52) at (3.25,-1.5) {};
      \node (V53) at (4.5,-1.5) {};
      \node (V54) at (5.75,-1.5) {$x$};

      \node (V55) at (8.5,-1.5) {} ;
      \node (V56) at (9.75,-1.5) {} ;
      \node (V57) at (11,-1.5) {} ;

      \node (B) at (5,3) {$b$};

  \end{scope}

  \begin{scope}[>={stealth},
                every edge/.style={draw=black,very thick}]

      \path [->] (V41) edge (V42);
      \path [->] (V42) edge node[style={fill=white,circle}] {$e$} (V43);
      \path [->] (V43) edge (V44);
      \path [->] (V44) edge  (V445);
      \path [->] (V445) edge  (V45);

      \path [->] (V45) edge (V46);
      \path [->] (V46) edge  (V47);
      \path [->] (V47) edge (V48);
      \path [->] (V48) edge (V49);

      \path [->] (V46) edge (V11);
      \path [->] (V11) edge (V21);
      \path [->] (V21) edge (V22);

      \path [->] (V11) edge (V12);
      \path [->] (V11) edge (V01);
      \path [->] (V46) edge (V55);

      \path [->] (V47) edge (V56);
      \path [->] (V56) edge (V57);

      \path [->] (V42) edge (V51);
      \path [->] (V43) edge (V52);
      \path [->] (V52) edge (V53);
      \path [->] (V53) edge (V54);

      \path [->] (V45) edge [dashed] (V53);k
      \path [->] (V42) edge[bend left=20,dashed] (B);
      \path [->] (B) edge[bend left=15,dashed] (V11);

      \path [->] (V11) edge [bend right=35.5, dashed] (V445);

  \begin{scope}[on background layer,every edge/.style=marked edge]

      \path (V41) edge (V42);
      \path (V42) edge[bend left=20] (B);
      \path  (B) edge[bend left=15] (V11);

      \path (V11) edge [bend right=35.5] (V445);
      \path (V445) edge (V45);
      \path (V45) edge (V53);
      \path (V53) edge (V54);

  \end{scope}

  \end{scope}
  \end{tikzpicture}
  \caption{Case 4.3: $R(s,x,H_w - e)$ is unweighted,
  it uses a node from $T$ in the sub-path from $v_i$ to $v_j$,
  and uses no nodes from $T$ after $v_j$. }
  \label{fig:jumpingB}
\end{figure}

This implies that the sub-path from $s$ to $v_j$ is edge disjoint to $P$
and so its length is $\D w s {v_j} P$.
So for every weight function $w\in W$,
we would have wished to define a new weight function $w|_P$
such that $w|_P(v_j) = \D w s {v_j} P$
for every $v_j \in P$.
We will then
ask the recursive call to estimate $\D[{H[S]}] {w|_P} s x e$.
This will indeed suffice in order to compute the length of the replacement path recursively,
as $R(s,x,H_w - e)$ uses no nodes from $T$ after $v_j$.
\snl
However, by doing so we increase the number of weight function passed to the
recursive call by a factor of $2$.
This sort of exponential growth will prevent us from achieving the desired running time.
So instead we define a new weight function $w|_S$ such that
$w|_S(x) = \D w s x P$ if $x \in P$ and $w|_S(x) = w(x)$ if $x \notin P$.
Note that for every $x$ it holds that $w|_S(x) \le w(x)$ ,
since the distance $\D w s x P$ is at most the weight of the edge $(s,x) \in H_w$
which is $w(x)$.
This implies that the $w|_S$ function preserves information from both $w$ and $w|_P$.
So instead of passing $w$ to the recursive call,
we pass $w|_S$.
Later in Claim \ref{claim:wS_complete} we prove that the new $w|_S$
function is truthful in the sense that for every
$e \in S, x \in S$ it holds that $\D[{H[S]}] {w|_S} s x e$ is at least
$\D w s x e$, meaning we do not create underestimations by
using $w|_S$ instead of $w$.
In the full version of the algorithm,
we prove the correctness of this case in Claim
\ref{claim:sound_PS_jumpingC}.
\snl
So as one can see, weight functions allow us to specifically choose
special nodes and decrease their weights in order to compress more information,
without sacrificing the truthfulness of the weight function.
\NL
\textbf{Case 4.3:} $R(s,x,H_w - e)$ is unweighted,
it uses no nodes from $T$ after $v_j$.
\snl
This is the most involved and interesting case our algorithm handles.
Note that since we assume $R(s,x,H_w - e)$ uses a node from $T$,
and since $R(s,x,H_w - e)$ uses no nodes from $T$ after $v_j$,
then the sub-path of $R(s,x,H_w - e)$ from $v_i$ to $v_j$ must contain a node from $T$.
An illustration for this case can be found in Figure \ref{fig:jumpingB}.
Similarly to Case 1.2,
we may assume that the edge failure is not among
the last $\kindaSqrtV$ edges of $P$ as otherwise we can use a brute force solution to compute the length
of the replacement path.
Since the sub-path from $v_i$ to $v_j$ uses a node from $T$,
its length is at least $\Dnm {v_i} {t}$ that is at least $\kindaSqrtV$.
So \whp\ we have sampled some pivot node $b \in B$ on this sub-path.
Note that the sub-path from $s$ to $b$ is departing as the replacement path returns to $P$ only at $v_j$.
So we can easily compute $\D {} s b e$
as stated before in Case 1.2.
\snl
To compress the sub-path from $b$ to $v_j$ we define a weight function
$w_b$ for every pivot node. We would have wished to define $w_b(v_j) = \D {} b {v_j} P$,
recursively compute
$\D[{H[S]}] {w_b} s x e$ and
add $\D {} s b e$ when receiving the answer from the recursion.
This will indeed suffice in order to compress the length of the sub-path from $v_i$ to $v_j$
as it is edge disjoint to $P$.
However this is not a valid weight function as it does not necessarily fulfill the weight requirements.
So instead we define
$w_b(v_j) = \Dnm s b + \D {} b {v_j} P$ which is a valid weight function,
and we fix the output of the recursion by replacing $\Dnm s b$ with $\D {} s b e$,
i.e. subtracting the former and adding the latter.
\snl
As in case 4.1, we need to prove that we never underestimate	
$\D {} s x e$. This is formally done in Claim \ref{claim:complete_PS_pivot}.
\snl
As we can see, while the weight function $w_b$ is untruthful,
in the sense that $w_b(v_j)$ is not necessarily the distance of a
path from $s$ to $v_j$ in $H-e$, we are able to fix this untruthfulness as we know
what pivot $b \in B$ is used in each weight function $w_b$.
In a sense if we could have used a different recursive call for every $b \in B$ we could have used edges from $s$ rather than weight functions but this would be very inefficient.
The weight functions allow us to compress all these recursive calls into one.
\snl
In the full version of the algorithm we denote the estimation
made using the $w_b$ functions by $\pivot e x$.
We show how to compute this estimation in step \ref{step:SS} of the algorithm and
prove the correctness of this case in Claim
\ref{claim:sound_PS_jumpingB}.

\subsection{Running Time Analysis}
First we note that the number of weight
functions in each recursive call increases by $\tilde{O}(\kindaSqrtV)$
in each level of the recursion, as we add the $c_S,c_T$ and $\{w_b\}_{b \in B}$
weight functions, and $|B| = \tilde{O}(\kindaSqrtV)$.
Since the number of the weight functions in the first call to the algorithm
is 1 (the function $w \equiv \infty$),
and since the depth of the recursion is logarithmic
we have that $|W| = \tilde{O}(\sqrt{n})$ at all times.
So if we simply analyze the non-recursive parts
of the algorithm, we can conclude the algorithm spends $\tilde{O}(n_H^2 \sqrt{n})$
times on the recursive call over the sub-graph $H$.
One can rather easily see that since the
BFS trees in each level of the recursion are edge disjoint
sub-trees of the original BFS tree $\widehat{K}$,
the total number of vertices in each level of the recursion is at most $2n$.
We prove this formally in Section \ref{subsec:running_time_rec}.
So the total time the algorithm spends on each level of the recursion is $\tilde{O}(n^{2.5})$.
Since the depth of the recursion is logarithmic
the running time of the algorithm is $\tilde{O}(n^{2.5})$.

\subsection{Going From \mathtitlewrapper{$\tilde{O}(n^{2.5})$} to \mathtitlewrapper{$\tilde{O}(m\sqrt{n} + n^2)$}}
In this section we sketch the ideas of improving the running time from
$\tilde{O}(n^{2.5})$ to $\tilde{O}(m\sqrt{n} + n^2)$.
\NL
We first note that as the number of weight functions in our algorithm is
$\tilde{O}(\sqrt{n})$ in each recursive call then even outputting $\D w s x e$
for every triplet $w \in W, x \in V(H), e \in E(K)$ is impossible (in the desired running time)
as there are $\tilde{O}(n_H^{2.5})$ such triplets.
In order to overcome this issue, the algorithm does not output distances to all such triplets but rather each recursive call
is given as input a set of queries $Q$ that is a small subset of all possible triplets (that is $Q \subseteq E(K) \times V(H) \times W$)
and the goal is to output the distances only for the given set of queries.
Initially, $Q$ is set to be $E(\widehat{K}) \times V(G) \times \{w\}$,
where $w \equiv \infty$, and so its size is $|Q| = \tilde{O}(n^2)$.
Each recursive call over a graph $H$ will make sure to ask only $\tilde{O}({n_H}^2)$
new queries (queries which it didn't received).
Since the total number of vertices in each level is at most $2n$,
the number of new queries added at each level of the recursion is $\tilde{O}(n^2)$.
Since the depth of the recursion is logarithmic, the total number
of queries asked is $\tilde{O}(n^2)$.
\NL
Secondly, recall that in Case 1.3,
where the edge failure is in $E(P)$ and the destination is $t$,
the algorithm computes the value of
$\min\limits_{u \text{ is after } e \text{ in } P} \{\D w s u P + \Dnm u t\}$
naïvely for every $e\in E(P), w\in W$.
This computation costs $\tilde{O}(|W| n_H^2)$ time.
However for a specific function $w \in W$,
this value can be computed for all $e \in E(P)$ in $\tilde{O}(n_H)$
time using a simple dynamic programming argument which will
be shown in Section \ref{subsec:DSTE} in the complete algorithm.
Hence, we can reduce the running time of this part to $\tilde{O}(|W|n_H)$.
\NL
Finally, and most importantly,
when handling departing unweighted paths  (Case 1.2)
and when using the $w_b$ weight function (Case 4.3)
the algorithm samples a set $B$ of pivots of size $\tilde{O}(\kindaSqrtV)$.
Then for every edge failure $e \in P$ and destination node $x \in V(H)$
we iterate over $B$
and find the pivot that provides the smallest distance estimation.
This implies that the algorithm spends $\tilde{O}(|B||P||V(H)|)$ time to find these pivots,
which is again $\tilde{O}(n_H^{2.5})$ time.
The problem is that our estimation for the distance between an edge
failure and the separator node $t$ is too loose.
On the one hand when sampling $B$ we say that this distance is
at least $\kindaSqrtV$, but on the other hand
when bounding $|P|$ we say that it is at most $O(n_H)$.
\snl
In order to solve this problem we use a standard scaling trick.
More specifically,
we consider a logarithmic number of sub-paths $\{P_k\}$,
where $P_k$ is the sub-path of $P$ induced by the vertices
$\{v \in V(P) : 2^{k+1} \kindaSqrtV \ge \Dnm v t \ge 2^ k \kindaSqrtV \}$.
$P_0$ is defined to be the sub-path of $P$ induced by the last $2\kindaSqrtV$ vertices of $P$.
Note that the set of paths $\{P_k\}$ is an edge disjoint partition of $P$,
and that $|P_k| = O(2^k \kindaSqrtV)$.
An illustration for this partition can be seen in Figure \ref{fig:pi_seperation}
in the appendix.
For every index $k \ne 0$
we then sample a random set $B_k$ of size $\tilde{O}(\frac{\kindaSqrtV}{2^k})$ using the sampling lemma \ref{lemma:sampling}.
Now, if we consider an edge failure $e \in P_k$ for $k \ne 0$,
we know that the distance from $e$ to $t$ is at least $2^k \kindaSqrtV$.
So when we wish to estimate the length of the departing replacement path in Case 1.2
or send the query $(e,x,w_b)$ to the recursive call over $H[S]$ in Case 4.3,
we only need consider pivot nodes $b$ that are from $B_k$.

%% file: sections/algorithm.tex
\section{ An \mathtitlewrapper{$\tilde{O}(m\sqrt{n} + n^2)$} Algorithm for SSRP in Unweighted Directed Graphs}
\label{sec:algorithm}
In this section we describe in details our $\tilde{O}(m\sqrt{n} + n^2)$ time algorithm for SSRP in unweighted directed graphs.
Our algorithm is recursive and uses a balance tree separator (formally stated in Lemma \ref{lemma:seperator}) in order to divide its input into two smaller inputs.
In our algorithm we utilize the following LCA data structure presented by Bender and Farach-Colton in \cite{bender2000lca}.
\begin{lemma}  [LCA Data Structure \cite{bender2000lca}]
\label{lemma:LCA}
Given a rooted tree $K$ containing $n$ vertices, one can construct an LCA data-structure in linear $O(n)$ time and answer LCA queries in constant time.
\end{lemma}
We now describe the algorithm
whose input is a sub-graph $H$ of the original graph $G$,
a BFS tree $K$ over $H$ rooted at a node $s$,
a set of weight functions $W$ and a set of queries $Q$.
The goal of the algorithm is to create an estimation $\Dhat w s x e$ for every query $(e,x,w) \in Q$,
such that the estimation matches the real distance $\D w s x e$.
We denote by $n_H = |V(H)| , m_H = |E(H)|$ and $n = |V(G)|, m = |E(G)|$.
The pseudo code code for the algorithm can be found in Algorithm \ref{alg_main}.
\begin{step}[Base case:]
\label{step:base}
If $n_H \le 6$ the algorithm constructs the graph $H_w-e$ for every weight function $w\in W$ and edge failure $e \in E(K)$.
The algorithm then runs Dijkstra's algorithm from ${s}$ in the graph $H_w - e$
to compute $\D w {{s}} x e$ for every $e \in E(K), x \in V(H), w \in W$.
The algorithm then returns $\D w {{s}} x e$ for every $(e,x,w) \in Q$.
\end{step}

\begin{step}[Tree Separation ($S,T$)]
\label{step:separation}
In this step, the algorithm finds (using Lemma \ref{lemma:seperator}) a balanced tree separator node $t\in V(H)$ that separates the
BFS tree $K$ into two edge disjoint sub-trees $S,T$,
such that $E(S)\cup  E(T)=E(K)$ and $|V(S)|,|V(T)| \le \frac{2n_H}{3}$.
An illustration of the separation can be seen in Figure \ref{fig:separation}.
Let $P$ denote the path from ${{s}}$ to $t$ in the BFS tree $K$.
The algorithm next computes the distances $\Dnm t x, \Dnm x t$ for every node $x \in V(H)$
by running the BFS algorithm from the node $t$ in the graphs $H,H^R$.
\end{step}
\begin{step}[Computing $\D w {{s}} x P$]
    \label{step:WSPX}
    The algorithm computes $\D w {{s}} x P$ for every node $x\in V(H)$ and weight function $w \in W$,
    by constructing the graph $H_w - P$ for every weight function $w \in W$,
    and running Dijkstra's algorithm from ${s}$ in the resulted graph.
\end{step}

\begin{step}[Sampling Pivots ($B_k$) and Defining Path Intervals ($P_k$)]
\label{step:pivots}
$\forall k \in [\lgV]$ the algorithm constructs a set $B_k$ by sampling every vertex in $V(H)$ independently at random with probability
$\frac{C\ln(n)}{2^k \sqrtV}$ (where $C\ge 3$ is a constant that will be fixed later on).
If the algorithm sampled too many vertices and it does not hold that
$|B_k| = \tilde{O}(\frac{\sqrtV}{2^k})$,
the algorithm re-samples $B_k$.
The algorithm sets $B=\bigcup_{k\in [\lgV]} B_k$ 
For every pivot
$\forall b\in B$ the algorithm then runs the BFS algorithm from $b$ in the graph $H-P$ and in the graph $(H-P)^R$.
\NL
Also,
$\forall k \in [\lgV]$ the algorithm sets the path $P_k$ to be the sub-path of $P$ induced by the vertices $\{v \in V(P) : 2^{k+1} \sqrtV \ge \Dnm v t \ge 2^ k \sqrtV \}$.
The algorithm then sets $P_0$ to be the sub-path of $P$ induced by the vertices, $\{v \in V(P) : 2 \sqrtV \ge \Dnm v t \}$.
\snl
An illustration of this separation can be seen in Figure \ref{fig:pi_seperation},
note that all the edges and nodes in $P_k$ are \textbf{after} all the edges and nodes in $P_{k+1}$,
and that the set of subpaths $\{P_k\} _{k\in [\lgV] \cup \{0\}}$ is an edge disjoint partition of $P$.
\end{step}

\begin{step}[Computing Departing Paths ($\depart e x$)]
\label{step:depart}
$\forall k \in \lgV, \forall b\in B_k, \forall e \in E(P_k)$ the algorithm naïvely computes
$\depart{e}{b} =  \min\limits_{u\in V(P) \text{ before } e}$
$\{\Dnm {{s}} u + \D {} u b P\}$.
Afterwards,
$\forall k \in \lgV,\forall e \in E(P_k), \forall x \in V(H) - B_k$,
the algorithm naïvely computes
$\depart{e}{x}=\min\limits_{b\in B_k} \{$
$\depart e b +$
$\D {} b x P \}$.
Also,
$\forall e\in P_0, \forall x \in V(H)$ the algorithm sets $\depart{e}{x} = \D {} {{s}} x e$, where
$\D {} {{s}} x e$ is computed by running the BFS algorithm from ${{s}}$ in the graph $H-e$ for every $e\in P_0$.
\end{step}

\begin{step}[Computing $\Dhat w {{s}} t e$ When $e\in E(P)$]
\label{step:DSTE}
The algorithm runs the replacement path algorithm from \cite{Roditty2005},
for the unweighted directed graph $H$ and the path $P$ (from $s$ to $t$),
to find (\whp) a replacement path for every edge failure $e\in E(P)$.
Let $\DRZ {{s}} t e$ be the length returned by the algorithm of \cite{Roditty2005}, for the edge failure $e\in E(P)$.
For every $w\in W$ and for every $e\in E(P)$ the algorithm computes
$A_w[e] \eqdef \min\limits_{u \text{ is after } e \text{ in } P} \{ {\D w {{s}} u P} + \Dnm u t \}$.
\snl
For efficiency reasons, the algorithm uses the following dynamic programming to compute the values $A_w[e]$ for $e\in P$.
Let $u_{|P|},...,u_0$ be the path from $s$ to $t$,
that is $u_{|P|}=s$ and $u_0=t$ and let $e_i=(u_{i+1}, u_{i})$.
For every $w \in W$ the algorithm
sets $A_w[e_0] \gets \D w {{s}} t P$ and then
for every $ 1 \le i \le |P|-1$  (in increasing order)
the algorithm sets
$A_w[e_i] \gets \min\{A_w[e_{i-1}], \D w {{s}} {u_{i-1}} P + \Dnm {u_{i-1}} t\}$.
\snl
$\forall w \in W, \; \forall e\in E(P)$,
the algorithm sets

\begin{center}
$\Dhat w {{s}} t e=$
$\min \{$
$\DRZ {{s}} t e,\;$
$A_w[e]\; \}$
\end{center}
\end{step}

\begin{step}[Computing $\Dhat w {{s}} x e$ When $e\in E(P),x\in V(T) - \{t\}$]
\label{step:PT}
$\forall (e,x,w) \in Q : e\in E(P), x \in V(T)- \{t\},w\in W$ the algorithm sets

\begin{center}
$\Dhat w {{s}} x e=$
$\min \{$
$\D w {{s}} x P,\;$
$\Dhat w {{s}} t e + \Dnm t x,\;$
$\depart{e}{x}\}$
\end{center}
\end{step}

\begin{step}[Computing $\Dhat w {{s}} x e$ When $e\in E(T),x\in V(T)- \{t\}$]
\label{step:TT}
\subsubsection*{Defining the Recursive Input:}
$\forall w \in W$ the algorithm defines
a new weight function
$w|_T : V(T) \rightarrow  \Nfty$ as follows
$\forall v \in V(T) :$
$w|_T(v) = w(v) - |P|$ .
The algorithm also defines a new weight function $c_T : V(T) \rightarrow \Nfty$ as follows
$\forall v \in V(T) :c_T(v)=$
$\min\limits_{u \in V(S) :  u\ne t , (u,v)\in E(H)} \{\Dnm {{s}} u$
$+ 1\} - |P|$.
\snl
The algorithm sets
the new set of weight functions to be $W_T = \{w|_T : w \in W \} \cup \{c_T\}$.
The algorithm sets the new query set $Q_T$ to be as follows:
$\forall (e,x,w)\in Q : e\in E(T) , x \in V(T) , w \in W$ the algorithm adds the query $(e,x,w|_T)$ to $Q_T$. Also $\forall e \in E(T), x \in V(T)$ the algorithm adds to $Q_T$ the query $(e,x,c_T) $.
\snl
The algorithm invokes recursively on
the following input: the induced graph $H[T]$, the BFS tree $T$,
the set of weight functions $W_T$ and the query set $Q_T$.
Let $\Dhat[{H[T]}]{w}{{s}}{x}{e}$ be the output values the recursive call returns.
\NL
\subsubsection*{Computing the Results}
$\forall (e,x,w)\in Q : e\in E(T), x\in V(T)- \{t\}, w\in W$ the algorithm sets
\begin{center}
$\Dhat w {{s}} x e=$
$\min \{$
$\Dhat[{H[T]}]{w|_T}{t}{x}{e} + |P|,\;$
$\Dhat[{H[T]}]{c_T}{t}{x}{e} + |P|\}$
\end{center}
\end{step}

\begin{step}[Computing $\Dhat w {{s}} x e$ When $e\in E(S),x\in V(S)- \{t\}$]
\label{step:SS}

\subsubsection*{Defining the Recursive Input}
$\forall w\in W$ the algorithm defines $w | _S : V(S) \rightarrow  \Nfty$ as follows:
$\forall v \in V(P)  : w|_S(v) = \D w {{s}} v P$ and $\forall v \in V(S) - V(P) : w|_S(v) = w(v)$.
The algorithm then defines the new weight function $c_S : V(S) \rightarrow \Nfty$ as follows: $ \forall v \in V(S) : c_S(v) = \min\limits_{u\in V(T) : u\ne t, (u,v) \in E(H)} \Dnm {{s}} u+ 1$.
$\forall k \in [\lgV], \forall b\in B_k$ the algorithm defines a weight function $w_b : V(S) \rightarrow \Nfty $ as follows:
$\forall v  \in V(S): $ $w_b(v) = \Dnm {{s}} b + \D {} b v P$.
The algorithm sets the new set of weight functions $W_S$ to be
$\{w|_S : w \in W\} \cup \{c_S\} \cup \{w_b : b \in B_k \text{ for some } k \in [\lgV] \}$.
\snl
The algorithm then sets the new set of queries $Q_S$ to be as follows:
$\forall e\in E(S) , \forall x\in V(S)$ the algorithm adds to $Q_S$ the query $(e,x,c_S)$.
Then
$\forall (e,x,w)\in Q : e\in E(S),x\in V(S),w\in W$ the algorithm adds to $Q_S$ the query $(e,x,w|_S)$.
Also,
$\forall k \in [\lgV ], \forall b\in B_k , \forall e \in P_k, \forall x \in V(S)$ the algorithm adds to $Q_S$ the query $(e,x,w_b)$.
\snl
The algorithm invokes recursively on
the following input: the induced graph $H[S]$, the BFS tree $S$,
the set of weight functions $W_S$ and the query set $Q_S$.
Let $\Dhat[{H[S]}]{w}{{s}}{x}{e}$ be the output values it received from the recursive call.
\NL
\subsubsection*{Computing \mathtitlewrapper{$\pivot e x$}}
$\forall k \in [\lgV],$
$\forall e\in E(P_k),$
$\forall x\in V(S)$
the algorithm computes
$\pivot e x =$ $\min\limits_{b\in B_k}$
$\{\Dhat[{H[S]}]{w_b}{{s}}{x}{e}-$ $\Dnm {{s}} b+$ $ \depart e b\}$.
\snl
Then
$\forall e \in E(P_0), \forall x \in V(S)$ the algorithm specially sets $\pivot e x = \depart e x = \D {} {{s}} x e$.
\NL
\subsubsection*{Computing the Results}
$\forall (e,x,w) \in Q : e\in E(S) - E(P), x \in V(S)- \{t\},w\in W$ the algorithm sets
\snl
\begin{center}
$\Dhat w {{s}} x e=$
$\min\{$
$\Dhat[{H[S]}]{w|_S}{{s}}{x}{e},\;$
$\Dhat[{H[S]}]{c_S}{{s}}{x}{e}\}$
\end{center}
\NL
$\forall (e,x,w) \in Q : e\in E(P), x \in V(S)- \{t\},w\in W$ the algorithm sets
\snl
\begin{center}
$
\Dhat w {{s}} x e= \min\{ \Dhat[{H[S]}]{w|_S}{{s}}{x}{e},
\;
$
$
\Dhat{w}{{s}}{x}{P},\;
$
$
\depart e x,\;
$
$
 \pivot e x,\;
$
$
 \Dhat[{H[S]}]{c_S}{{s}}{x}{e} + \Dhat w {{s}} t e - \Dnm {{s}} t\}
$
\end{center}

\end{step}

\begin{step}[Outputting the results] \label{step:final}
For every query $(e,x,w) \in Q$,
the algorithm checks if $e$ is on the shortest path from $s$ to $x$ in $K$ (this can be done easily by computing an
LCA data structure (see e.g. \ref{lemma:LCA}) on the BFS tree $K$).
If $e$ is not on the shortest path from $s$ to $x$ in $K$ then the algorithm simply returns $\Dnm {{s}} x$.
Otherwise the algorithm returns $\Dhat w s x e$.
\NL
Note that if $e\in E(T), x \in V(S)$,
then $e$ is not on the path from $s$ to $x$ in $K$
since this path is contained in the BFS tree $S$.
Similarly if $e\in E(S)-E(P)$ and $x \in V(T)$
then $e$ is not on the path from $s$ to $x$ in $K$
since this path uses only edges from $E(P) \cup E(T)$.
For intuition see Figure \ref{fig:separation}.
So we ensured that algorithm has computed all the $\widehat{d}$ values it needs to.
\end{step}

\begin{center}
    \captionof{algorithm}{The SSRP algorithm}\label{alg_main}
    \begin{algorithmic}[1]
    \Function{Generalized SSRP}{$H,K,W,Q$}
    \If{$|V(H)| \le 6$}
    \Comment{Naïvely computing the SSRP for every edge failure and weight function}
    \For{$e \in E(K)$}
    \For{$w \in W$}
    \State{$\widehat{d}(s,\circ,H_w - e) \gets $ \Call{Dijkstra}{$s,H_w-e$}}
    \EndFor
    \EndFor
    \State{\Return{$\widehat{d}$}}
    \EndIf
    \\
    \State{$S,T \gets$ \Call{Balanced Separation}{$K$}}
    \Comment{See Lemma \ref{lemma:seperator}}
    \\\\
    \Comment{Recall That $s$ is the root of $S$ and $K$,
    $t$ is the root of $T$ and $P$ is the path from $s$ to $t$ in $K$}
    \\
    \State{$d(t,\circ,H) \gets $ \Call{Dijkstra}{$t,H$}}
    \State{$d(\circ,t,H) \gets $ \Call{Dijkstra}{$t,H^R$}}
    \\
    \For{$w \in W$}
    \State{$d(s,\circ,H_w - P) \gets $ \Call{Dijkstra}{$s,H_w-P$}}
    \EndFor
    \\
    \For{$k := 1$ \textbf{to} $\lgV$}
    \Comment{The pivot sampling}
    \State{$B_k \gets$ Sample\Big{(}$V(H),\; \frac{C \cdot\ln n}{2^k\sqrt{|V(H)|}}$\Big{)}}
    \Comment{Here $C \ge 3$ is a constant, see Lemma \ref{lemma:sampling}}
    \For{$b \in B_k$}
    \State{$d(b,\circ,H-P) \gets$ \Call{Dijkstra}{$b,H-P$}}
    \State{$d(\circ,b,H-P) \gets$ \Call{Dijkstra}{$b,(H-P)^R$}}
    \EndFor
    \EndFor
    \partTitle{Computation of $\depart e x$}
    \For{$k := 1$ \textbf{to} $\lgV$}
    \For{$e \in P_k$}
    \For{$b \in B_k$}
    \Comment{Computing depart for the pivots}
    \State $\depart e b \gets \min\limits_{u \text{ before } e \text{ in } P} \{\Dnm s u + d(u,b,H-P)\}$
    \EndFor
    \For{$x \in V(H) - \{B_k\}$}
    \Comment{Computing depart for the non-pivots}
    \State $\depart e x \gets \min\limits_{b \in B_k} \{\depart e b + d(b,x,H-P)\}$
    \EndFor
    \EndFor
    \EndFor
    \For{$e \in P_0$}
    \State {$\depart e \circ \gets $ \Call{Dijkstra}{$s,H-e$}}
    \EndFor
    \partTitle{The case when the failure is in $P$ and the destination is $t$}
    \State $\langle u_0,u_1,...,u_{|P|} \rangle \gets P^R$
    \Comment{Note that $u_0 = t$ and $u_{|P|} = s$}
    \For{ $i := 0$ \textbf{to}  $|P|-1$}
        \State $e_i \gets (u_{i+1},u_i)$
    \EndFor
    \\
    \For{ $w \in W$ }
    \State $A_w[e_0] \gets d(s,t,H_w-P)$
    \For{ $i := 1$ \textbf{to}  $|P|-1$}
        \State $A_w[e_i] \gets \min\{ A_w[e_{i-1}] , d(s,u_i, H_w - P) + d(u_i,t,H)\}$
    \EndFor
    \EndFor
    \\
    \State $\widehat{d}_{RZ} \gets $ RP($H,P$)
    \Comment{RP is the algorithm from \cite{Roditty2005}}
    \For{$e \in P, w \in W$}
    \State $\widehat{d}(s,t,H_w - e) \gets \min\{\DRZ s t e, A_w[e]\}$
    \label{lst:line:DSTE_set}
    \EndFor
    \partTitle{The case when the failure is in $P$ and the destination is in $V(T) - \{t\}$}
    \For{$(e,x,w) \in Q$}
    \If {$e \in P$ \textbf{and} $x \in V(T) - \{t\}$}
    \State$\Dhat{w}{s}{x}{e} \gets \min\{
        \Dhat w s t e + \Dnm t x,
        \D w s x P,
        \depart e x
    \}$
    \label{lst:line:PT_set}
    \EndIf
    \EndFor
    \partTitle{The case when the failure is in $T$ and the destination is in $V(T) - \{t\}$}
    \State $W_T \gets \varnothing$
    \For{$w \in W$}
    \Comment{Defining the restricted weight functions}
    \State Let $w|_T : V(T) \rightarrow \Nfty$ be a new function
    \For{$v \in V(T)$}
    \State $w|_T(v) \gets w(v) - \Dnm s t$
    \EndFor
    \State $W_T \gets W_T \cup \{ w|_T \}$
    \EndFor
    \\
    \Comment{Defining the "help from above" weight function}
    \State Let $c_T : V(T) \rightarrow \Nfty$ be a new function
    \For{$v \in V(T)$}
    \State $c_T(v) \gets \min\limits_{u \in V(S) : (u,v) \in E(H),u \ne t}
    \{\Dnm s u + 1 \} - \Dnm s t$
    \EndFor
    \State $W_T \gets W_T \cup \{ c_T \}$
    \\
    \State $Q_T \gets E(T) \times V(T) \times  \{c_T\}$
    \Comment{Defining the new set of queries}
    \For{$(e,x,w) \in Q$}
    \If {$e \in V(T)$ \textbf{and} $x \in V(T) - \{t\}$}
    \State $Q_T \gets Q_T \cup \{ (e,x,w|_T)\}$
    \EndIf
    \EndFor
    \\
    \State $\widehat{d}(\circ,\circ,\circ) \gets $ \Call{Generalized SSRP}{$H[T],T,W_T,Q_T$}
    \label{lst:line:recursive_call_T}
    \For{$(e,x,w) \in Q$}
    \If {$e \in T$ \textbf{and} $x \in V(T) - \{t\}$}
    \State $\Dhat w s x e \gets \min \{ \Dhat[{H[T]}] {w|_T} t x e + \Dnm s t, \;
    \Dhat[{H[T]}] {c_T} t x e + \Dnm s t
    \}$
    \label{lst:line:TT_set}
    \EndIf
    \EndFor
    \partTitle{The case when the failure is in $S$ and the destination is in $V(S) - \{t\}$}
    \State $W_S \gets \varnothing$
    \For{$w \in W$}
    \Comment{Defining the restricted weight functions}
    \State Let $w|_S : V(S) \rightarrow \Nfty$ be a new function
    \For{$v \in V(S)$}
    \If{$v \in P$}
    \State $w|_S(v) \gets \D w s v P$
    \Else
    \State $w|_S(v) \gets w(v)$
    \EndIf
    \EndFor
    \State $W_S \gets W_S \cup \{ w|_S \}$
    \EndFor
    \\
    \Comment{Defining the "help from below" weight function}
    \State Let $c_S : V(S) \rightarrow \Nfty$ be a new function
    \For{$v \in V(S)$}
    \State $c_S(v) \gets \min\limits_{u \in V(T) : (u,v) \in E(H),u \ne t}
    \{\Dnm s u + 1 \}$
    \EndFor
    \State $W_S \gets W_S \cup \{ c_S \}$
    \\
    \State $Q_S \gets E(S) \times V(S) \times  \{c_S\}$
    \Comment{Defining the new set of queries}
    \For{$(e,x,w) \in Q$}
    \If {$e \in V(S)$ \textbf{and} $x \in V(S) - \{t\}$}
    \State $Q_S \gets Q_S \cup \{ (e,x,w|_S)\}$
    \EndIf
    \EndFor
    \\
    \For{$k := 1$ \textbf{to} $\lgV$}
    \Comment{Defining the pivots weight functions}
    \For{$b \in B_k$}
    \State Let $w_b : V(S) \rightarrow \Nfty$ be a new function
    \For{$v \in V(S)$}
    \State $w_b(v) \gets \Dnm s b + \D {} b v P$
    \EndFor
    \State $W_S \gets W_S \cup \{w_b\}$
    \State $Q_S \gets Q_S \cup E(P_k) \times V(S) \times \{w_b\}$
    \Comment{Adding the relavent queries to $Q_S$}
    \EndFor
    \EndFor
    \\
    \State $\widehat{d}(\circ,\circ,\circ) \gets $ \Call{Generalized SSRP}{$H[S],S,W_S,Q_S$}
    \label{lst:line:recursive_call_S}
    \\
    \For{$k := 1$ \textbf{to} $\lgV$}
    \Comment{Computing the $\pivot ex$ estimation}
    \For{$e \in E(P_k)$}
    \For{$x \in V(S)$}
    \State { $\pivot e x = \min\limits_{b\in B_k}$
    $\{\Dhat[{H[S]}]{w_b}{{s}}{x}{e}-$ $\Dnm {{s}} b+$ $ \depart e b\}$}
    \EndFor
    \EndFor
    \EndFor
    \For{$e \in P_0$}
    \For{$x \in V(S)$}
    \State{$\pivot e x = \depart e x$}
    \EndFor
    \EndFor
    \\
    \For{$(e,x,w) \in Q$}
    \Comment{Computing results}
    \If {$e \in E(S) - E(P)$ \textbf{and} $x \in V(S) - \{t\}$}
    \State $\Dhat w s x e \gets \min \{ \Dhat[{H[S]}] {w|_S} s x e, \;
    \Dhat[{H[S]}] {c_S} s x e
    \}$
    \label{lst:line:SS_set}
    \\
    \ElsIf{$e \in P$ \textbf{and} $x \in V(S) - \{t\}$}
    \State{$\Dhat w {{s}} x e \gets \min\{ \Dhat[{H[S]}]{w|_S}{{s}}{x}{e}, \; \D{w}{{s}}{x}{P}, \;
    \depart e x, \; $}
    \label{lst:line:PS_set_begin}
    \State{
        \hspace{33.2mm}
        $\pivot e x, \; \Dhat[{H[S]}]{c_S}{{s}}{x}{e} + \Dhat w {{s}} t e - \Dnm {{s}} t\}$}
    \label{lst:line:PS_set_end}
    \EndIf
    \EndFor
    \partTitle{Handling the trivial case and outputting the results}
    \State{Construct an LCA data structure for the tree $K$}
    \Comment{See Lemma \ref{lemma:LCA}}
    \For{$(e,x,w) \in Q$}
    \If{$e$ is not on the path from $s$ to $x$ in $K$}
    \Comment{Using the LCA data structure}
    \State{$\Dhat{w}{s}{x}{e} \gets \Dnm s x$}
    \Comment{Taking care of the "trivial" case}
    \EndIf
    \EndFor
    \\
    \State{\Return{$\widehat{d}$}}
    \EndFunction
\end{algorithmic}
\myendalg
\end{center}

%% file: sections/pre_analysis.tex
\section{Proof of correctness}
\label{sec:correctness}
We now prove the correctness of our algorithm.
To do so we  define the following properties for a call to our algorithm.
\begin{definition}[Complete and Sound calls]
    We say that a call to our algorithm is \textit{complete} if for every
    query $(e,x,w) \in Q$ the output of the algorithm to the query is \textbf{at least}
    $\D w s x e$.
    Similarly we say that the call is \textit{sound} if the answer for every
    query $(e,x,w) \in Q$ is \textbf{at most}  $\D w s x e$. 
\end{definition}
We will prove the correctness of our algorithm
by showing that it is complete and that \whp\
it is also sound.
\NL
The proof of both the soundness and completeness will be done by induction on the height
of the recursive calls.
The base case of the induction is a call to the algorithm that makes no recursive calls,
this happens if and only if $n_H$ is at most $6$.
In this case the completeness and soundness of our algorithm holds trivially as the algorithm
computes all of the distances naïvely.
In the induction step,
we assume that both recursive calls the algorithm makes
(in steps \ref{step:TT} and \ref{step:SS} of the algorithm or 
lines \ref{lst:line:recursive_call_T} and \ref{lst:line:recursive_call_S} in the pseudo code 
)
are complete and sound,
and show that the current call is also complete and sound,
as stated in claims \ref{claim:complete_induction},\ref{claim:sound_induction}.
The proof of these claims will be rather involved
and is comprised of an extensive case analysis
that depends on the relation between $e,x$ (the edge failure and destination node)
and $S,T$ (the tree separation).
The different cases are presented in Sections
\ref{subsec:DSTE},
\ref{subsec:PS}
\ref{subsec:PT},
\ref{subsec:TT},
and
\ref{subsec:SS}.
\NL
For the proof of soundness, we define the following property of paths.
\begin{definition}[\simple\ path]
    Let $e \in E(K), x \in V(H), w \in W$ and let
    $R = ({s}=v_1,v_2,...,v_l = x)$ be a path in the graph $H_w-e$.
    $R$ will be called \simple\ if $\forall i \in [l]$, the following holds:
    \snl
    If the path from ${s}$ to $v_i$ in the BFS tree $K$ does not contain $e$,
    then the path $(v_1 , v_2,... v_i)$ is the path from ${s}$ to $v_i$ in the BFS tree $K$.
    \end{definition}
Note that by the weight requirement,
every \RP{${s}$}{$x$}{$e$}{$H_w$}
can be easily transformed
into a \simple\ \RP{${s}$}{$x$}{$e$}{$H_w$}
without increasing it is length.
And so we can assume (\WLOG) that for
every $e \in E(K), x \in V(H), w \in W$,
the path $R(s,x,H_w-e)$ is \simple.
\NL
We remind the reader that a path $R(s,x,H_w-e)$ will be called unweighted if 
it is fully contained in $H-e$ and weighted otherwise.
We note that if a path $R(s,x,H_w-e)$ is weighted
then only the first edge of $R(s,x,H_w-e)$
may be from $E(H_w) - E(H)$, since all edges of $E(H_w) - E(H)$
begin at $s$, the rest of the path will be fully contained in $H$.
\NL
We also remind the reader of the definition of jumping and departing replacement paths,
which is defined only for edge failures from $P$, and was first defined in the overview.
For an edge failure $e \in P$ and a destination node $x \in H$
the path $R(s,x,H_w-e)$ will be called jumping if it uses some node $u$
such that $u \in P$ and $u$ is \textbf{after} the edge
failure $e$ in the path $P$.
A path which is not jumping will be called departing.
\NL

%% file: sections/DEPART.tex
\subsection{The \mathtitlewrapper{$\depart e x$} value}
\label{subsec:DEPART}
The $\depart \circ \circ$ values are used by the algorithm in several cases,
so we now state some auxiliary claims regarding these values
that will be used later on.
The proofs of these claims are deferred to the appendix.

\begin{restatable}[Proof of Completness]{claim}{completeDEPART}
    \label{claim:depart_compleness}
    Let $e \in P$
    be an edge failure
    and let $x \in V$ be a destination,
    then $\depart e x \ge \D {} s x e$.
\end{restatable}
\begin{restatable}[Proof of Soundness]{claim}{soundDEPART}
    \label{claim:sound_depart_unweighted}
    Let $e \in E(P), x \in V(H), w \in W$,
    assume $\R w s x e$ is departing and unweighted,
    \textbf{and} assume that
    $\R w s x e$ contains some node $v_r \in V(T) - \{t\}$.
    Then \whp\ the length of $\R w s x e$ is at least $\depart e x$.
\end{restatable}

%% file: sections/DSTE.tex
\subsection{The case when 
\mathtitlewrapper{$e \in P$} and 
\mathtitlewrapper{$x = t$}
}
\label{subsec:DSTE}
The proof of correctness of this case can be found in the appendix.
We state here the specific claims proved in the appendix as they will be 
used in later cases.

\begin{restatable}[Proof of Completeness]{claim}{completeDSTE}
    \label{claim:completeDSTE}
    Let $e \in P$ be an edge failure and $w \in W$ a weight function,
    then $\Dhat wste \ge \D wste$.
\end{restatable}

\begin{restatable}[Proof of Soundness]{claim}{soundDSTE}
    \label{claim:sound_DSTE}
    Let $e \in E(P), w \in W$, \whp\ $\D w s t e \ge \Dhat w s t e$.
\end{restatable}

%% file: sections/PS.tex
\subsection{The case when
\mathtitlewrapper{$e \in P$} and
\mathtitlewrapper{$x \in V(S) - \{t\}$}
}
\label{subsec:PS}
We now consider the most involved case, in which
the edge failure is from $P$ and the destination node is from $V(S)-\{t\}$.
Recall that in this case
the algorithm sets
$\Dhat w {{s}} x e$
to be the minimum between:
$\Dhat[{H[S]}]{w|_S}{{s}}x e,$ $ \D{w}{{s}}x{P},$
$ \depart e x,$
$\pivot e x$
and
$\Dhat[{H[S]}]{c_S}e{{s}}x + \Dhat w {{s}} t e - \Dnm {{s}} t$,
in step \ref{step:SS} of the algorithm (lines \ref{lst:line:PS_set_begin}-\ref{lst:line:PS_set_end}
of the pseudocode).
\NL
\subsubsection*{Proof of Completeness}
We begin by
showing that the weight functions $w|_S$ are truthful for
edge failures from both $P$ \textbf{and} $E(S) - E(P)$.
More formally we prove the following claim: 
\begin{claim}
    \label{claim:wS_complete}
    Let $w \in W , e \in E(S), x \in V(S) - \{t\}$,
    then $\D [{H[S]}] {w|_S} s x e \ge \D w s x e$.
\end{claim}
\begin{proof}
Let us denote the shortest \RP{$s$}{$x$}{$e$}{$H[S]_{w|_S}$} by $R$.
Note that if $R \subseteq H[S]-e$ then
we have that $R$ is a \RP{$s$}{$x$}{$e$}{$H$} $\subseteq H_w - e$.
This implies that $\D w s x e \le d(R) = \D[{H[S]}] {w|_S} s x e$ as required.
So we may assume that $R$ uses some weighted edge $(s,v) \in E(H[S]_{w|_S}) - E(H[S])$,
that is of weight $w|_S(v)$.
Note that other than this edge, the path $R$ is fully contained in the graph $H[S] - e$.
So we wish to replace the edge $(s,v)$
with some path from $s$ to $v$ in $H_w-e$ of length at most $w|_S(v)$.
In order to do so we distinguish between 3 cases depending on the relation between
$e,v$ and the path $P$.
\NL
If $v \notin P$ then $w|_S(v) = w(v)$,
and so we can simply replace the edge $(s,v)$ with the edge $(s,v) \in E(H_w) - E(H)$.
\snl
If $v \in P$ then 
$w|_S(v) = \D w s v P$.
If $e \in P$ we can simply replace the edge $(s,v)$ with shortest
\RP{$s$}{$v$}{$P$}{$H_w$}.
Note that we can do that since $e \in P$ and so $H_w - P$ does not contain $e$.
\snl
If $e \notin P$ we can replace the edge $(s,v)$ with the subpath
of $P$ from $s$ to $v$. Since $P$ is a shortest path in $H$
the length of this path is $\Dnm s v = \Dnm [{H_w}] s v \le \D w s v P = w|_S(v)$,
where the first equality holds by the weight requirements.
\NL
Let $\overline{R}$ denote the path resulted after one of these replacements.
The path $\overline{R}$ is a \RP{$s$}{$x$}{$e$}{$H_w$},
this implies that $\D w s x e \le d(\overline{R})$.
Since we have shown that in all of the cases $d(\overline{R}) \le d(R)$,
and since $d(R) = \D[{H[S]}] {w|_S} s x e$,
the claim holds.
\end{proof}

\begin{claim}
    \label{claim:complete_PS_cS}
    Let $e\in E(P), x \in V(S) - \{t\}, w \in W$.
    Assuming that the recursive call over $H[S]$ is complete, then
    $\Dhat [{H[S]}] {c_S} s x e + \Dhat w s t e - \Dnm s t
    \ge \D w s x e$.
\end{claim}
\begin{proof}
    A crucial observation is that $\Dhat w s t e \ge \Dnm s t$.
    To see this note that
    $\Dhat w s t e \ge \D w s t e \ge \Dnm [{H_w}] s t = \Dnm s t$,
    where the first inequality holds by Claim \ref{claim:completeDSTE}
    and the last equality holds by the weight requirements.
    \NL
    Recall that $\Dhat [{H[S]}] {c_S} s x e$ is the result
    obtained by the recursive call of the algorithm over
    the sub-graph $H[S]$ and the query $(e,x,c_S)$.
    Let $R$ be the shortest \RP{$s$}{$x$}{$e$}{$H[S]_{c_S}$}.
    By the assumption that the recursive call over $H[S]$ is complete, we have that
    $d(R) \le \Dhat [{H[S]}] {c_S} s x e$.
    \snl
    If $R$ uses no edges from $E(H[S]_{c_S}) - E(H[S])$,
    then $R \subseteq H[S] - e \subseteq H - e \subseteq H_w - e$.
    So $R$ is a \RP{$s$}{$x$}{$e$}{$H_w$}, 
    meaning that $\D w s x e \le d(R)$.
    Since we have shown that $\Dhat w s t e \ge \Dnm s t$
    we get that
    $\D w s x e \le d(R) \le d(R) + \Dhat w s t e - \Dnm s t
    \le \Dhat [{H[S]}] {c_S} s x e + \Dhat w s t e - \Dnm s t$
    which implies the claim.
    \NL
    So we may assume that $R$ uses some edge
    $(s,v) \in E(H[S]_{c_S}) - E(H[S])$.
    As in the proof of Claim \ref{claim:wS_complete}, we wish to slightly fix $R$ by replacing $(s,v)$ with some
    \RP{$s$}{$v$}{$e$}{$H_w$}.
    Recall that the weight of the edge $(s,v)$ is $c_S(v)$.
    By definition of $c_S$ we have that $c_S(v) = \Dnm s u + 1$
    for some $u \in V(T)$ such that $u\ne t$ and $(u,v) \in E$.
    \snl
    Since $u \in T$ we can denote by $R(t,u,H)$ the path from $t$ to $u$ in
    the BFS tree $T$.
    Since $T$ is a BFS tree the length of $R(t,u,H)$ is $\Dnm t u$.
    Note that since $e \in P \subseteq S$ we have that
    $e \notin R(t,u,H)$.
    Also, since $u \in V(T), u \ne t$ we have that $u\notin S$
    and so $(u,v) \notin S$, and so $(u,v) \ne e$.
    Let $\R w s t e$ denote the shortest \RP{$s$}{$t$}{$e$}{$H_w$}.
    Note that by Claim \ref{claim:completeDSTE} we have that
    $d(\R w s t e) \le \Dhat w s t e$.
    Overall we have that $\overline{R} = \R w s t e \circ R(t,u,H) \circ (u,v)$
    is a \RP{$s$}{$v$}{$e$}{$H_w$}
    of length at most $\Dhat w s t e + \Dnm t u + 1$.
    \snl
    Note that since $u \in V(T)$,
    we have that $s$ is an ancestor of $t$ which is an ancestor
    of $u$ in the BFS tree $K$, which implies
    that $\Dnm s u = \Dnm s t + \Dnm u t$.
    This implies that $\Dnm t u + 1 = \Dnm s u - \Dnm s t + 1 = c_S(v) + 1$.
    So we have that $d(\overline{R}) \le 
    c_S(v) + \Dhat w s t e - \Dnm s t$.
    So if we replace the edge $(s,v)$ with the path $\overline{R}$
    we get a \RP{$s$}{$x$}{$e$}{$H_w$} of length at most
    $d(R) + \Dhat w s t e - \Dnm s t$
    (since we increased the length of $R$ by $\Dhat w s t e - \Dnm s t$).
    This implies that
    $\D w s x e \le d(R) + \Dhat w s t e - \Dnm s t \le
    \Dhat [{H[S]}] {c_S} s x e + \Dhat w s t e - \Dnm s t$.
\end{proof}

\begin{claim}
    \label{claim:complete_PS_pivot}
    Let $e\in E(P), x \in V(S) - \{t\}$.
    Assuming that the recursive call over $H[S]$ is complete,
    then $\pivot e b \ge \D {} s x e$.
\end{claim}
\begin{proof}
    Let $k\in [\lgV] \cup \{0\}$
    be the unique integer such that $e \in P_k$.
    If $k=0$ we have that $\pivot e x = \depart e x = \D {} s x e$
    which implies the claim.
    Otherwise $k \ne 0$,
    and so $\pivot e x = \Dhat[{H[S]}] {w_b} s x e - \Dnm s b + \depart e b$ for some
    $b \in B_k$.
    \NL
    Similarly to the proof of Claim \ref{claim:complete_PS_cS}, a crucial observation is that $\depart e b \ge \Dnm s b$.
    To see this note that
    $\depart e b \ge \D {} s b e \ge \Dnm s b$,
    where the first inequality is by Claim \ref{claim:depart_compleness}.
    \NL
    Recall that $\Dhat[{H[S]}] {w_b} s x e$ is the result
    obtained by the recursive call of the algorithm over
    the subgraph $H[S]$ and the query $(e,x,w_b)$.
    Let $R$ be the shortest \RP{$s$}{$x$}{$e$}{$H[S]_{w_b}$}.
    Since the recursive call over $H[S]$ is complete we have that
    $d(R) \le \Dhat[{H[S]}] {w_b} s x e$.
    If $R$ uses no edges from $E(H[S]_{w_b}) - E(H[S])$,
    then $R \subseteq H[S] - e \subseteq H - e$.
    So $R$ is a \RP{$s$}{$x$}{$e$}{$H$},
    meaning that $d(R) \ge \D {} s x e$.
    Since we have shown that $\depart e b \ge \Dnm s b$ we get that
    $\D {} s x e \le
    d(R) \le d(R) + \depart e b  - \Dnm s b \le
    \Dhat[{H[S]}] {w_b} s x e +  \depart e b  - \Dnm s b = \pivot{e}{x}$
    as required.
    \snl
    So we may assume $R$ uses some edge
    $(s,v) \in E(H[S]_{w_b}) - E(H[S])$.
    As in previous cases we wish to slightly fix $R$ by replacing $(s,v)$ with some
    \RP{$s$}{$v$}{$e$}{$H$}.
    Recall that the weight of the edge $(s,v)$ is $w_b(v) = \Dnm s b + \D {} b v P$.
    Let $\R {} s b e$ denote the shortest \RP{$s$}{$b$}{$e$}{$H$},
    note that by Claim \ref{claim:depart_compleness},
    $d(\R {} s b e) \le \depart e b$.
    Let $\R {} b v P$ denote the shortest \RP{$b$}{$v$}{$P$}{$H$},
    note that since $e \in P$ we have that $e\notin \R {} b v P$.
    So the path $\overline{R} = \R {} s b e \circ \R {} b v P$
    is a \RP{$s$}{$v$}{$e$}{$H$} of length at most
    $\depart e b + \D {} b v P$.
    Note that $d(\overline{R}) \le \depart e b + \D {} b v P
    = \Dnm s b + \D {} b v P + \depart e b - \Dnm s b = w_b(v) + \depart e b - \Dnm s b$.
    So if we replace the edge $(s,v)$ with the path $\overline{R}$
    we get a \RP{$s$}{$x$}{$e$}{$H$} of length at most
    $d(R) + \depart e b - \Dnm s b$
    (since we increased the length of $R$ by at most $\depart e b - \Dnm s b$).
    This implies that
    $\D {} s x e \le d(R) + \depart e b - \Dnm s b \le
    \Dhat [{H[S]}] {w_b} s x e + \depart e b  - \Dnm s b = \pivot e x$ as required.
\end{proof}

\begin{claim}
    \label{claim:complete_PS}
    Let $(e,x,w) \in Q$ be a query such that $e \in P$
    and $x \in V(S) - \{t\}$.
    Assuming that the recursive call over $H[S]$
    is complete,
    then $\Dhat w s x e \ge \D w s x e$.
\end{claim}
\begin{proof}
    We show that each one of the elements in the minimum that defines
    $\Dhat w s x e $ is at least $ \D w s x e$.
    This will suffice to show that  $\Dhat w s x e $ is at least $ \D w s x e$.
    \snl
    Note that since $H-e \subseteq H_w - e$ we have that
    $\D {} s x e \ge \D w s x e$.
    So for the cases of
    $ \depart e x,$ $\Dhat[{H[S]}]{c_S}e{{s}}x + \Dhat w {{s}} t e - \Dnm {{s}} t$
    and $\pivot e x$ we have by Claims \ref{claim:depart_compleness}, \ref{claim:complete_PS_cS}
    and \ref{claim:complete_PS_pivot} (correspondingly) that each one of these values
    is at least $\D w s x e$.
    \snl
    For the case of $\D{w}{{s}}x{P}$, since $e \in P$ we have that $H_w - P \subseteq H_w - e$
    and so $\D{w}{{s}}x{P} \ge \D w s x e$.
    \snl
    For the case of $\Dhat[{H[S]}]{w|_S}{{s}} x e$
    since the recursive call over $H[S]$ is complete we have that
    $\Dhat[{H[S]}]{w|_S}{{s}} x e \ge \D[{H[S]}]{w|_S}{{s}} x e$
    and by Claim \ref{claim:wS_complete} we have that
    $\D[{H[S]}]{w|_S}{{s}} x e \ge \D w s x e$,
    these implies the claim.
\end{proof}

\mynewpage
\subsubsection*{Proof of Soundness - Departing Paths}
We again start with a couple of general claim which will be useful
in future cases.
\begin{claim}
    \label{claim:sound_PS_noT}
    Let $e \in E(S), x \in V(S) - \{t\}, w \in W$,
    assume $\R w {{s}} x e$ \textbf{uses no nodes} from $V(T)-\{t\}$,
    then its length is at least $\D [{H[S]}] {w|_S} s x e$.
\end{claim}
\begin{proof}
    If $\R w {{s}} x e$ is unweighted then since it does not use any
    edges from $V(T)-\{t\}$ it is a
    \RP{$s$}{$x$}{$e$}{$H[S]$}.
    Since $H[S]-e \subseteq H[S]_{w|_S} - e$
    it is also a path in $H[S]_{w|_S} - e$ and so its length must be at least
    $\D[{H[S]}] {w|_S} s x e $ as required.
    \NL
    The more interesting case is when $\R w {{s}} x e$ is weighted,
    meaning the first edge is the weighted edge $(s,v) \in E(H_w) - E(H)$.
    Since $\R w {{s}} x e$ uses no nodes from $V(T) - \{t\}$
    the rest of the edges are from $H[S]$.
    So if we replace the edge $(s,v) \in E(H_w) - E(H)$
    with the edge $(s,v) \in E(H[S]_{w|_S}) - E(H[S])$,
    we get a new \RP{$s$}{$x$}{$e$}{$H[S]_{w|_S}$},
    and so its length is at least $\D[{H[S]}] {w|_S} s x e $.
    We claim that by replacing the edge we \textbf{do not increase}
    the length of $\R w {{s}} x e$, this will imply that the length of $\R w {{s}} x e$
    is at least $\D[{H[S]}] {w|_S} s x e $ as well, which implies the claim.
    \snl
    To see why the length \textbf{does not increase} note that the length difference
    between $\R w {{s}} x e$ and the new path is exactly $w|_S(v) - w(v)$,
    so we only need to show that $\forall v \in V(S) : w|_S(v) \le w(v)$.
    To see this note that is $v \notin P$ we have that $w|_S(v) = w(v)$ by definition and
    as required.
    If $v \in P$ the we have $w|_S(v) = \D w s v P$.
    Since $(s,v) \in E(H_w) - E(H)$ is a path of length $w(v)$
    from $s$ to $v$ in the graph $H_w - P$, we have that $\D w s v P \le w(v)$,
    as required.
\end{proof}

\begin{claim}
    \label{claim:sound_depart_weighted}
    Let $e \in E(P),x \in V(H),w \in W$,
    assume $\R w s x e$ is departing and weighted.
    Then the length of $\R w s x e$ is $\D w s x P$.
\end{claim}
\begin{proof}
We claim that $V(\R w s x e) \cap V(P) = \{s\}$.
To see this assume for the sake of contradiction that there is some node $v \in \R w s x e$
such that $v \in P - \{s\}$.
If $v$ is \textbf{after} the edge failure $e$ in $P$ then $\R w s x e$
is jumping and hence not departing.
So $v$ must be \textbf{before} the edge failure $e$ in $P$,
this implies that the path from $s$ to $v$ in the BFS tree $K$
does not contain $e$.
Since we assume $\R w s x e$ is \simple, this implies that the subpath
of $\R w s x e$ from $s$ to $v$ is the path from $s$ to $v$ in $K$.
However since $\R w s x e$ is weighted, the very first edge of  $\R w s x e$
is not from $H$, and hence not from $K$, contradiction.
\snl
So we conclude that $V(\R w s x e) \cap V(P) = \{s\}$,
and so $E(\R w s x e) \cap E(P) = \varnothing$.
Since $e \in P$ this implies that $\R w s x e$ is a shortest
path from $s$ to $x$ in the graph $H_w-P$, and so
its length is $\D w s x P$.
\end{proof}

Now, using previously made claims,
we show that if a replacement path is departing,
we have successfully computed its weight (\whp).
We show this by proving the following claim:
\begin{claim}
    \label{claim:sound_PS_departing}
    Let $(e,x,w) \in Q$ be a query such that $e \in P$ and $x \in V(S) - \{t\}$.
    Assume that the recursive call over the subgraph $H[S]$ is sound and assume that
    $\R w s x e$ is departing.
    Then \whp\ $\D w s x e \ge \Dhat w s x e$.
\end{claim}
\begin{proof}
    Note that if $\R w s x e$ uses no nodes from $V(T)-\{t\}$
    then we fall in the case of Claim \ref{claim:sound_PS_noT},
    and so its length is at least $\D [{H[S]}] {w|_S} s x e$.
    Since the recursive call over $H[S]$ is sound we have that
    $\D [{H[S]}] {w|_S} s x e \ge \Dhat [{H[S]}] {w|_S} s x e$
    which is at least $\Dhat w {{s}} x e$.
    This implies the claim.
    \snl
    Otherwise if $\R w s x e$ does use a node from $V(T)-\{t\}$ and is unweighted,
    then we fall in the case of Claim \ref{claim:sound_depart_unweighted}, and so
    the length of $\R w s x e$ is \whp\ at least $\depart e x$
    which is at least $\Dhat w {{s}} x e$.
    This implies the claim.
    \snl
    Finally if $\R w s x e$ is weighted
    then we fall in the case of Claim \ref{claim:sound_depart_weighted}, and so
    the length of $\R w s x e$ is at least $\D w s x P$
    which is at least $\Dhat w {{s}} x e$.
    This implies the claim.
\end{proof}

\subsubsection*{Proof of Soundness - Jumping Paths}
So we may assume that the path $\R w s x e$ is jumping.
We define the following notation:
\snl
Since $\R w s x e$ is jumping it must first leave the path $P$
at some node $v_i \in P$ before the edge failure ($e$)
and then return to $P$
at some node $v_j \in P$.
Let us denote the subpath of $\R w s x e$ from $v_i$ to $v_j$ by $R[v_i,v_j]$ .
Note that $V(R[v_i,v_j]) \cap P = \{v_i,v_j\}$ (since $\R w s x e$ returns to $P$ at $v_j$)
and that the path $R[v_i,v_j]$ is edge disjoint to $P$.
We will use this notation for the rest of this section.
So it is important to visually grasp it.
Illustrations for this notation can be seen in Figures
\ref{fig:jumpingA},\ref{fig:jumpingB},\ref{fig:jumpingC}.
\NL
Note that all the nodes from $\R w s x e$ which after $v_i$ and
are from $P$ (in particular $v_j$) must be after the edge failure $e$
as otherwise $\R w s x e$ was not \simple.

\mynewpage
\begin{claim}
    \label{claim:sound_PS_jumpingA}
    Under the above notation, assume that
    $\R w s x e$ \textbf{does} use a node $v_r \in V(T) -\{t\}$ \textbf{outside} of the
    subpath $R[v_i,v_j]$.
    Then the length of $\R w s x e$
    is at least $\D [{H[S]}] {c_S} s x e - \Dnm s t + \D w s t e$.
\end{claim}
We would like to clarify some important characterizations of the current case.
Note that $\R w s x e$ uses $v_r$ is after it uses $v_j$
(it cannot use it before $v_i$ since the subpath from $s$ to $v_i$ is a subpath of $P$).
Note also that since $v_j$ is after the edge failure $e$ in $P$,
and since $v_r \in T$,
then when $\R w s x e$ goes from $v_j$ to $v_r$ it
passes through (\WLOG) the separator $t$.
An illustration of this case can be seen in Figure \ref{fig:jumpingA}.
\begin{proof}
    We can split the path $\R w s x e$
    into two subpath:
    $R[s,t]$ - the subpath from $s$ to $t$,
    and $R[t,x]$ - the subpath from $t$ to $x$.
    Note that $R[s,t]$ is a replacement \RP{$s$}{$t$}{$e$}{$H_w$},
    and so its length is
    $\D w s t e$.
    \NL
    We claim that the length of $R[t,x]$ is at least
    $\D [{H[S]}] {c_S} s x e - \Dnm s t$. Clearly this will imply the claim.
    To see why this is true, recall that $R[t,x]$ contains a node from $V(T)-\{t\}$
    (namely $v_r$).
    Let $u$ be the last node in $R[t,x]$ which is from $V(T)-\{t\}$
    and let $v$ be the node right after it.
    This implies that $(u,v) \in E$,
    and so by definition $c_S$ we have that $c_S(v) \le \Dnm s u + 1$.
    Let us denote the subpath of $R[t,x]$ from $t$ to $v$ by $R[t,v]$.
    The length of $R[t,v]$ is at least
    $\Dnm t u + 1 = \Dnm s t + \Dnm t u + 1 - \Dnm s t
    \ge \Dnm s u + 1 - \Dnm s t \ge c_S(v) - \Dnm s t$.
    So we conclude that
    $d(R[t,v]) \ge c_S(v) - \Dnm s t$.
    \NL
    Let us denote by $R[v,x]$ the subpath of $R[t,x]$ from $v$ to $x$.
    Note that since $u$ was the last node from $V(T)-\{t\}$,
    then $R[v,x]$ is fully contained in $H[S]$.
    Consider the edge $(s,v) \in E(H[S])_{c_S} - E(H)$, its weight is $c_S(v)$.
    The path $(s,v) \circ R[v,x]$ is a \RP{$s$}{$x$}{$e$}{$H[S]_{c_S}$}
    and so its length is at least $\D[{H[S]}] {c_S} s x e$.
    This implies that $d(R[v,x]) + c_S(v) \ge \D [{H[S]}] {c_S} s x e$.
    Since we have shown that $d(R[t,v]) \ge c_S(v) - \Dnm s t$
    by simple arithmetics we get that
    $d(R[t,v]) + d(R[v,x]) \ge \D [{H[S]}] {c_S} s x e - \Dnm s t$,
    which concludes the proof.
\end{proof}

\begin{claim}
    \label{claim:sound_PS_jumpingC}
    Under the above notation, assume that
    $\R w s x e$ \textbf{does not use} nodes from $V(T) -\{t\}$ \textbf{outside} of the
    subpath $R[v_i,v_j]$.
    Also assume that $\R w s x e$ is \textbf{weighted}.
    Then the length of $\R w s x e$
    is at least $\D[{H[S]}] {w|_S} s x e$.
\end{claim}
Note that in this case, $\R w s x e$ leaves the path $P$
at $s$ (since all weighted edges begin at $s$) and so $v_i=s$ by definition.
An illustration of this case can be seen in Figure \ref{fig:jumpingC}.
\begin{proof}
    Let us rename $R[v_i,v_j]$ to $R[s,v_j]$ for clarity.
    Recall that by a previously made observation $R[s,v_j]$ is edge disjoint to
    the path $P$.
    Since $e \in P$, this implies that the length of
    $R[s,v_j]$ is exactly $\D w s {v_j} P$
    which is exactly $w|_S(v_j)$
    since $v_j \in P$.
    So we have that $d(R[s,v_j]) = w|_S(v_j)$.
    \snl
    Let us denote by $R[v_j,x]$ the subpath of $\R w s x e$ from $v_j$ to $x$.
    We claim that the length of $R[v_j,x]$ is at least $\D[{H[S]}] {w|_S} s x e - w|_S(v_j)$.
    In-fact, showing this will suffice in order to prove the claim,
    since the length of $\R w s x e$ is $d(R[s,v_j]) + d(R[v_j,x])$
    and we have shown that $d(R[s,v_j]) = w|_S(v_j)$.
    \snl
    So we aim to show that the length of $R[v_j,x]$ is at least $\D[{H[S]}] {w|_S} s x e - w|_S(v_j)$,
    to see this consider the edge $(s,v_j) \in E(H[S]_{w|_S}) - E(H[S])$,
    its weight is $w|_S(v_j)$.
    Since $\R w s x e$ \textbf{does not} use any nodes from $V(T)-\{t\}$
    outside of the subpath $R[s,v_j]$, the path
    $R[v_j,x]$ is fully contained in $H[S]$.
    And so the path $(s,v_j) \circ R[v_j,x]$ is a \RP{$s$}{$x$}{$e$}{$H[S]_{w|_S}$},
    and so its length is at least $\D[{H[S]}] {w|_S} s x e$.
    This implies that the length of $R[v_j,x]$ is at least $\D[{H[S]}] {w|_S} s x e - w|_S(v_j)$.

\end{proof}

\begin{claim}
    \label{claim:sound_PS_jumpingB}
    Under the above notation, assume that
    $\R w s x e$ \textbf{does not use} nodes from $V(T) -\{t\}$ \textbf{outside} of the
    subpath $R[v_i,v_j]$, but \textbf{does} use a node $v_r \in V(T) -\{t\}$
    \textbf{inside} the subpath $R[v_i,v_j]$.
    Also, assume that  $\R w s x e$ is \textbf{unweighted}
    and that the recursive call over the subgraph $H[S]$ is sound.
    Then \whp, the length of $\R w s x e$
    is at least $\pivot e x$.
\end{claim}
An illustration of this case can be seen in Figure \ref{fig:jumpingB}
\begin{proof}
    The proof for this claim is rather similar
    to the proofs for Claims \ref{claim:sound_PS_jumpingA} and \ref{claim:sound_depart_unweighted}.
    Note that since $\R w s x e$ is unweighted its length is exactly
    $\D {} s x e$.
    And so if $e \in P_0$ we have by the definition of $\pivot e x$
    that $\pivot e x = \depart e x = \D {} s x e$ and so the claim holds.
    The more interesting case would be when $e \in P_k$ for some $k \in [\lgV]$.
    By similar observations to those made in the proof of Claim
    \ref{claim:sound_depart_unweighted} we can show that
    the length of the path from $v_i \in P$ to $v_r \in T$ is at least $2^k \sqrtV$,
    and so \whp\ we have sampled some $b \in B_k$ such that $b \in R[v_i,v_j]$.
    So we can split the path $\R w s x e$ into two subpaths:
    $R[s,b]$ - the subpath of $\R w s x e$ from $s$ to $b$,
    and $R[b,x]$ - the subpath of $\R w s x e$ from $b$ to $x$.
    \NL
    Note that since the subpath $R[v_i,v_j]$ is edge-disjoint to $P$,
    we have that the path from $v_i$ to $b$ uses no edges from $P$.
    Since $e \in P$, this implies that the length of $R[s,b]$ is exactly
    $\Dnm s {v_i} + \D {} {v_i} b P$.
    Since $v_i$ is before $e$,
    by the definition of $\depart e b$ we get that
    the length of $R[s,b]$ is at least $\depart e b$.
    \NL
    We now show that the length of $R[b,x]$ is at least
    $\D [{H[S]}] {w_b} s x e - \Dnm s b$.
    In-fact we claim that showing this will suffice in order to prove the claim.
    To see this note that by the assumption that the recursive
    call over the subgraph $H[S]$ is sound we have that
    $\Dhat [{H[S]}] {w_b} s x e \le \D [{H[S]}] {w_b} s x e$.
    Hence, the length of $\R w s x e$ is
    $d(R[s,b]) + d(R[b,x])$
    which is at least
    $
    \Dhat [{H[S]}] {w_b} s x e -  \Dnm s b + \depart e b$
    which is at least $\pivot e x$ by its definition.
    \NL
    So we aim to prove that the length of $R[b,x]$ is at least
    $\D [{H[S]}] {w_b} s x e - \Dnm s b$.
    Recall that $v_j \in R[b,x]$, and so
    let us denote by $R[b,v_j]$ the subpath of $R[b,x]$
    from $b$ to $v_j$,
    and by $R[v_j,x]$ the subpath of $R[b,x]$
    from $v_j$ to $x$.
    Note that since the path $R[v_i,v_j]$ is edge-disjoint to $P$,
    and since $e \in P$, the length of
    $R[b,v_j]$ is exactly $\D {} b {v_j} P = \Dnm s b + \D {} b {v_j} P - \Dnm s b = w_b(v_j) - \Dnm s b$.
    \snl
    We now aim to lower bound the length of $R[v_j,x]$.
    Since $\R w s x e$ uses no nodes from $V(T) - \{t\}$
    outside of the subpath $R[v_i,v_j]$, the path $R[v_j,x]$
    is fully contained in $H[S]$.
    Consider the edge $(s,v_j) \in H[S]_{w_b}$, its weight is $w_b(v_j)$.
    The path $(s,v_j) \circ R[v_j,x]$ is then a \RP{$s$}{$x$}{$e$}{$H[S]_{w_b}$},
    and so its length is at least $\D[{H[S]}] {w_b} s x e$.
    This implies that $d(R[v_j,x]) + w_b(v_j) \ge \D[{H[S]}] {w_b} s x e$.
    Since we have shown that $d(R[b,v_j]) = w_b(v_j) - \Dnm s b$
    by simple arithmetics we get that
    $d(R[b,v_j]) + d(R[v_j,x]) \ge \D [{H[S]}] {w_b} s x e - \Dnm s b$.
    Since the length of $R[b,x]$ is $d(R[b,v_j]) + d(R[v_j,x])$
    this concludes the proof.
\end{proof}

\begin{claim}
    \label{claim:sound_PS_jumping}
    Let $(e,x,w) \in Q$ be a query such that $e \in P, x \in V(S) -\{t\}$.
    Assume that the recursive call over $H[S]$ is sound
    and assume $\R w {{s}} x e$ is jumping.
    Then \whp\
    $\D w s x e \ge \Dhat w s x e$.
\end{claim}
\begin{proof}
    Note that if $\R w s x e$ uses no nodes from $V(T)-\{t\}$
    then we fall in the case of Claim \ref{claim:sound_PS_noT},
    meaning that the length
    of $\R w s x e$ is at least $\D[{H[S]}] {w|_S} s x e$.
    Since the recursive call over $H[S]$ is sound, we have that
    $\D[{H[S]}] {w|_S} s x e \ge \Dhat[{H[S]}] {w|_S} s x e$,
    and so the length of $\R w {{s}} x e$ is at least $\Dhat w s x e$.
    \NL
    If $\R w s x e$ does use a node from $V(T)-\{t\}$ we must fall in the one of the cases of Claims
    \ref{claim:sound_PS_jumpingA},
    \ref{claim:sound_PS_jumpingC} or
    \ref{claim:sound_PS_jumpingB}.
    In total we have proven in each case that the length of $\R w s x e$
    is \whp\ at least one of the elements in the minimum defining  $\Dhat w s x e$.
    And so the length of $\R w {{s}} x e$ is \whp\ at least $\Dhat w s x e$.
\end{proof}

%% file: sections/PT.tex
\subsection{The case when 
\mathtitlewrapper{$e \in P$} and 
\mathtitlewrapper{$x \in V(T) - \{t\}$}
}
\label{subsec:PT}
Recall that in the case when $e \in E(P)$ and $x \in V(T)$,
the algorithm sets $\Dhat w s x e$
to be the minimum between 
$\D w s x P$, 
$\depart e x$ and 
$\Dhat w s t e + \Dnm t x$
in step \ref{step:PT} of the algorithm (line \ref{lst:line:PT_set} of the pseudocode).
\NL
\subsubsection*{Proof of Completeness}
\begin{claim}
    \label{claim:completePT}
    Let $(e,x,w)\in Q$ be a query that $e \in P$ and $x \in V(T) - \{t\}$,
    then $\Dhat w s x e \ge \D w s x e$ 
\end{claim}
\begin{proof}
    We again use the proof technique used in the proof of Claim
    \ref{claim:complete_PS}, and show that each one of the elements 
    in the minimum defining  $\Dhat w s x e$ is at least $\D w s x e$.
    For the case of $\depart e x$ we have by Claim
    \ref{claim:depart_compleness} that $\depart e x \ge \D {} s x e$
    which is at least $\D w s x e$ since $H-e \subseteq H_w-e$.
    \snl
    For the case of $\D w s x P$, note that since 
    $e \in P$ we have that $H_w - P \subseteq H_w - e$
    and so $\D w s x P \ge \D w s x e$.
    \snl
    It remains to take care of the case of $\Dhat w s t e + \Dnm t x$.
    Note that the path from $t$ to $x$ in $K$
    is fully contained in $T$ and so does not contain $e$
    (since $e \in P \subseteq S$).
    Since by the weight requirements $K$ is a BFS tree of $H_w$ 
    this implies that $\D w t x e = \Dnm t x$.
    By the triangle inequality we get that 
    $\D w s x e \le \D w s t e + \D w t x e = \D w s t e + \Dnm t x$.
    Since by Claim \ref{claim:completeDSTE}
    $\D w s t e \le \Dhat w s t e$
    we get that $\D w s x e \le \Dhat w s t e + \Dnm t x$.
\end{proof}
\subsubsection*{Proof of Soundness}
\begin{claim}
    \label{claim:sound_PT}
    Let $(e,x,w) \in Q$ be a query such that $e \in P$ and $x \in V(T) - \{t\}$.
    Then \whp\ $\D w s x e \ge \Dhat w s x e$.
\end{claim}
\begin{proof}
    If $\R w s x e$ is departing and weighted then by Claim \ref{claim:sound_depart_weighted}
    its length is $\D w {{s}} x P$ and so its length is at least  $\Dhat w {{s}} x e$.
    If $\R w s x e$ is departing and unweighted, since it contains a node from $V(T) - \{t\}$
    ($x \in V(T) - \{t\}$), by Claim \ref{claim:sound_depart_unweighted} its length is \whp\
    at least $\depart{e}{x}$ and so its length is \whp\ at least $\Dhat w {{s}} x e$.
    \NL
    We claim that if $\R w s x e$ is jumping then its length is at least $\D w {{s}} t e + \Dnm t x$.
    In-fact showing this will suffice in order to prove the claim since by Claim
    \ref{claim:sound_DSTE} we have that $\Dhat w {{s}} t e \le \D w {{s}} t e$ \whp\
    and so the length of $\R w s x e$ is \whp\ at least $\Dhat w {{s}} t e + \Dnm s t \ge \Dhat w {{s}} x e$.
    To see this note that if $\R w s x e$ is jumping it uses a node $u \in V(P)$
    which is after the edge failure $e$.
    Note that since $u$ is after the edge failure $e$ in $P$, the path from $u$ to $x$ in $K$
    does not contain $e$.
    Since $K$ is a BFS tree in $H_w$ (by the weight requirements) we can assume
    (\WLOG) that when  $\R w s x e$ goes from $u$ to $x$ it takes the path from $u$ to $x$ in $K$.
    Since this path passes through the separator $t$ we can split $\R w s x e$
    into two subpaths: 
    $R[s,t]$ a subpath of $\R w s x e$ from $s$ to $t$,
    and $R[t,x]$ the subpath of $\R w s x e$ from $t$ to $x$.
    The first path $R[s,t]$ is a \RP{$s$}{$t$}{$e$}{$H_w$},
    and so its length is at least $\D w s t e$,
    and the length of the second path is at least $\Dnm[{H_w}] t x$ 
    which is $\Dnm t x$
    (by the weight requirements).
    So the length of the path 
    $\R w s x e$ is at least $\D w {{s}} t e + \Dnm t x$ as required.
\end{proof}

%% file: sections/TT.tex
\subsection{The case when 
\mathtitlewrapper{$e \in T$} and 
\mathtitlewrapper{$x \in V(T) - \{t\}$}
}
\label{subsec:TT}
Recall that in the case when $e \in E(T)$ and $x \in V(T) - \{t\}$,
the algorithm sets $\Dhat w s x e$
to be the minimum between 
$\Dhat[{H[T]}] {w|_T} s x e$ and 
$\Dhat[{H[T]}] {c_T} s x e $
in step \ref{step:TT} of the algorithm (line \ref{lst:line:TT_set} of the pseudocode).
\snl
Recall that $\Dhat[{H[T]}] {w|_T} s x e$ and 
$\Dhat[{H[T]}] {c_T} s x e $ are the result obtained by 
the recursive calls over the subgraph $H[T]$ for the queries
$(e,x,w|_T)$ and $(e,x,c_T)$ correspondingly.
\NL
\subsubsection*{Proof of Completeness}
\begin{claim}
    \label{claim:completeTT}
    Let $(e,x,w)\in Q$ be a query such that $e \in T, x\in V(T) - \{t\}$.
    Assuming that the recursive call over the subgraph $H[T]$ is complete,
    then $\Dhat w s x e \ge \D w s x e$.
\end{claim}
\begin{proof}
As in previous cases, we again show that both $\Dhat[{H[T]}]{w|_T} t x e + \Dnm s t$
and $\Dhat[{H[T]}]{c_T} t x e +\Dnm s t$ are at least $\D w s x e$.
\NL
We now handle the term $\Dhat[{H[T]}]{w|_T} t x e + \Dnm s t$.
Recall that $\Dhat[{H[T]}]{w|_T} t x e$ is the result obtained by the recursive call
over the graph $H[T]$ for the query $(e,x,w|_T)$.
Let $R$ be the shortest \RP{$t$}{$x$}{$e$}{$H[T]_{w|_T}$}.
Since the recursive call over $H[T]$ is complete, we have that
$d(R) \le \Dhat[{H[T]}]{w|_T} t x e$.
Note that if $R \subseteq H[T] - e$,
then $R \subseteq H - e$.
Since $e \notin P$, the path $P \circ R$ is a
\RP{$s$}{$x$}{$e$}{$H$} $\subseteq H_w - e$.
This implies that $\D w s x e \le d(R) + d(P) \le \Dhat[{H[T]}]{w|_T} t x e + \Dnm s t$, as 
required.
\snl
If $R \not\subset H[T] - e$ then $R$ uses some weighted
edge $(t,v) \in E(H[T]_{w|_T}) - E(H[T])$.
Recall that other than this edge, the path $R$ is fully contained in $H[T]-e$.
Let $\overline{R}$ be the path obtained
by replacing the edge $(t,v)$ in $R$
with the edge $(s,v) \in E(H_w)$.
The length of $\overline{R}$ is then larger then the length of $R$
by $w(v) - w|_T(v)$.
Since by definition $w|_T(v) = w(v) - \Dnm s t$
we have that $d(\overline{R}) = d(R) + \Dnm s t \le \Dhat[{H[T]}]{w|_T} t x e + \Dnm s t$.
And obviously $\overline{R}$
is a \RP{$s$}{$x$}{$e$}{$H_w$},
meaning that $\D w s x e \le d(\overline{R}) \le \Dhat[{H[T]}]{w|_T} t x e + \Dnm s t$
as required.
\NL
We now handle the term $\Dhat[{H[T]}]{c_T} t x e + \Dnm s t$.
Again we denote by $R$ the shortest \RP{$t$}{$x$}{$e$}{$H[T]_{w|_T}$},
and by the assumption of completeness we have that
$d(R) \le \Dhat[{H[T]}]{w|_T} t x e$.
Again if $R \subseteq H[T] - e$ we can concatenate $P$ before $R$
and show that $\D w s x e \le d(R) + \Dnm s t \le \Dhat[{H[T]}]{c_S} t x e + \Dnm s t$.
So we may assume that $R$ uses some weighted
edge $(t,v) \in E(H[T]_{c_T}) - E(H[T])$.
We again want to change $R$ by replacing the edge $(t,v)$
with some \RP{$s$}{$v$}{$e$}{$H_w$}.
\snl
By definition of $c_T$ there is some vertex $u \in V(S) , u \ne t, (u,v) \in E$
such that $c_T(v) = \Dnm s u + 1 - \Dnm s t$.
Note that the path from $s$ to $u$ in the BFS tree $K$ is contained in $S$
and so does not contain $e$.
Let us denote this path by $R(s,u,H)$.
Also since $u \notin T$ we have that
$(u,v) \notin T$ and so $e \ne (u,v)$.
We can conclude that  $e \notin R(s,u,H) \circ (u,v)$.
Let $\overline{R}$ be the path obtained by replacing the edge $(s,v)$ in $R$ with the
path $R(s,u,H) \circ (u,v)$.
We have that $\overline{R}$
is a \RP{$s$}{$x$}{$e$}{$H$} $\subseteq H_w -e$.
Meaning that $\D w s x e \le d(\overline{R})$.
The length of $\overline{R}$ is larger than the length of $R$
by $d(R(s,u,H)) + 1 - c_T(v) = \Dnm s t$.
So we have that $d(\overline{R}) = d(R) + \Dnm s t
\le \Dhat[{H[T]}]{c_S} t x e + \Dnm s t$
which implies the claim.
\end{proof}

\subsubsection*{Proof of soundness}
\begin{claim}
    \label{claim:sound_TT_unweighted}
    Let $e \in E(T), x \in V(T) - \{t\}, w \in W$,
    if $R(s,x,H_w - e)$ is unweighted then
    its length is at least $\D[{H[T]}] {c_T} t x e + \Dnm s t$.
\end{claim}
\begin{proof}
    Note that $R(s,x,H_w - e)$ uses nodes from $V(S)$ as it uses $s$.
    Let $u$ be the last node in $R(s,x,H_w - e)$ which is from $V(S)$.
    We fall in to two cases:
    \snl
    If $u = t$ then since $R(s,x,H_w - e)$ is \simple, 
    the path from $s$ to $t$ in $R(s,x,H_w - e)$
    is the shortest path $P$.
    Meaning we can split the path $R(s,x,H_w - e)$ into two parts:
    $P$ - the path from $s$ to $t$, and $R[t,x]$ the sub-path of $R(s,x,H_w - e)$ from $t$ to $x$.
    Note that since $t=u$ was the last node in $R(s,x,H_w - e)$ which is from 
    $V(S)$, and since  $R(s,x,H_w - e)$ is unweighted we have that $R[t,x]$ is fully contained
    in $H[T]$.
    This implies that 
    $R[t,x]$ is a \RP{$t$}{$x$}{$e$}{$H[T]$}
    and so 
    $d(t,x,H[T]-e) \le d(R[t,x])$.
    Since $H[T]-e \subseteq H[T]_{c_T} - e$ we have that 
    $\D[{H[T]}] {c_T} t x e \le d(t,x,H[T]-e)$.
    Since the length of $R(s,x,H_w - e)$ is $d(P) + d(R[t,x])$
    we can conclude that the length of $R(s,x,H_w - e)$ is at least
    $\D[{H[T]}] {c_T} t x e + \Dnm s t$ as required.
    \snl
    Otherwise we have that $u \ne t$, and so $u \in V(S) - \{t\}$.
    An illustration of this case can be seen in Figure \ref{fig:help_above}.
    Let $v$ denote the node right after $u$ in $R(s,x,H_w - e)$.
    We can split the path $R(s,x,H_w - e)$ into two parts:
    $R[s,v]$ the subpath of $R(s,x,H_w - e)$ from $s$ to $v$,
    and $R[v,x]$ the subpath of $R(s,x,H_w - e)$ from $v$ to $x$.
    Note that the length of $R[s,v]$ is at least $d(s,u,H) + 1$.
    Since $(u,v) \in E$ and $u \in V(S) - \{t\}$
    by the definition of $c_T$ we have that  $c_T(v) \le d(s,u,H) + 1 - \Dnm s t$. 
    So we can conclude that $d(R[s,v]) - \Dnm s t \ge c_T(v) $.
    Consider the edge $(t,v) \in E(H[T]_{c_T}) - E(H[T])$, its weight is 
    $c_T(v)$.
    Since $u$ was the last node in $\R w s x e$ that is from $V(S)$,
    the path $(t,v) \circ R[v,x]$ is a \RP{$t$}{$x$}{$e$}{$H[T]_{c_T}$}.
    So the length of $(t,v) \circ R[v,x]$ is at least $\D[{H[T]}] {c_T} t x e$.
    So we have that $c_T(v) + d(R[v,x]) \ge \D[{H[T]}] {c_T} t x e$
    and that $d(R[s,v]) - \Dnm s t \ge c_T(v) $.
    By simple arithmetics this implies that 
    $d(R[s,v]) + d(R[v,x]) \ge \D[{H[T]}] {c_T} t x e + \Dnm s t$ which implies the claim.
\end{proof}

\begin{claim}
    \label{claim:sound_TT_weighted}
    Let $e \in E(T), x \in V(T) - \{t\}, w \in W$,
    if $R(s,x,H_w - e)$ is weighted then
    its length is at least $\D[{H[T]}] {w|_T} t x e + \Dnm s t$.
\end{claim}
\begin{proof}
We claim that in this case the only node from $V(S)$ that $R(s,x,H_w - e)$ uses is $s$.
To see this  assume towards contradiction that $R(s,x,H_w - e)$ uses a node $u \in V(S)$
such that $u \ne s$.
Since $u \in S$ the path from $s$ to $u$ in the BFS tree is contained
in $S$ and so does not contain the edge failure $e \in T$.
Since $R(s,x,H_w - e)$ is \simple\ this implies that the subpath
of $R(s,x,H_w - e)$ from $s$ to $u$ is the path from $s$ to $u$ in $K$.
However since $R(s,x,H_w - e)$ is weighted the first edge of $R(s,x,H_w - e)$ 
is from $E(H_w) - E(H)$ and hence not from $K$, contradiction.
\NL
So we conclude that $V(R(s,x,H_w - e)) \cap V(S) = \{s\}$.
Let $(s,v)$ be the first edge of $R(s,x,H_w - e)$.
Note that $(s,v) \in E(H_w) - E(H)$ since $R(s,x,H_w - e)$ is weighted,
and that the weight of $(s,v)$ is $w(v)$.
Consider the path $\overline{R}$ obtained by replacing the edge $(s,v) \in E(H_w) - E(H)$ 
with the edge $(t,v) \in E(H[T]_{w|_T}) - E(H[T])$
(which is of weight $w|_T(v)$).
Since $V(R(s,x,H_w - e)) \cap V(S) = \{s\}$ the path 
$\overline{R}$ is a \RP{$t$}{$x$}{$e$}{$H[T]_{w|_T}$}.
This implies that $d(\overline{R}) \ge \D[{H[T]}] {w|_T} t x e$.
Note that the length of $\overline{R}$
is smaller then the length of $R(s,x,H_w - e)$ by exactly 
$w(v) - w|_T(v)$.
By the definition of $w|_T$ we have that $w(v) - w|_T(v) = \Dnm s t$.
So we have that $d(\overline{R}) = \D w s x e - \Dnm s t$
and that $d(\overline{R}) \ge \D[{H[T]}] {w|_T} t x e$,
these implies the claim.

\end{proof}

\begin{claim}
    \label{claim:sound_TT}
    Let $(e,x,w) \in Q$ be a query such that $e \in T$ and $x \in V(T) - \{t\}$.
    Assuming that the recursive call over the subgraph $H[T]$ is sound,
    then $\D w s x e \ge \Dhat w s x e$.
\end{claim}
\begin{proof}
    Recall that in the case when $e \in T$ and $x \in V(T) - \{t\}$
    the algorithm sets $\Dhat w s x e$
    to be the minimum between  $\Dhat[{H[T]}] {c_T} t x e + \Dnm s t$
    and 
    $\Dhat[{H[T]}] {c_T} t x e + \Dnm s t$.
    \NL
    Let us look at the replacement path $\R w s x e$.
    If $R(s,x,H_w - e)$ is unweighted then by Claim \ref{claim:sound_TT_unweighted}
    its length is at least $\D[{H[T]}] {c_T} t x e + \Dnm s t$.
    Since the recursive call over $H[T]$ is sound we have that 
    $\D[{H[T]}] {c_T} t x e \ge \Dhat[{H[T]}] {c_T} t x e$
    and so $\D w s x e \ge \Dhat[{H[T]}] {c_T} t x e + \Dnm s t \ge \Dhat w s x e$.
    This implies the claim.
    \NL
    Similarly we can use Claim \ref{claim:sound_TT_weighted}
    to show that if $R(s,x,H_w - e)$ is weighted it holds that
    $\D w s x e \ge \Dhat[{H[T]}] {w|_T} t x e + \Dnm s t \ge \Dhat w s x e$
    which implies the claim.
\end{proof}

%% file: sections/SS.tex
\subsection{The case when 
\mathtitlewrapper{$e \in E(S) - E(P)$} and 
\mathtitlewrapper{$x \in V(S) - \{t\}$}
}
\label{subsec:SS}
Recall that in the case when $e \in E(S) - E(P)$ and $x \in V(S)$,
the algorithm sets $\Dhat w s x e$
to be the minimum between 
$\Dhat[{H[S]}] {w|_S} s x e$ and 
$\Dhat[{H[S]}] {c_S} s x e $
in step \ref{step:SS} of the algorithm (line \ref{lst:line:SS_set} of the pseudocode).
\snl
Recall that $\Dhat[{H[S]}] {w|_S} s x e$ and 
$\Dhat[{H[S]}] {c_S} s x e $ are the result obtained by 
the recursive calls over the subgraph $H[S]$ for the queries
$(e,x,w|_S)$ and $(e,x,c_S)$ correspondingly.
\NL
\subsubsection*{Proof of Completeness}

\begin{claim}
    \label{claim:completeSS}
    Let $(e,x,w)\in Q$ be a query such that $e \in E(S)-E(P)$ and $x \in V(S) - \{t\}$.
    Assuming that the recursive call over $H[S]$
    is complete,
    then $\Dhat w s x e\ge \D w s x e$.
\end{claim}
\begin{proof}
    Recall that in the case when $e\in E(S)-E(P)$ and $x \in V(S) - \{t\}$,
    the algorithm sets
    $\Dhat w {{s}} t e$
    to be the minimum between
    $\Dhat[{H[S]}] {w|_S} s x e$
    and $\Dhat[{H[S]}] {c_S} s x e$.
    As in previous cases,
    we show that both terms are at least $\D w s x e$.
    \NL
    We now handle the term $\Dhat[{H[S]}] {w|_S} s x e$.
    Note that since the recursive call over $H[S]$ is complete we have that
    $\Dhat[{H[S]}] {w|_S} s x e \ge \D[{H[S]}] {w|_S} s x e$,
    and by Claim \ref{claim:wS_complete} we have that 
    $\D[{H[S]}] {w|_S} s x e \ge \D w s x e$,
    these implies the claim.
    \NL
    We now handle the term $\Dhat[{H[S]}] {c_S} s x e$.
    We denote the shortest \RP{$s$}{$x$}{$e$}{$H[S]_{c_S}$} by $R$.
    Since the recursive call over $H[S]$ is complete we have that
    $d(R) \le \Dhat[{H[S]}]{c_S} t x e$.
    If $R \subseteq H[S]-e$ then since 
    $H[S] - e \subseteq H - e \subseteq H_w - e$ we have that $R$ is a 
    \RP{$s$}{$x$}{$e$}{$H_w$}.
    This implies that 
    $\D w s x e \le d(R) \le \Dhat[{H[S]}] {c_S} s x e$ as required.
    \NL
    So we may assume $R$ uses some edge $(s,v) \in E(H[S]_{c_S})
    - E(H[S])$.
    As in previous claims, we wish
    to replace the edge $(s,v)$ with some \RP{$s$}{$v$}{$e$}{$H_w$} without changing
    the length of $R$.
    \snl
    Recall that the weight of the edge $(s,v)$ is $c_S(v)$ which is
    equal to $\Dnm s u + 1$ for some
    $u \in V(T) - \{t\}$ such that $(u,v) \in E(H)$.
    Let $R(s,u,H)$ be the path from $s$ to $u$ in the BFS tree $K$.
    Since $K$ is a BFS tree the length of $R(s,u,H)$ is $\Dnm s u$.
    Note that since $u \in V(T)$ the path  $R(s,u,H)$ contains
    only edges from $E(P) \cup E(T)$ and so \textbf{does not} contain
    the edge $e \in E(S) - E(P)$.
    Also note that since $u \in V(T) - \{t\}$
    we have that $u \notin S$ and so the edge $(u,v)$ is not from $E(S)$
    and hence is not $e$.
    So we have that the path $R(s,u,H) \circ (u,v)$ is a
    \RP{$s$}{$v$}{$e$}{$H$} of length
    $\Dnm s u + 1 = c_S(v)$.
    \snl
    So we can replace the edge $(s,v)$ in the path $R$ with the path 
    $R(s,u,H) \circ (u,v)$ and receive a \RP{$s$}{$x$}{$e$}{$H$} $\subseteq H_w - e$.
    The length of this path is $d(R)$ which implies that 
    $\D w s x e \le d(R) \le \Dhat[{H[S]}] {c_S} s x e$ as required.
\end{proof}
\subsubsection*{Proof of Soundness}

\begin{claim}
    \label{claim:sound_SS_yesT}
    Let $e \in E(S) - E(P), x \in V(S) - \{t\}, w \in W$,
    assume $\R w {{s}} x e$ \textbf{uses a node} from $V(T)-\{t\}$,
    then its length is at least $\D [{H[S]}] {c_S} s x e$.
\end{claim}
\begin{proof}
    Let us denote by $u$ \textbf{the last} node in $\R w s x e$ which is also in $V(T)-\{t\}$
    and let $v$ be the node right after it.
    We can then split the path  $\R w s x e$ into two subpaths:
    $R[s,v]$ - the subpath of $\R w s x e$ from $s$ to $v$,
    and $R[v,x]$ - the subpath of $\R w s x e$ from $v$ to $x$.
    We claim that the length of $R[s,v]$ is at least $c_S(v)$.
    To see this note that $R[s,v]$ goes from $s$ to $u$, and 
    then from $u$ to $v$ through the edge $(u,v) \in E(H)$.
    Meaning the length of  $R[s,v]$ is at least $\Dnm s u + 1$
    (recall that $\Dnm s u = \Dnm [{H_w}] s u$ by the weight requirement).
    By the definition of $c_S$ we have that $c_S(v) \le \Dnm s u + 1$.
    So we can conclude that $d(R[s,v]) \ge c_S(v)$.
    \snl
    We now attempt to lower bound the length of $R[v,x]$.
    Note that since $u$ was the last node in $\R w s x e$ which  
    is from $V(T)-\{t\}$, the subpath $R[v,x]$ is fully contained in $H[S]$.
    Consider the edge $(s,v) \in E(H[S]_{c_S}) - E(H[S])$, its weight is $c_S(v)$.
    The path $(s,v) \circ R[v,x]$ is then a
    \RP{$s$}{$x$}{$e$}{$H[S]_{c_S}$} and so its length is at least $\D [{H[S]}] {c_S} s x e$.
    So we have that $c_S(v) + d(R[v,x]) \ge \D [{H[S]}] {c_S} s x e$
    and that $d(R[s,v]) \ge c_S(v)$.
    By simple arithmetics this implies
    that the length of $\R w s x e$ is at least $\D [{H[S]}] {c_S} s x e$.
\end{proof}

\begin{claim}
    \label{claim:sound_SS}
    Let $(e,x,w) \in Q$ be a query such that $e \in E(S)-E(P), x \in V(S) - \{t\}$.
    Assuming that the recursive call over the subgraph $H[S]$ is sound,
    then $\D w s x e \ge \Dhat w s x e$.
\end{claim}
\begin{proof}
    Note that if $\R w s x e$ uses no nodes from $V(T)-\{t\}$
    then we fall in the case of Claim \ref{claim:sound_PS_noT},
    and so the length of $\R w s x e$ is at least $\D[{H[S]}]{w|_S}{{s}}{x}{e}$.
    Since the recursive call over $H[S]$ is sound
    (by assumption) we have that  
    $\D[{H[S]}]{w|_S}{{s}}{x}{e} \ge \Dhat[{H[S]}]{w|_S}{{s}}{x}{e}$,
    which is at least $\Dhat w {{s}} x e$,
    this implies the claim.
    Otherwise $\R w s x e$ does use a node from $V(T)-\{t\}$,
    meaning we fall to the case of Claim \ref{claim:sound_SS_yesT},
    and so by the same arguments we get that the length of 
    $\R w s x e$ is at least $\Dhat[{H[S]}]{c_S}{{s}}{x}{e}$ which is at least $\Dhat w s x e$,
    which again implies the claim.
\end{proof}

\mynewpage

%% file: sections/post_analysis.tex
Combined together claims 
\ref{claim:completeDSTE},
\ref{claim:completePT} 
\ref{claim:completeTT},
\ref{claim:completeSS},
 and
\ref{claim:complete_PS}
imply the following claim:
\begin{claim}
    \label{claim:complete_induction}
    Assuming that $|V(H)| > 6$ and assuming that the recursive calls over $H[T]$ and $H[S]$ are complete,
    the call over $H$ is also complete.
\end{claim}

Similarly, combined together 
claims 
\ref{claim:sound_DSTE},
\ref{claim:sound_PT} 
\ref{claim:sound_TT},
\ref{claim:sound_SS},
\ref{claim:sound_PS_departing}
and
\ref{claim:sound_PS_jumping}
imply the following claim:
\begin{claim}
    \label{claim:sound_induction}
    Assuming that $|V(H)| > 6$ and assuming that the recursive calls over $H[T]$ and $H[S]$ are sound,
    the call over $H$ is \whp\ also sound.
\end{claim}

Since the algorithm is both sound and complete for graphs of size at most $6$
we can use these to claims to prove by induction 
that the algorithm is complete, and \whp\ is also sound.

%% file: sections/running_time.tex
\section{ Running Time analysis}
\label{sec:running_time_rec}
We now prove that the running time of the above algorithm is indeed $\tilde{O}(m\sqrt{n} + n ^2)$.
In order to upper bound the algorithm's running time,
we prove the following claim.
\begin{claim}
    \label{claim:master_formula}
    Let $T(H,W,Q)$ denote the running time of the algorithm 
    over the sub-graph $H$ of the original graph $G$,
    weight functions set $W$ and queries set $Q$.
    Denote $n_H = |V(H)|, m_H = |E(H)|$.
    \snl
    Then $T(H,W,Q) = T(H[S],W_S,Q_S) + T(H[T],W_T,Q_T) + 
    \tilde{O}(m_H |W| + m_H \sqrt{n_H} + n_H^2 + |Q|)$.
    \snl
    And it holds that 
    $|Q_S|+|Q_T| = |Q| + \tilde{O}(n_H^2)$
    and $|W_S| = |W| + \tilde{O}(\sqrt{n_H}), |W_T| = |W| + 1$.
\end{claim}
\begin{proof}
The proof of this claim will be done by simply going through the different steps of the algorithm,
 upper bounding their running time and upper bounding the size of 
the recursive inputs constructed.
\NL
\subsubsection*{Step \ref{step:separation}: Tree separation  (\mathtitlewrapper{$S,T$}):\;}
Since by Lemma \ref{lemma:seperator}
one can find a balanced tree separator in linear time, the algorithm takes $\tilde{O}(n_H+m_H)$
to find the separator and preform the two BFS invocations described in this section. 
\NL
\subsubsection*{Step \ref{step:WSPX}: Computing \mathtitlewrapper{$\D w {{s}} x P$}:\;}
In this section for every $w\in W$  the algorithm constructs the graph
$H_w-E(P)$, this obviously can be done in $\tilde{O}(n_H+m_H)$ time.
Then the algorithm runs Dijkstra's algorithm on the generated graph.
For each $w \in W$ this take $\tilde{O}(n_H+m_H) = \tilde{O}(m_H)$ time so
in total this section of the algorithm will take $\tilde{O}(|W| \cdot m_H)$ time.
\NL
\subsubsection*{Step \ref{step:pivots}: Sampling Pivots (\mathtitlewrapper{$B_k$}) and Defining Path Intervals (\mathtitlewrapper{$P_k$}):\;}
We show that the running time of this step of the algorithm is $\tilde{O}(m_H\sqrt{n_H})$ time.
Note that by Lemma \ref{lemma:sampling},
$|B_k| = \tilde{O}(\frac{\sqrt{n_H}}{2^k})$ \whp.
and so \whp\ the algorithm samples each $B_k$ only once
- resulting in $\tilde{O}(n)$ time for the sampling step.
Also note that  $|B|= \tilde{O}(\sqrt{n_H})$
and that $\forall k \in [\lgV]: |V(P_k)| = O(2^k \sqrt{n_H})$.
\snl
Invoking the BFS algorithm for every $b \in B$ then takes 
$\tilde{O}(m |B|)$ time which is $\tilde{O}(m\sqrt{n})$ time.
\NL
\subsubsection*{Step \ref{step:depart}: Computing Departing Paths ($\depart e x$)}
We show that the running time of this step of the algorithm is $\tilde{O}(n_H^2+m_H\sqrt{n_H})$ time.
To see that let us fix some $k \in [\lgV]$
and $e \in P_k$.
It takes $\tilde{O}(n_H)$ time to compute
$\depart e b$ for every pivot node $b \in B_k$g66
and $\tilde{O}(|B_k|)$ time to compute $\depart e x$ 
for every non-pivot node $x \in V(H) - B_k$.
This implies that computing $\depart e v$ 
for every node $v \in V(H)$ takes 
$O(n_H \cdot |B_k|)$ time.
So for some $k \in [\lgV]$
it takes $O(n_H \cdot |B_k| |P_k|)$ 
to take care of edge failures from $P_k$.
Note that $|P_k| = O(2^k \sqrt{n_H})$ and that
$|B_k| = \tilde{O}(\frac{\sqrt{n_H}}{2^k})$.
This implies that $|P_k| |B_k| = \tilde{O}(n_H)$.
So it takes $\tilde{O}(n_H^2)$ to care of some $k \in [\lgV]$,
meaning that it takes $\tilde{O}(n_H^2)$ time to take 
care of edge failures from $E(P) - E(P_0)$
\snl
Taking care of an edge failure from $P_0$ 
involves invoking Dijkstra's algorithm a single time meaning it takes
$\tilde{O}(m_H |P_0|)$ time to take care of edge failures from $P_0$,
which is $\tilde{O}(m_H \sqrt{n_H})$ time since $|P_0| = O(\sqrt{n_H})$.
\NL
\subsubsection*{Step \ref{step:DSTE}: Computing $\Dhat w {{s}} t e$ When $e\in E(P)$}
We show that the running time of this step of the algorithm is $\tilde{O}(n_H |W| + m\sqrt{n_H})$ time.
To see this note that defining 
the nodes $u_i$ and edges $e_i$ takes $\tilde{O}(n_H)$ time.
For every $w \in W$ the algorithm then computes $A_w$ using dynamic
programming, by iterating over the set $\{e_i\}_{i=0}^{|P|-1}$ once.
This implies that the algorithm takes $\tilde{O}(|P|) = \tilde{O}(n_H)$ time 
to compute each $A_w$, meaning it takes 
$\tilde{O}(n_H |W|)$ to compute all $A_w$.
Invoking the algorithm from \cite{Roditty2005} takes $\tilde{O}(m\sqrt{n_H})$
time, and computing the final results 
takes $\tilde{O}(n_H |W|)$ time
\NL
\subsubsection*{Step \ref{step:PT}: Computing $\Dhat w {{s}} x e$ When $e\in E(P),x\in V(T) - \{t\}$}
It can be easily verified that the running time of this step of the algorithm is 
$\tilde{O}(|Q|)$ time.
\NL
\subsubsection*{Step \ref{step:TT}: Computing $\Dhat w {{s}} x e$ When $e\in E(T),x\in V(T)- \{t\}$}
We now show that the non recursive part of this step runs in 
$\tilde{O}(n_H|W| + n_H^2 + |Q|)$ time.
To see this 
note that computing the restricted weight function $w|_T$
for every $w \in W$ takes $\tilde{O}(n_H)$ time, resulting in a 
total $\tilde{O}(n_H |W|)$ time to define all of them.
Computing $c_T$ naïvely takes $\tilde{O}(n_H^2)$ time
and constructing $Q_T$ takes $|Q| + \tilde{O}(n_H^2)$ time.
Finally computing the final distance estimations takes $\tilde{O}(|Q|)$ time. 
\NL
We also upper bound the size of the recursive input this step constructs
(as stated by the claim).
It can be easily seen that the new set of weight functions $W_T$
is of size $|W|+1$ (we only added the weight function $c_T$).
Also the new queries added to $Q_T$ 
(queries that do not originate from a query in $Q$)
are the queries $E(T) \times V(T) \times \{c_T\}$.
Meaning we only add $O(n_H^2)$ new queries. 
\NL
\subsubsection*{Step \ref{step:SS}: Computing $\Dhat w {{s}} x e$ When $e\in E(S),x\in V(S)- \{t\}$}
We now show that the non recursive part of this step runs in 
$\tilde{O}(n_H|W| + n_H^2 + |Q|)$ time.
We will also upper bound the size of the recursive input this step constructs.
\NL
\textbf{Computing the new weight functions set: }
Computing the restricted weight function $w|_S$
for every $w \in W$ takes $\tilde{O}(n_H)$ time, resulting in a 
total $\tilde{O}(n_H |W|)$ time to define all of them.
Computing $c_S$ naïvely takes $\tilde{O}(n_H^2)$ time.
Defining and computing the new weight functions 
$w_b$ takes $\tilde{O}(n_H)$ time for each $b \in B$.
Since $|B| = \tilde{O}(\sqrt{n_H})$ defining and computing
all the $w_b$ weight functions take $\tilde{O}(n_H^{1.5})$ time.
So it takes $\tilde{O}(n_H^2 + n_H |W|)$ time to define
the new weight functions set $W_S$ and it is of size 
$|W| + 1 + |B| = |W| + \tilde{O}(\sqrt{n_H})$
\NL
\textbf{Computing the new queries set: }
The algorithm first add the query $(e,x,w|_S)$ for every $(e,x,w) \in Q$ 
such that $e \in S, x \in V(S)- \{t\}$, this takes $\tilde{O}(|Q|)$ time.
Then it adds the queries $E(S) \times V(S) \times \{c_S\}$, 
which takes $\tilde{O}(n_H^2)$ time.
Then, for every $k \in [\lgV]$ the algorithm adds to 
$Q_S$ the queries $(e,x,w_b)$ for every $(e,x,b) \in E(P_k) \times V(S) \times B_k$.
This implies that the algorithm adds $n_H \cdot |P_k||B_k|$ new queries to $Q_S$.
Since $|P_k| = O(2^k \sqrt{n_H})$ and 
$|B_k| = \tilde{O}(\frac{\sqrt{n_H}}{2^k})$
this implies that for every $k \in [\lgV]$ the algorithm adds  $\tilde{O}(n_H^2)$
new queries to $Q_S$.
So it takes $\tilde{O}(n_H^2 + |Q|)$ time to compute the new set of queries
$Q_S$ and it contains $\tilde{O}(n_H^2)$ new queries 
(queries that do not originate from a query in $Q$). 
\NL
\textbf{Computing $\pivot{e}{x}$:}
Let us fix some $k \in [\lgV]$ and $e \in P_k$.
Computing $\pivot{e}{x}$ for each $x \in V(S)$
takes $\tilde{O}(|B_k|)$ time.
Meaning that computing $\pivot{e}{x}$ for every $e \in E(P) - E(P_0)$
and $x \in V(S)$ takes
$\sum_{k\in [\lgV]} \tilde{O}(|P_k| \cdot n_H \cdot |B_k|) = 
\sum_{k\in [\lgV]} \tilde{O}(\sqrt{n_H} 2^k \cdot n_H \cdot \frac{\sqrt{n_H}}{2^k}) 
= \sum_{k\in [\lgV]} \tilde{O}(n_H^2) = \tilde{O}(n_H^2)$.
Computing  $\pivot{e}{x}$ for every $e \in E(P_0)$
and $x \in V(S)$ takes $\tilde{O}(|P_0|n_H) = \tilde{O}(n_H^{1.5})$ time.
\NL
Finally computing the final distance estimations takes $\tilde{O}(|Q|)$ time.
\NL
\subsubsection*{Step \ref{step:final}: Outputting the Results:\;}
In this section the algorithm constructs
a LCA data structure over the BFS tree $K$, which takes $O(n_H)$
time by Lemma \ref{lemma:LCA}.
Then for every $(e,x,w) \in Q$ the algorithm
checks in $O(1)$ time if $e$ is on the path from $s$
to $x$ in the BFS tree $K$,
if so the algorithm sets the estimated
value $\Dhat w {{s}} x e$ as the result for the query $(e,x,w)$,
otherwise it sets $\Dnm s x$ as the result for the query.
Overall the running time for this section is $\tilde{O}(|Q| + n_H)$.

\end{proof}
\NL
\subsection{ Running Time Analysis of the Recursion}
\label{subsec:running_time_rec}
We now use Claim \ref{claim:master_formula} to preform an analysis of our recursive algorithm.
The input to the algorithm is a graph $G$,
a BFS tree $\widehat{K}$,
a weight functions set of size 1 $W_0 = \{w_\infty\}$
(where $w_\infty$ assigns $\infty$ to each vertex),
and the complete set of queries $Q_0 = E(\widehat{K}) \times V(G) \times W_0$.
\NL
Let $L$ denote the number of levels of the recursion, note that $L=O(\log n)$
since the size of the graphs decays exponentially.
Let $\mathcal{H}_i$ denote the set of
graphs in the $i$'th level of the recursion.
Note that  $\sum_{H \in \mathcal{H}_i} |E(H)| \le m$
since the sets $\{E(H)\}_{H \in \mathcal{H}_i}$ are
disjoint subsets of $E(G)$.
We also claim that $\sum_{H \in \mathcal{H}_i} |V(H)| \le 2n$.
\NL
To see this note that every graph $H \in \mathcal{H}_i$
arrives to the recursive call with a BFS tree $K$.
Let $\mathcal{K}_i = \{K \text{ the BFS tree of } H : H \in \mathcal{H}_i \}$.
Note that by construction $\mathcal{K}_i$
is an edge disjoint family of sub-trees of $\widehat{K}$ -
the original BFS tree of $G$.
Note also that every tree in $\mathcal{K}_i$
has at least 2 vertices as otherwise the tree
which called it recursively $K' \in \mathcal{K}_{i-1}$,
had at most $6$ vertices (since
$\frac{|V(K')|}{3} - 1 \le |V(K)|$
by Lemma \ref{lemma:seperator}).
If $K'$ had at most $6$ vertices
it would not have preform any recursive call,
since it is in the base case of the recursion.
Let $v \in V(G)$ be some node.
We claim that the number of $K \in \mathcal{K}_i$ such that
$v \in K$  is at most $\deg_{\widehat{K}}(v)$ - where $\deg_{\widehat{K}}(v)$ is the
degree of $v$ in the BFS tree ${\widehat{K}}$.
To see this note that if $v\in K$
then since $K$ is a tree of size at least 2,
$v$ has some neighbor $u \in K$.
Since $\mathcal{K}_i$ is an edge disjoint family of subtrees,
no other $K' \in \mathcal{K}_i$ has both $v$ and $u$ as nodes
as otherwise $K$ and $K'$ will share the edge $(u,v)$ or $(v,u)$.
And so by the pigeonhole principle there are at most
$\deg_{\widehat{K}}(v)$ trees $K\in \mathcal{K}_i$ such that $v \in K$.
And so
$\sum_{H \in \mathcal{H}_i} |V(H)|=
\sum_{K \in \mathcal{K}_i} |V(K)|=
\sum_{v\in V(G)} |\{K \in \mathcal{K}_i : v \in K\}| \le
\sum_{v\in V(G)} \deg_{\widehat{K}}(v) =
2\cdot |E({\widehat{K}})| = 2 \cdot (n-1) < 2n$.
\NL
Let $\mathcal{Q}_i$ denote the set of all query sets
in the $i$'th level.
Let $Q_i = |\bigcup_{Q\in \mathcal{Q}_i} Q|$ denote the total number of
queries in the $i$'th level.
Note that by Claim \ref{claim:master_formula},
every recursive call of over a sub-graph $H \in \mathcal{H}_i$
only increase the \textbf{total} number of queries in it is recursive calls
by at most $\tilde{O}(|V(H)|^2)$.
So we have that $Q_{i+1} = Q_i + \tilde{O}\Big(\sum_{H\in \mathcal{H}_i} |V(H)|^2 \Big)$.
Note that $\sum_{H\in \mathcal{H}_i} |V(H)|^2 \le \Big(\sum_{H\in \mathcal{H}_i} |V(H)| \Big)^2 \le
(2n)^2$.
And so $Q_{i+1} = Q_i + \tilde{O}(n^2)$.
Since $Q_0 = (n-1) \cdot n \cdot 1 \le n^2$ and $L = O(\log(n))$,
we have that $\forall i \in [L] : Q_i = \tilde{O}(n^2)$.
Note that this implies that the total number of queries
asked by the algorithm is $\sum_{i=0}^L Q_i = \sum_{i=0}^L \tilde{O}(n^2) = \tilde{O}(n^2)$.
\NL
Let $\mathcal{W}_i$ denote the set of all weight functions sets
in the $i$'th level.
Let $W_i = \max_{W\in \mathcal{W}_i} |W|$.
Note that by Claim \ref{claim:master_formula},
every recursive call of over a sub-graph $H \in \mathcal{H}_i$
only increase the number of weight functions for
\textbf{each} of its recursive calls by at most
$\tilde{O}(\sqrt{|V(H)|}) = \tilde{O}(\sqrt{n})$.
And so $W_{i+1} = W_i + \tilde{O}(\sqrt{n})$,
since $Q_0 = 1$ we have that $\forall i \in [L] : W_i = \tilde{O}(\sqrt{n})$.
\NL
Let us fix some $i \in [L]$.
Note that by Claim \ref{claim:master_formula},
the total time the algorithm spends for some recursive
call over a graph $H \in \mathcal{H}_i$ with a query set $Q \in \mathcal{Q}_i$
can be bounded by
$\tilde{O}(|E(H)|\cdot\sqrt{|V(H)|} + |E(H)| \cdot W_i + |V(H)|^2 + |Q|) =
\tilde{O}(|E(H)|\cdot\sqrt{n} + |V(H)|^2 + |Q|)$.
And so the total time the algorithm spends for the $i$'th
level of the recursion is
$
\tilde{O}\Big(\sum_{H \in \mathcal{H}_i}|E(H)|\sqrt{n} \Big)
+ \tilde{O}\Big(\sum_{H \in \mathcal{H}_i}|V(H)|^2 \Big)
+ \tilde{O}\Big(\sum_{Q\in \mathcal{Q}_i} |Q| \Big)
=
$
$
\tilde{O}\Big(\sqrt{n} \cdot \sum_{H \in \mathcal{H}_i}|E(H)| \Big)
+ \tilde{O}\bigg( \Big(\sum_{H \in \mathcal{H}_i}|V(H)| \Big)^2 \bigg)
+ \tilde{O}(Q_i)
=
$
$
\tilde{O}(\sqrt{n} \cdot m)
+  \tilde{O} (n^2)
+ \tilde{O}(n^2)
=
$
$
\tilde{O}(m\sqrt{n} + n^2)
$.
Since the total number of layers is
$O(\log(n))$
we have that the total running time of our algorithm is
$\tilde{O}(m\sqrt{n} + n^2)$.

%% file: sections/reductions.tex
\section{ Conditional Lower Bound for SSRP With Rational Weights}
\label{sec:reductions}
In this section we present our conditional lower bound for the SSRP problem for graphs with rational weights in $[1,2)$.
Let $A$ and $B$ be two $n\times n$ matrices with entries from $\mathbb{R} \cup \{\infty\}$. 
The \textit{min-plus} product of these two matrices $A \star B$ is defined to be an
$n \times n$ matrix $C$ such that $C_{i,j} = \min\limits_{k=1}^n \{ A_{i,k} + B_{k,j} \}$.
Finding efficient algorithms for computing
the min-plus product of two matrices is a well-studied and active area of work
(see \cite{AGM97,G76,williams2019truly}).
For a subset of numbers $S \subseteq \mathbb{R}$ we denote by $MP(n,m,S)$ the problem of computing the min-plus product of
two $n$ by $n$ matrices with entries from $S \cup \{ \infty\}$, such that there is
a total of at most $m$ entries to the two matrices which are not $\infty$.
\NL
In order to represent rational numbers,
we assume that we work in the word-RAM model with $w$-bit size words.
Let $\mathbb{N}_w$ denote the the set of integers representable as a single $w$-bit computer word.
Let $\mathbb{Q}_w = \{\frac{a}{2^k} : a,2^k \in \mathbb{N}_w \text{ and } k \in \mathbb{N} \}$
be the set of (efficiently) representable rationals.
Finally, let $[1,2)_{w} = \mathbb{Q}_w \cap [1,2)$
be the set of representable rationals between $1$ and $2$.
\NL
We denote by $A_{MP[1,2)}(n,m)$ an algorithm for the $MP(n,m,[1,2)_{w})$ problem.
We prove the following claim.
\begin{claim}
\label{claim:normalization_claim}
Given an algorithm $A_{MP[1,2)}(n,m)$ whose running time is $T(n,m)$,
there is a $T(n,m) + O(n^2)$ time algorithm for the $MP(n,m,\mathbb{N}_w)$ problem.
\end{claim}
\begin{proof}
The proof of this claim is rather trivial.
Let $A,B$ be two matrices that are our input to the $MP(n,m,\mathbb{N}_w)$ problem.
Assume at least one of the matrices has an entry which is not $\infty$,
as otherwise the min-plus product is trivial (a matrix whose all entries are $\infty$).
Let $M$ denote the maximum entry \textbf{which is not $\infty$} among all entries of both $A$ and $B$.
Let $\overline{M}$ denote the smallest power of $2$ which is greater then $M$.
\snl
We normalize $A$
by
dividing it by $\overline{M}$
and adding $1$ to all entries.
We denote the resulted matrix by $\overline{A}$.
We normalize $B$ in the same way to get the normalized matrix $\overline{B}$.
It can be easily verified that $\overline{A},\overline{B}$ is a valid input to the
$MP(n,m,[1,2)_w)$ problem.
\snl
We invoke the algorithm $A_{MP[1,2)}$ over the normalized matrices, and obtain a result
$\overline{C}$.
We then subtract a value of $2$ from each entry of $\overline{C}$ and multiply the resulted matrix
by $\overline{M}$.
Let $C$ be the resulted matrix.
One can easily verify that $C$ is indeed the distance product of the two matrices
$A$ and $B$. 
\end{proof}
We denote by $A_{SSRP[1,2)}(n,m)$ an algorithm for the \textbf{undirected} SSRP
problem, for a graph with at most $n$ nodes and $m$ edges, and edge weights from $[1,2)_w$.
The following claim is obtained by a slight generalization of the construction presented
in \cite{CC19}.
\begin{claim}
\label{claim:reduction_claim}
Given an algorithm $A_{SSRP[1,2)}(n,m)$ whose running time is
$T(n,m)$ there is a $O(\sqrt{n}\cdot T(O(n),O(m))$
time algorithm for the $MP(n,m,[1,2)_w)$ problem
\end{claim}
\begin{proof}
    Let $X,Y$ be two $n \times n$ matrices with entries
    from $[1,2)_w \cup \{\infty\}$ such that there is a total
    of $m$ entries that are not $\infty$.
    Denote $Z = X \star Y$.
    We will show how using a single invocation of $A_{SSRP[1,2)}$
    over a graph with $O(n)$ vertices and $O(m)$ edges and edge weights from $[1,2)_w$,
    we can compute $Z_{i,j}$ for every $1 \le i \le \sqrt{n}$ and
    $1 \le j \le n$.
    We can then use $O(\sqrt{n})$ invocations of $A_{SSRP[1,2)}$
    and "shift" the rows of the matrix $X$ in each invocation
    to compute the entire matrix $Z$.
    Hence showing this will imply the claim.
    \NL
    Let us denote $L = \sqrt{n} + 1$.
    To compute $Z_{i,j}$ for every $1 \le i < L$ and
    $1 \le j \le n$ we construct the following undirected graph $G$.
    Let $A = \{a_1,a_2,...,a_n\}$,
    $B = \{b_1,b_2,...,b_n\}$,
    $C = \{c_1,c_2,...,c_n\}$
    be three independent sets of vertices.
    For all $1 \le i,k \le n$
    if $X_{i,k} \ne \infty$ we add the edge
    $(a_i,b_k)$ with weight $X_{i,k}$.
    Similarly for all $1 \le j,k \le n$
    if $Y_{k,j} \ne \infty$ we add the edge
    $(b_k,c_j)$ with weight $Y_{k,j}$.
    We then add a path of new vertices $P = \{x_1,x_2,...,x_L\}$.
    For every $1 \le i \le L$
    we add a path from $x_i$ to $a_i$ of length $8 \cdot (L - i) + 1$
    using auxiliary vertices and edges of length $1$.
    Note that for every $1 \le i \le L$ it holds that
    $\Dnm[G] {x_1} {a_i} = i + 8 \cdot (L - i) + 1 =
    8L - 7i + 1$.
    An illustration of this construction can be seen in Figure \ref{fig:reduction} in the appendix.
    \snl
    We now invoke the algorithm $A_{SSRP[1,2)}$ over $G$ with source node $x_1$.
    For every $1 \le i < L$ and $1 \le j \le n$
    let us denote by $\alpha_{i,j}$
    the distance from $x_1$ to $c_j$ with the edge failure $(x_i,x_{i+1})$.
    \NL
    We claim that if  $\alpha_{i,j} <  8L - 7i + 5$ then
    $Z_{i,j} = \alpha_{i,j} - (8L - 7i + 1)$,
    and otherwise $Z_{i,j} = \infty$.
    To see why this is true note that
    the distance in $G$ from $x_1$ to $a_t$ such that $t < i$
    is at least $8L - 7i + 8$.
    Also, note that any path from $x_1$ to $c_j$ 
    in the graph $G- (x_i, x_{i+1})$, that passes through $x_t$ such that 
    $t > i$ is of length at least
    $\Dnm [G] {x_1} {a_i} + 4 =  8L - 7i + 5$.
    So
    we have that $\alpha_{i,j} <  8L - 7i + 5$ \textbf{if and only if}
    the replacement path from $x_1$ to $c_j$ with edge failure $(x_i,x_{i+1})$
    goes from $x_1$ to $a_i$ and then preforms a 3 vertex path:
    $a_i \rightarrow b_k \rightarrow c_j$ for some index $1 \le k \le n$.
    \snl
    If $\alpha_{i,j} <  8L - 7i + 5$ then
    $k$ must be a index which minimizes the length of this 3 vertex path.
    Since this length is $X_{i,k} + Y_{k,j}$
    we must have that the length of this 3 vertex path is $Z_{i,j}$.
    Meaning that
    $Z_{i,j} = \alpha_{i,j} - (8L - 7i + 1)$.
    Otherwise $\alpha_{i,j} \ge  8L - 7i + 5$, so we have that for every index $k$ there is
    no such  3 vertex path, meaning that for every $1 \le k \le n$
    it holds that $X_{i,k} + Y_{k,j} = \infty$,
    meaning that indeed $Z_{i,j} = \infty$
\end{proof}

We denote by $APSP(n,m,\mathbb{N}_w)$ the problem of computing
the APSP of a graph with $n$ nodes an $m$ edges with edge weights from $\mathbb{N}_w$.
We denote by $A_{MP}(n,m)$ an algorithm for the $MP(n,m,\mathbb{N}_w)$
problem.
The following claim is a well known reduction, presented in \cite{AGM97}.

\begin{claim}
    \label{claim:apsp_via_minplus}
    Let $A_{MP}(n,m)$ be an algorithm whose running time is
    $T(n,m)$.
    Then there is an algorithm for the $APSP(n,m,\mathbb{N}_w)$ problem that runs in time
    ${O}(T(n,n^2) \cdot \log(n))$.
\end{claim}
We now turn to prove the conditional lower bound, using the above 3 reductions.
\begin{theorem}
\label{theorem:lowerbound}
Let $A_{SSRP[1,2)}(n,m)$ be an algorithm whose running time is
$T(n,m)$. If $T(n,m) = O(m \cdot n^{1/2-\epsilon} )$ for some $0 < \epsilon \le \frac{1}{2}$
then there is a $O(n^{3-\epsilon} \log(n))$ time algorithm for the $APSP(n,m,\mathbb{N}_w)$ problem.
\end{theorem}
\begin{proof}
    Note that by a combination of Claims \ref{claim:normalization_claim} and \ref{claim:reduction_claim}
    we can use the SSRP algorithm $A_{SSRP[1,2)}(n,m)$ to construct an algorithm $A_{MP}(n,m)$ that solves the
    $MP(n,m,\mathbb{N}_w)$ problem in $O(n^{3-\epsilon})$ time.
    And so by Claim \ref{claim:apsp_via_minplus} we can construct an algorithm
    for the $APSP(n,m,\mathbb{N}_w)$ problem in ${O}(n^{3-\epsilon} \cdot \log(n))$ time.
\end{proof}

%% file: sections/appendix_figures.tex
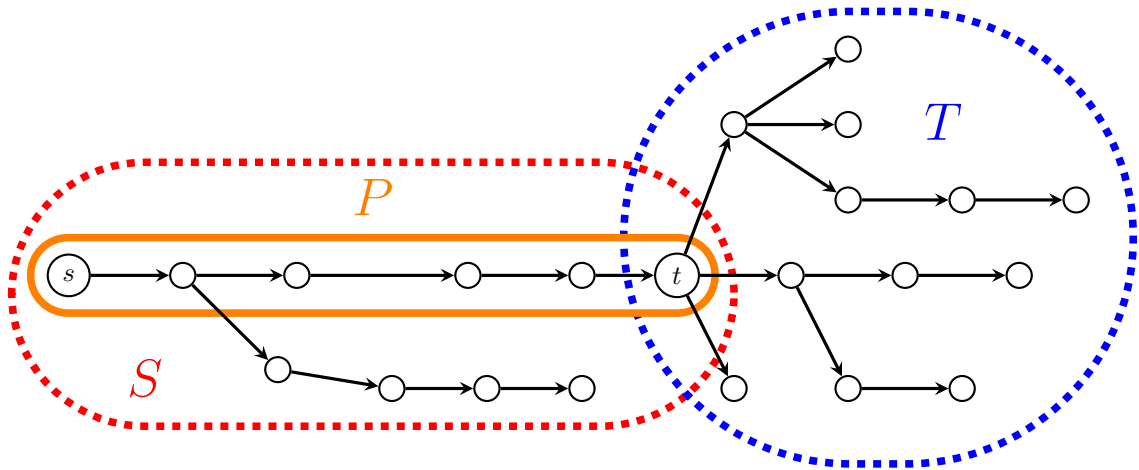
\begin{figure}[H]
  \centering
  \captionsetup{justification=centering}

  \begin{tikzpicture}

  \begin{scope}
    \fontsize{20}{20}
    \node[below, red] at (-0.5,-0.9) {$S$};
    \node[below, blue] at (10,2.5) {$T$};
    \node[below, orange] at (2.5,1.5) {$P$};

      \draw [red,line width=1mm,dashed, rounded corners=50] (-2.25,-2) rectangle  (7.25, 1.5);
      \draw [blue,line width=1mm, dashed,rounded corners=85] (5.8,-2.5) rectangle  (12.5, 3.5);
      \draw [orange,line width=1mm, rounded corners=14] (-2,-0.5) rectangle  (7, 0.5);
  \end{scope}

  \begin{scope}[every node/.style={circle,thick,draw}]
      \node (V41) at (-1.5,0) {${{{s}}}$} ;
      \node (V42) at (0,0) {} ;
      \node (V43) at (1.5,0) {};
      \node (V44) at (3.75,0) {};
      \node (V45) at (5.25,0) {};
      \node (V46) at (6.5,0) {$t$};
      \node (V47) at (8,0) {};
      \node (V48) at (9.5,0) {};
      \node (V49) at (11,0) {} ;

      \node (V21) at (8.75,1) {} ;
      \node (V22) at (10.25,1) {} ;
      \node (V23) at (11.75,1) {} ;

      \node (V11) at (7.25,2) {} ;
      \node (V12) at (8.75,2) {} ;

      \node (V01) at (8.75,3) {} ;

      \node (V51) at (1.25,-1.25) {};
      \node (V52) at (2.75,-1.5) {};
      \node (V53) at (4,-1.5) {};
      \node (V54) at (5.25,-1.5) {};
      \node (V55) at (7.25,-1.5) {} ;
      \node (V56) at (8.75,-1.5) {} ;
      \node (V57) at (10.25,-1.5) {} ;

  \end{scope}

  \begin{scope}[>={stealth},
                every edge/.style={draw=black,very thick}]

      \path [->] (V41) edge (V42);
      \path [->] (V42) edge (V43);
      \path [->] (V43) edge  (V44);
      \path [->] (V44) edge  (V45);

      \path [->] (V45) edge (V46);
      \path [->] (V46) edge  (V47);
      \path [->] (V47) edge (V48);
      \path [->] (V48) edge (V49);

      \path [->] (V46) edge (V11);
      \path [->] (V11) edge (V21);
      \path [->] (V21) edge (V22);
      \path [->] (V22) edge (V23);
      \path [->] (V11) edge (V12);
      \path [->] (V11) edge (V01);
      \path [->] (V46) edge (V55);

      \path [->] (V47) edge (V56);
      \path [->] (V56) edge (V57);

       \path [->] (V42) edge (V51);
       \path [->] (V51) edge (V52);
       \path [->] (V52) edge (V53);
       \path [->] (V53) edge (V54);

  \end{scope}
  \end{tikzpicture}
  \caption{Tree separation}
  \label{fig:separation}
\end{figure}
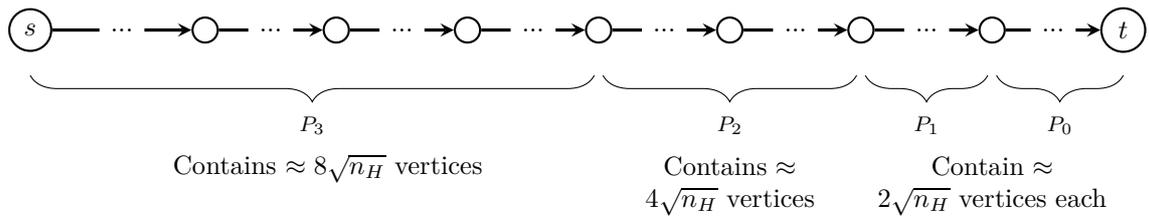
\begin{figure}[H]
    \centering
    \captionsetup{justification=centering}
  
    \begin{tikzpicture}[scale=1.15]
    \begin{scope}
    \draw [decorate,decoration={brace,amplitude=10pt,mirror,raise=4pt},yshift=0pt]
    (12.05,-0.4) -- (13.5,-0.4) node [black,midway,yshift=-0.8cm] {\footnotesize $P_0$};
    \node[align=center] at (12,-1.8) { Contain $\approx$ \\ $2\kindaSqrtV$ vertices each};
    \draw [decorate,decoration={brace,amplitude=10pt,mirror,raise=4pt},yshift=0pt]
    (10.55,-0.4) -- (11.95,-0.4) node [black,midway,yshift=-0.8cm] {\footnotesize $P_1$};
    \node[align=center] at (9,-1.8) { Contains $\approx$ \\ $4\kindaSqrtV$ vertices};
    \draw [decorate,decoration={brace,amplitude=10pt,mirror,raise=4pt},yshift=0pt]
    (7.55,-0.4) -- (10.45,-0.4) node [black,midway,yshift=-0.8cm] {\footnotesize $P_2$};
    [decorate,decoration={brace,amplitude=10pt,mirror,raise=4pt},yshift=0pt]
    \draw [decorate,decoration={brace,amplitude=10pt,mirror,raise=4pt},yshift=0pt]
    (1,-0.4) -- (7.45,-0.4) node [black,midway,yshift=-0.8cm] {\footnotesize $P_3$};
    \node[align=center] at (4.4,-1.6) { Contains $\approx 8\kindaSqrtV$ vertices};
    \end{scope}
  
    \begin{scope}[every node/.style={circle,thick,draw}]
        \node (V4) at (1,0) {${{{s}}}$};
        \node (V5) at (3,0) {};
        \node (V6) at (4.5,0) {};
        \node (V7) at (6,0) {};
        \node (V8) at (7.5,0) {};
        \node (V9) at (9,0) {};
        \node (V10) at (10.5,0) {};
        \node (V11) at (12,0) {};
        \node (V12) at (13.5,0) {$t$};
  
    \end{scope}
  
    \begin{scope}[>={stealth},
                   every node/.style={fill=white,circle},
                  every edge/.style={draw=black,very thick}]

        \path [->] (V4) edge node {$...$}  (V5);
        \path [->] (V5)  edge node {$...$} (V6);
        \path [->] (V6)  edge node {$...$}  (V7);
        \path [->] (V7)  edge node {$...$} (V8);
        \path [->] (V8)  edge node {$...$} (V9);
        \path [->] (V9)  edge node {$...$} (V10);
        \path [->] (V10)  edge node {$...$} (V11);
        \path [->] (V11)  edge node {$...$} (V12);

    \end{scope}
    \end{tikzpicture}
    \caption{The $P_k$ partition }
    \label{fig:pi_seperation}
\end{figure}

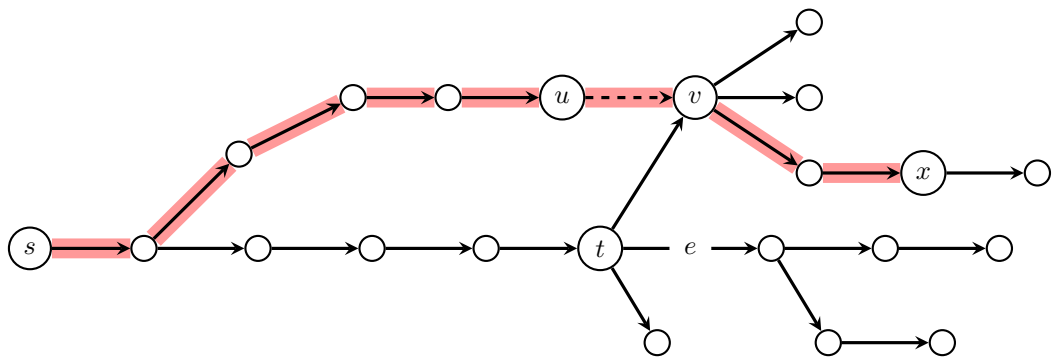
\begin{figure}[H]
    \centering
    \captionsetup{justification=centering}
    \begin{tikzpicture}
    \begin{scope}[every node/.style={circle,thick,draw}]
        \node (V41) at (-1.5,0) {${{s}}$} ;
        \node (V42) at (0,0) {} ;
        \node (V43) at (1.5,0) {};
        \node (V44) at (3,0) {};
        \node (V45) at (4.5,0) {};
        \node (V46) at (6,0) {$t$};
        \node (V47) at (8.25,0) {};
        \node (V48) at (9.75,0) {};
        \node (V49) at (11.25,0) {} ;

        \node (V21) at (8.75,1) {} ;
        \node (V22) at (10.25,1) {$x$} ;
        \node (V23) at (11.75,1) {} ;

        \node (V11) at (7.25,2) {$v$} ;
        \node (V12) at (8.75,2) {} ;

        \node (V01) at (8.75,3) {} ;

        \node (V51) at (1.25,1.25) {};
        \node (V52) at (2.75,2) {};
        \node (V53) at (4,2) {};
        \node (V54) at (5.5,2) {$u$};

        \node (V55) at (6.75,-1.25) {} ;
        \node (V56) at (9,-1.25) {} ;
        \node (V57) at (10.5,-1.25) {} ;

    \end{scope}

    \begin{scope}[>={stealth},
                  every edge/.style={draw=black,very thick}]

        \path [->] (V41) edge (V42);
        \path [->] (V42) edge (V43);
        \path [->] (V43) edge (V44);
        \path [->] (V44) edge  (V45);

        \path [->] (V45) edge (V46);
        \path [->] (V46) edge  node[style={fill=white,circle}] {$e$} (V47);
        \path [->] (V47) edge (V48);
        \path [->] (V48) edge (V49);

        \path [->] (V46) edge (V11);
        \path [->] (V11) edge (V21);
        \path [->] (V21) edge (V22);
        \path [->] (V22) edge (V23);
        \path [->] (V11) edge (V12);
        \path [->] (V11) edge (V01);
        \path [->] (V46) edge (V55);

        \path [->] (V47) edge (V56);
        \path [->] (V56) edge  (V57);

         \path [->] (V42) edge (V51);
         \path [->] (V51) edge  (V52);
         \path [->] (V52) edge (V53);
         \path [->] (V53) edge (V54);

        \path [->] (V54) edge[dashed] (V11);

    \begin{scope}[on background layer,every edge/.style=marked edge]

        \path (V41) edge (V42);
        \path (V42) edge (V51);
        \path (V51) edge (V52);
        \path (V52) edge (V53);
        \path (V53) edge (V54);
        \path [->] (V54) edge (V11);
        \path [->] (V11) edge (V21);
        \path [->] (V21) edge (V22);

    \end{scope}

    \end{scope}
    \end{tikzpicture}
    \caption{$R(s,x,H_w - e)$ is a "help from above" path}
    \label{fig:help_above}
\end{figure}
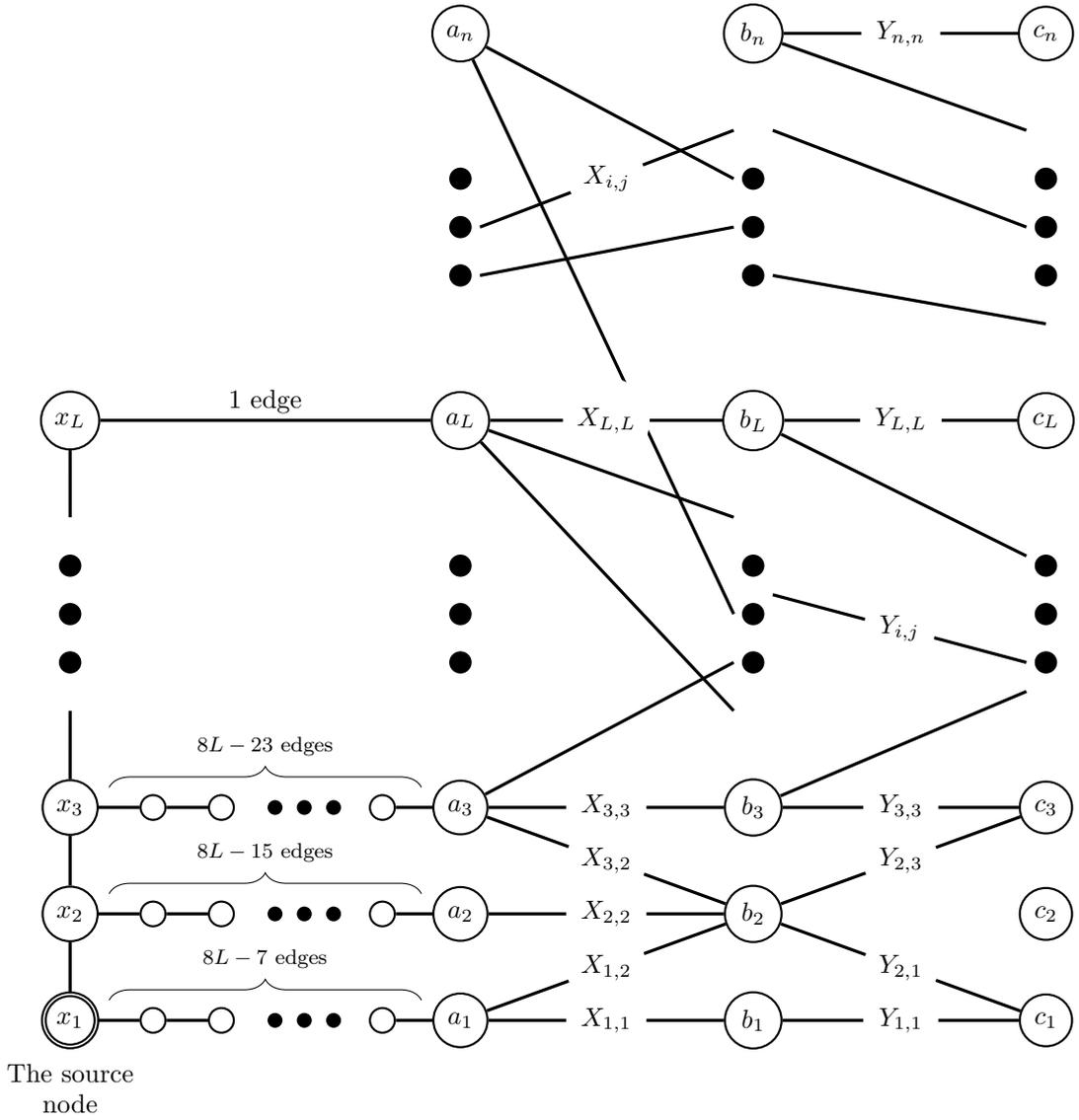
\begin{figure}[H]
    \centering
    \captionsetup{justification=centering}  
    \begin{tikzpicture}[scale=1.3]

    \begin{scope}[every node/.style={circle,thick,draw}]
    \node (x1) [draw=black,double=white, circle, inner sep=3pt] at (-1,1.8) {$x_1$};
    \node (x2) at (-1,2.9) {$x_2$};
    \node (x3) at (-1,4) {$x_3$};
    \node (xL) at (-1,8) {$x_L$};
    
    \node (a1) at (3,1.8) {$a_1$};
    \node (a2) at (3,2.9) {$a_2$};
    \node (a3) at (3,4) {$a_3$};
    \node (aL) at (3,8) {$a_L$};
    \node (an) at (3,12) {$a_n$};
    
    \node (b1) at (6,1.8) {$b_1$};
    \node (b2) at (6,2.9) {$b_2$};
    \node (b3) at (6,4) {$b_3$};
    \node (bL) at (6,8) {$b_L$};
    \node (bn) at (6,12) {$b_n$};
    
    \node (c1) at (9,1.8) {$c_1$};
    \node (c2) at (9,2.9) {$c_2$};
    \node (c3) at (9,4) {$c_3$};
    \node (cL) at (9,8) {$c_L$};
    \node (cn) at (9,12) {$c_n$};
    \end{scope}
    \begin{scope}[>={stealth},
        every edge/.style={draw=black,very thick}]
        \path  (3.2,10) edge  node[style={fill=white,circle}] {$X_{i,j}$} (5.8,11);
        \path  (3.2,9.5) edge  (5.8,10);
        \path  (an) edge  (5.8,10.5);
    
        \path  (an) edge (5.8,6);
        \path  (aL) edge (5.8,7);
        \path  (aL) edge  node[style={fill=white,circle}] {$X_{L,L}$} (bL);
        \path  (aL) edge (5.8,5);
        \path  (a3) edge  node[style={fill=white,circle}] {$X_{3,2}$} (b2);
        \path  (a2) edge  node[style={fill=white,circle}] {$X_{2,2}$} (b2);
        \path  (a1) edge  node[style={fill=white,circle}] {$X_{1,2}$} (b2);
        \path  (a1) edge  node[style={fill=white,circle}] {$X_{1,1}$} (b1);
        \path  (a3)  edge  node[style={fill=white,circle}] {$X_{3,3}$} (b3);
        \path  (a3)  edge (5.8,5.5);
    \end{scope}
    \begin{scope}[>={stealth},
        every edge/.style={draw=black,very thick}]
        \path  (6.2,11) edge  (8.8,10);
        \path  (6.2,9.5) edge (9,9);
        \path  (bn) edge  node[style={fill=white,circle}] {$Y_{n,n}$} (cn);
        \path  (bn) edge  (8.8,11);
        
        \path  (bL) edge  (8.8,6.6);
        \path  (bL) edge  node[style={fill=white,circle}] {$Y_{L,L}$} (cL);
        \path  (6.2,6.2) edge  node[style={fill=white,circle}] {$Y_{i,j}$} (8.8,5.5);
        \path  (b3)  edge  (8.8,5.2);

        \path  (b3) edge  node[style={fill=white,circle}] {$Y_{3,3}$} (c3);
        \path  (b2) edge  node[style={fill=white,circle}] {$Y_{2,3}$} (c3);
        \path  (b2) edge  node[style={fill=white,circle}] {$Y_{2,1}$} (c1);
        \path  (b1) edge  node[style={fill=white,circle}] {$Y_{1,1}$} (c1);
    \end{scope}
    
    \begin{scope}[every node/.style={circle,fill,inner sep=3pt}]
        \node at (-1,5.5){};
        \node at (-1,6){};
        \node at (-1,6.5){};
        \node at (3,5.5){};
        \node at (3,6){};
        \node at (3,6.5){};
        \node at (3,9.5){};
        \node at (3,10){};
        \node at (3,10.5){};

        \node at (6,5.5){};
        \node at (6,6){};
        \node at (6,6.5){};
        \node at (6,9.5){};
        \node at (6,10){};
        \node at (6,10.5){};

        \node at (9,5.5){};
        \node at (9,6){};
        \node at (9,6.5){};
        \node at (9,9.5){};
        \node at (9,10){};
        \node at (9,10.5){};
    \end{scope}
    \begin{scope}[>={stealth},
        every edge/.style={draw=black,very thick}]
    
        \path  (x1) edge (x2);
        \path  (x2) edge (x3);
        \path  (x3) edge  (-1,5);
        \path  (-1,7) edge (xL);
    \end{scope}

    \begin{scope}
        \node[align=center] at (-1,1.1) {The source\\ node};
        \draw [decorate,decoration={brace,amplitude=8pt,raise=4pt},yshift=0pt]

    (-0.6,1.9) -- (2.6,1.9) node [black,midway,yshift=0.7cm] {\footnotesize $8L - 7 \text{ edges}$};
        \draw [decorate,decoration={brace,amplitude=8pt,raise=4pt},yshift=0pt]
    (-0.6,3.0) -- (2.6,3.0) node [black,midway,yshift=0.7cm] {\footnotesize $8L - 15\text{ edges}$};

        \draw [decorate,decoration={brace,amplitude=8pt,raise=4pt},yshift=0pt]
    (-0.6,4.1) -- (2.6,4.1) node [black,midway,yshift=0.7cm] {\footnotesize $8L - 23 \text{ edges}$};
        
    \node [align=center] at (1,8.2){$1 \text{ edge}$};
    \end{scope}
    \begin{scope}[every node/.style={circle,thick,draw}]
        \node (x11) at (-0.15,1.8) {};
        \node (x12) at (0.55,1.8) {};
        \node (x13) at (2.2,1.8) {};

        \node (x21) at (-0.15,2.9) {};
        \node (x22) at (0.55,2.9) {};
        \node (x23) at (2.2,2.9) {};

        \node (x31) at (-0.15,4) {};
        \node (x32) at (0.55,4) {};
        \node (x33) at (2.2,4) {};
    \end{scope}
    \begin{scope}[every node/.style={circle,fill,inner sep=2pt}]
        \node at (1.1,1.8){};
        \node at (1.4,1.8){};
        \node at (1.7,1.8){};

        \node at (1.1,2.9){};
        \node at (1.4,2.9){};
        \node at (1.7,2.9){};

        \node at (1.1,4){};
        \node at (1.4,4){};
        \node at (1.7,4){};
    \end{scope}
    \begin{scope}[>={stealth},
        every edge/.style={draw=black,very thick}]
        \path  (x1) edge  (x11);
        \path  (x11) edge  (x12);
        \path  (x13) edge  (a1);

        \path  (x2) edge  (x21);
        \path  (x21) edge  (x22);
        \path  (x23) edge  (a2);

        \path  (x3) edge  (x31);
        \path  (x31) edge  (x32);
        \path  (x33) edge  (a3);

        \path  (xL) edge  (aL);
    \end{scope}
\end{tikzpicture}

\caption{The reduction from min-plus product to rationally weighted SSRP. Using a single invocation of SSRP
from the source $x_1$
over a graph with $O(n)$ vertices and $O(m)$ edges,
we can compute $Z_{i,j}$ for every $1 \le i < L$ and $1 \le j \le n$ 
(where $L = \sqrt{n} + 1$)
}
\label{fig:reduction}
\end{figure}

%% file: sections/DEPART_proofs.tex
\newpage
\subsection*{The \mathtitlewrapper{$\depart e x$} value - Proof of Correctness}
\subsubsection*{Proof of Completeness}

We first prove the following claim
\begin{claim}[Proof of Completness for Pivots]
    \label{claim:compleness_depart_eb}
    Consider some $k \in [\lgV]$.
    Let $b \in B_k$ be a pivot node and let $e \in P_k$
    be an edge failure, then $\depart e b \ge \D {} s b e$
\end{claim}
\begin{proof}
    The proof is rather trivial.
    Recall that by definition (Step \ref{step:depart})
    we have that $\depart{e}{b} =  \min\limits_{u\in V(P) \text{ before } e}\{$
    $\Dnm s u + \D {} u b P\}$.
    Let $u \in V(P)$ be that vertex such that $u$ is before $e$ in $P$ that minimizes
    the expression $\Dnm s u + \D {} u b P$.
    \snl
    Note that since $u$ is before $e$
    the subpath of $P$ from $s$ to $u$ (denoted $P(s,u)$) does not contain $e$.
    Note also that since $e \in P$,
    the shortest \RP{$u$}{$b$}{$P$}{$H$}
    (denoted $\R {} u b P$) does not contain $e$.
    And so $P(s,u) \circ \R {} u b P $ is a \RP{$s$}{$b$}{$e$}{$H$}
    of length  $\Dnm s u + \D {} u b P = \depart{e}{b}$.
    This implies that
    $\depart{e}{b} \ge \D {} s b e$.
\end{proof}
We now turn to prove the completeness for non-pivot nodes as well. 
\completeDEPART*
\begin{proof}
    Let $k \in [\lgV] \cup \{0\}$ be the unique integer such that $e \in P_k$.
    Note that $k$ exists and is unique since the family $\{P_k\}_{[\lgV] \cup \{0\}}$ is an edge disjoint partition of $P$.
    If $k = 0$ then
    $\depart e x = \D {} s x e$ by definition and so we are done.
    Otherwise $k \ne 0$ and so
    $\depart e x = \depart e b + \D {} b x P$ for some $b \in B_k$.
    By Claim \ref{claim:compleness_depart_eb} we have that 
    $\depart e b \ge \D {} s b e$ and 
    since $e \in P$ we have that $\D {} b x P \ge \D {} b x e$,
    so we get that $\depart e x \ge  \D {} s b e + \D {} b x e$.
    By the triangle inequality we have that 
    $\D {} s b e + \D {} b x e \ge \D {} s x e$ 
    and so $\depart e x \ge  \D {} s x e$.
\end{proof}

\subsubsection*{Proof of Soundness}
\soundDEPART*
An illustration for this case can be seen in Figure \ref{fig:departing}
\begin{proof}
    Note that since $\R w s x e$ is an unweighted replacement path,
    its length is $\D {} s x e$.
    If $e \in P_0$ we have by the definition of $\depart e x$
    that $\depart e x = \D {} s x e$ and so the claim holds.
    \snl
    The more interesting case would be where $e \in P_k$ for some $k \in [\lgV]$.
    Note that $\R w s x e$ must leave the path $P$ at some node $u \in P$ \textbf{before} $e$.
    Let us denote by $R[u,x]$ the subpath of $\R w s x e$ from $u$ to $x$,
    and let us denote by $R[s,u]$ the subpath of $\R w s x e$ from $s$ to $u$.
    \snl
    Note that since $u$ is \textbf{before} $e$ in $P$ the path from $s$ to $u$ in the BFS tree $K$
    does not contain $e$.
    Since we assume $\R w s x e$ is \simple\ this implies that $R[s,u]$ is the subpath of $P$
    from $s$ to $u$. In particular its length is $\Dnm s u$ and it is fully contained in $S$.
    \NL 
    Since $v_r \in V(T) - \{t\}$ we have that $v_r \notin S$.
    And so it must be the case that $v_r \in R[u,x]$.
    Since $v_r \in R[u,x]$ its length is at least $\Dnm u {v_r}$.
    Note that since $u \in P$ and $v_r \in T$ we have that $u$ is an ancestor of $t$
    which is an ancestor of $v_r$ in the BFS tree $K$.
    And so $d(R[u,x]) \ge \Dnm u {v_r} = \Dnm u t + \Dnm t {v_r} \ge \Dnm u t$.
    Since $u$ is before $e$ in $P$, and since $e \in P_k$ the distance from $u$ to $t$
    is at least $2^k \sqrtV$.
    So the length of $R[u,x]$ is at least $2^k \sqrtV$,
    and since $R[u,x]$ is unweighted it must contain at least $2^k \sqrtV$ vertices.
    And so by the sampling Lemma 
    \ref{lemma:sampling},
    \whp\ we have sampled some node $b \in B_k$ such that $b \in R[u,x]$.
    \snl
    Note that since $\R w s x e$ is departing it will not return to $P$ after it leaves it at the node $u$.
    In other words $V(R[u,x]) \cap V(P) = \{u\}$,
    and so $E(R[u,x]) \cap E(P) = \varnothing$.
    Since $\R w s x e$ is unweighted 
    this implies that $R[u,x]$ is a shortest path from $u$ to $x$ in $H-P$.
    Since we have shown that $b \in R[u,x]$ we have that 
    the length of $R[u,x]$ is $\D {} u b P + \D {} b x P$.
    \NL
    To conclude, the length of $\R w s x e$ is exactly 
    $\Dnm s u + \D {} u b P +  \D {} b x P$.
    By the definition of $\depart e b$  we have that 
    $\depart e b \le \Dnm s u + \D {} u b P$ and by the definition of 
    $\depart e x$ we have that $\depart e x \le \depart e b + \D {} b x P$.
    So we get that the length of $\R w s x e$ is \whp\ at least $\depart e x$.
\end{proof}

%% file: sections/DSTE_proofs.tex
\subsection*{The case when 
\mathtitlewrapper{$e \in P$} and 
\mathtitlewrapper{$x = t$}  - Proof of correctness
}
Recall that by the definition of $\Dhat w s t e$
in step \ref{step:DSTE} of the algorithm
(line \ref{lst:line:DSTE_set} in the pseudocode),
$\Dhat w s t e$ is set to be the minimum between 
$A_w[e]$ and $\DRZ s t e$.
Where $\DRZ {{s}} t e$ is the distance
estimation obtained by the RP algorithm from \cite{Roditty2005}
and $A_w[e] = \min\limits_{u \text{ is after } e \text{ in } P} \{ {\D w {{s}} u P} + \Dnm u t \}$.

\subsubsection*{Proof of Completeness}

\completeDSTE*
\begin{proof}
    We wish to prove that each one of the two values $\Dhat wste$ can receive 
    ($A_w[e]$ and $\DRZ ste$) is
    at least $\D wste$.
    This will suffice to show that $\Dhat wste \ge \D wste$.
    \NL
    Recall that  $\DRZ ste$ is the result obtained by the
    algorithm from \cite{Roditty2005},
    over the unweighted directed graph $H$ and the edge failure $e$.
    By the one-sided error property
    of the algorithm from \cite{Roditty2005}
    we have that $\DRZ ste \ge d(s,t,H-e)$,
    and since $H - e \subseteq H_w - e$ we get that $ d(s,t,H-e) \ge  d(s,t,H_w-e)$,
    and so $\DRZ ste \ge d(s,t,H_w - e)$ as required.
    \NL
    We now handle the term 
    $A_w[e]$.
    Let $u \in V(P)$ be some vertex such that $u \text{ is after } e \text{ in } P$.
    Note that since $u$ is after $e$ in $P$,
    the subpath of $P$ from $u$ to $t$ (denoted by $P(u,t)$) does not contain $e$.
    Since $P$ is a shortest path in $H$ we have
    that the length of $P(u,t)$ is $d(u,t,H)$.
    \snl
    Since $e \in P$ we can by concatenate the shortest path from $s$ to $u$ in the graph $H_w-P$ 
    with the path $P(u,t)$ and get a \RP{$s$}{$t$}{$e$}{$H_w$}
    of length $d(s,u,H_w-P) + d(u,t,H)$.
    This implies that 
    $d(s,u,H_w-P) + d(u,t,H) \ge d(s,t,H_w-e)$
    which implies in particular that
    $A_w[e]
    \ge
    d(s,t,H_w-e)$ as required.
\end{proof}

\subsubsection*{Proof of Soundness}
\begin{claim}
    \label{claim:sound_DSTE_weighted}
    Let $e \in E(P), w \in W$, 
    if $\R w s t e$ is weighted 
    then its length is \textbf{at least}
    $A_w[e]$
\end{claim}
\begin{proof}
    The proof for this case is rather simple.
    Let $u$
    be the \textbf{first} node in $\R w {{s}} t e$ such that $u \in P$ and $u$ is \textbf{after} $e$.
    Note that $u$ exists since $t \in P$ and $t$ is \textbf{after} $e$.
    \NL
    Let $R[s,u]$ denote the subpath of $\R w {{s}} t e$ from $s$ to $u$.
    We claim that $V(R[s,u]) \cap P = \{s,u\}$.
    To see this assume for the sake of contradiction there is some node $v \in R[s,u]$
    such that $v \in P -\{s,u\}$.
    If $v$ is \textbf{after} $e$, then $u$ is not the first node in $\R w {{s}} t e$ which is 
    \textbf{after} $e$, contradiction.
    If $v$ is \textbf{before} $e$, then the path from $s$ to $v$ in $K$
    does not contain $e$. Since we assume $\R w {{s}} t e$ is \simple\
    this means that the subpath of $\R w {{s}} t e$ from $s$ to $v$ is contained 
    in $K$. However since $\R w {{s}} t e$ is weighted the very first edge in it 
    is from $E(H_w) - E(H)$ and so is not from $K$, contradiction.
    We conclude that $V(R[s,u]) \cap P = \{s,u\}$,
    meaning that $E(R[s,u]) \cap P \subseteq (s,u)$.
    However if $(s,u) \in R[s,u]$ then $(s,u)$ is the first edge in  $\R w {{s}} t e$,
    and so (since $\R w {{s}} t e$ is weighted) $(s,u) \in E(H_w) - E(H)$,
    meaning in particular that $(s,u) \notin P$. 
    So we can conclude that $E(R[s,u]) \cap P = \varnothing$. Since $R[s,u]$ is a shortest
    path from $s$ to $u$ in $H_w-e$ and $e \in P$ 
    we can conclude that the length of 
    $R[s,u]$ is exactly $\D w s u P$.
    \NL
    Let $R[u,t]$ denote the subpath of $\R w {{s}} t e$ from $u$ to $t$.
    Its length is $\D w u t e$ which is at least $\Dnm [H_w] u t$
    which is equal to (by the weight requirements)  $\Dnm u t$.
    So the we can conclude that the length of $\R w {{s}} t e$ is at least
    $\D w s u P + \Dnm u t$.
    Since $A_w[e] = \min\limits_{u \text{ is after } e \text{ in } P}
    \{d(s,u, H_w - P) + \Dnm u t\}$ this implies the claim.
\end{proof}

\soundDSTE*
\begin{proof}
    If $\R w s x e$ is weighted then by Claim \ref{claim:sound_DSTE_weighted}
    we have that its length is at least  $A_w[e]$ meaning that its length is at least $\Dhat w s t e$.
    Otherwise $\R w s x e$ is unweighted, and so $\R w s x e$ 
    is a replacement \RP{$s$}{$x$}{$e$}{$H$}.
    So by the proof of correctness of the algorithm from \cite{Roditty2005}
    we have that the length of $\R w s x e$ is equal to $\DRZ s x e$ \whp.
    Meaning that \whp\ the length of $\R w s x e$ is at least $\Dhat w s t e$.

\end{proof}